\documentclass[useAMS,usenatbib]{mn2e}
\usepackage{graphicx}
\usepackage{natbib}

\title[Footprints of Sgr in the Galaxy]{Footprints of the Sagittarius dwarf galaxy in the {\it Gaia} data set}
\author[Laporte et al.]{
\parbox[t]{\textwidth}{Chervin F. P. Laporte,$^{1}$\thanks{E-mail:cfpl@uvic.ca}\thanks{CITA National Fellow} Ivan Minchev,$^{2}$ Kathryn V. Johnston,$^{3}$, Facundo A. G\'omez$^{4,5}$\\
}
\\
$^{1}$Department of Physics \& Astronomy, University of Victoria, 3800 Finnerty Road, Victoria BC, Canada V8P 5C2\\
$^{2}$ Leibniz-Institut f\"ur Astrophysik Potsdam (AIP), An der Sternwarte 16, 14482 Potsdam, Germany\\
$^{3}$ Department of Astronomy, Columbia University, 550 West 120th Street, New York, NY, 10027, U.S.A\\
$^{4}$Instituto de Investigaci\'on Multidisciplinar en Ciencia y Tecnolog\'ia, Universidad de La Serena, Ra\'ul Bitr\'an 1305, La Serena, Chile\\
$^{5}$Departamento de F\'isica y Astronom\'ia, Universidad de La Serena, Norte, Av. Juan Cisternas 1200, La Serena, Chile\\
}
\date{Accepted . Received ; in original form }
\pubyear{2018}
\begin{document}
\label{firstpage}
\pagerange{\pageref{firstpage}--\pageref{lastpage}}
\maketitle

\label{firstpage}
\begin{abstract}

We analyse an N-body simulation of the interaction of the Milky Way (MW) with a Sagittarius-like dSph (Sgr), looking for signatures which may be attributed to its orbital history in the phase space volume around the Sun in light of {\it Gaia} DR2 discoveries. The repeated impacts of Sgr excite coupled vertical and radial oscillations in the disc which qualitatively, and to a large degree quantitatively are able to reproduce many features in the 6D {\it Gaia} DR2 samples, from the median $V_{R},V_{\phi}, V_{z} $ velocity maps to the local $\delta\rho(v_{z},z)$ phase-space spiral which is a manifestation of the global disc response to coupled oscillations within a given volume. The patterns in the {\it large-scale} velocity field are well described by tightly wound spirals and vertical corrugations excited from Sgr's impacts. We show that the last pericentric passage of Sgr resets the formation of the {\it local} present-day $\delta\rho(v_z,z)$ spiral and situate its formation around 500-800 Myr. As expected $\delta\rho(vz,z)$ grows in size and decreases in woundedness as a function of radius in both the {\it Gaia} DR2 data and simulations. This is the first N-body model able to explain so many of the features in the data on different scales. We demonstrate how to use the full extent of the Galactic disc to date perturbations dating from Myr to Gyr, probe the underlying potential and constrain the mass-loss history of Sgr. $\delta\rho(vz,z)$ looks the same in all stellar populations age bins down to the youngest ages which rules out a bar buckling origin.

\end{abstract}

\section{Introduction}

Gaia's second data release has brilliantly revealed the richness of substructure in our Galactic disc in unprecedented and unforeseen ways \citep{katz18,antoja18}. The existence of substructure in the form of moving groups of stars around the Sun was first pointed out many decades ago and interpreted as a signature of disrupted clusters \citep{eggen69}. Analysis of  the {\it Hipparcos} data \citep{dehnen98} offered additional insight that such groups could be excited by asymmetries in the disc, such as the bar or spiral arms \citep{dehnen00, quillen05}.

However, other out-of-equilibrium features might be excited through external agents as well, such as mergers, which produce ringing in the $U-V$ plane \citep{minchev09}. These arching features become clearly visible in the space of integrals $E-Lz$, \footnote{These can be loosely understood as integrals on intuitive grounds despite the fact Milky Way potential is non-axisymmetric and time-dependent.} around solar neighbourhood-like (SNhd) regions \citep{gomez12a}. We also note that recent measurements of asymmetries in the $U-V$ plane as a function of height, such as for the Coma Berenices seen in GALAH \citep{quillen18a} and in RAVE \citep{monari18} have also indicated another hint for a merger origin of this moving group. The Sagittarius dwarf galaxy \citep{ibata94}, has long been suggested to be a potential perturber to the Milky Way Galactic disc \citep{ibata98, dehnen98,bailin04}. Indeed, using test particle simulations \cite{quillen09} showed that a Sgr-like satellite could induce perturbations to a disc in the form of a warp, spiral structure as well as streams in the $U-V$ plane. However, only recently has it been possible to assess the impacts of Sagittarius on the Galactic disc  with high-resolution live N-body simulations \citep{purcell11,gomez12,gomez13,laporte18a,laporte18b}. These are necessary in order to study the excitation of vertical density waves which are known to be sustained by self-gravitating discs \citep{weinberg1991}, an aspect which cannot be captured by simple phase-mixing models of tracers subject to impulsive perturbations.

Direct first evidence of ringing in the Milky Way using the \cite{schuster06} and SEGUE F/G dwarf stellar sample was measured by \cite{gomez12} who attributed the signal to a passage with the Sgr dSph. Co-incidentally, signatures of vertical oscillations in the disc became apparent as local asymmetries in density and velocity seen in spectroscopic surveys \citep{widrow12,williams13,carlin13,schonerich18,carrillo18} and large scale arc-like features above and below the disc plane in the outer disc \citep{newberg02,crane03,grillmair06,martin07,sharma10}.

Since then, \cite{antoja18} used the Gaia second data release to uncover a series of substructures in and around the solar neighbourhood, confirming unambiguously that the galaxy is out-of-equilibrium and in a middle of an ongoing phase-mixing process. The authors used a toy model from \cite{minchev09} to estimate the time of impact/disturbance in the solar neighbourhood finding timescales coincident with Sgr's orbital timescales from its last pericentric passage \citep{laporte18b}. This was presented as evidence that the source of perturbation could have been related to Sgr.

In this paper, we explore whether any of the various features in the recently reported substructures in velocity space of {\it Gaia} DR2 \citep{antoja18} can be attributed as clear signals to Sgr's orbital history. We do so, using simulations from the suite of numerical N-body experiments in \citep{laporte18b}. These simulations follow the orbit of Sgr-like galaxies around a Milky Way-like host from the first pericentric passage to the present-day. These were the first sets of numerical experiments to successfully reproduce the excitation of the Monoceros Ring and more distant stellar overdensities while producing vertical fluctuations in the solar neighbourhood in good agreement with the constraints set by \cite{widrow12} (see section 2.2 for more details).

This paper is organised as follow. In section 2, we describe the simulations used in this contribution. In section 3 we present maps of velocity in regions around the Sun which we compare to the recent results of \cite{katz18} and what these may reflect in the Galaxy. In section 4, we compare models to the recent results of \citep{antoja18}, \cite{monari18} on substructure in the solar neighbourhood. In particular, we look for signs of substructure and phase-mixing which can be related to the Sgr dwarf galaxy. In section 5, we discuss why the Sgr scenario is currently the only one supported by data spanning 10 scale lengths of the disc ruling out alternative models. In section 6, we review various successes of our models and highlight a number of predictions which can be explored with current capabilities beyond Galactic structure/kinematics studies (e.g. through the CMD diagram) as well with future observational campaigns to probe larger volumes of the disc's phase-space. We end the discussion by highlighting a number of ways by which the models could be refined further in the future.

\section{Prior work}
\subsection{Numerical methods}

The simulations used in this paper are taken from \cite{laporte18b}. These were a series of live N-body experiments of the interaction Sgr with the Galactic disc as well as a set combined with the effect of the LMC on a first infall orbit \citep{besla07} for which separate N-body simulations were also presented in \cite{laporte18a}. We note that these experiments significantly improved on earlier works on the impact of Sgr on the disc \citep[e.g.][]{purcell11}. Indeed, all prior works used initial conditions which were too simplistic ignoring the effect of earlier passages of Sgr in exciting vertical perturbations to the disc through the dark matter halo wake \citep{weinberg89,weinberg98} and thus were unable to reproduce outer disc structures similar to the Monoceros Ring. One exception to this, is the live N-body simulations of \citet[][DL17]{dierickx17} and hydrodynamical simulations of \citet[][using the ICs derived by DL17]{tepper18} which also considered the full orbit of a low-mass Sgr progenitor from the virial radius to its present-day location, but these authors focused solely on the properties of the stellar remnant stream or fate of the gas locked in the progenitor (respectively), not the reaction of the disc.

Here, we give a brief description of our simulations suite, but the interested reader is referred to relevant contributions for more details. In these runs, the host MW had the following properties: a dark halo of $M_{h} = 10^{12} \,\rm{M_{\odot}}$, an exponential disc of $M_{disc} = 6\times10^{10}\,\rm{M_{\odot}}$, with a scale length of $R_{d} = 3.5 \,\rm{kpc}$ and scale height of $h_{d} = 0.53\, \rm{kpc}$ and a central bulge with a mass of $M_{bulge} = 10^{10}\, \rm{M_{\odot}}$. The mass model adiabatically contracted following \cite{blumenthal86}, steepening the dark matter profile to lead to a final mass model with a circular velocity of $V_{circ} = 239 \rm{km\,s^{-1}}$ at $R_{0} = 8\,\rm{kpc}$. Four different models for Sgr are considered with varying masses ($M\sim6\times10^{10}-10^{11}\rm{M_{\odot}}$) and concentrations within the scatter of the mass-concentration relations \citep{gao08,ludlow14}. The simulations end with a final remnants masses in agreement with current estimates \citep{frinchaboy12} producing streams in agreement with the M-giants from \cite{majewski03} and orbital timescales for the Sgr dwarf varying between $t_{orb}\sim0.4\,\rm{Gyr}$ and $t_{orb}\sim1.0\,\rm{Gyr}$ in the last stages. The final disc properties do not significantly change  except for the fact they are no longer axisymmetric, for which we review some of their properties below.

\subsection{Comparison of simulations and prior observations}

\begin{figure*}
\includegraphics[width=1.0\textwidth,trim=89mm 0mm 170mm 0mm, clip]{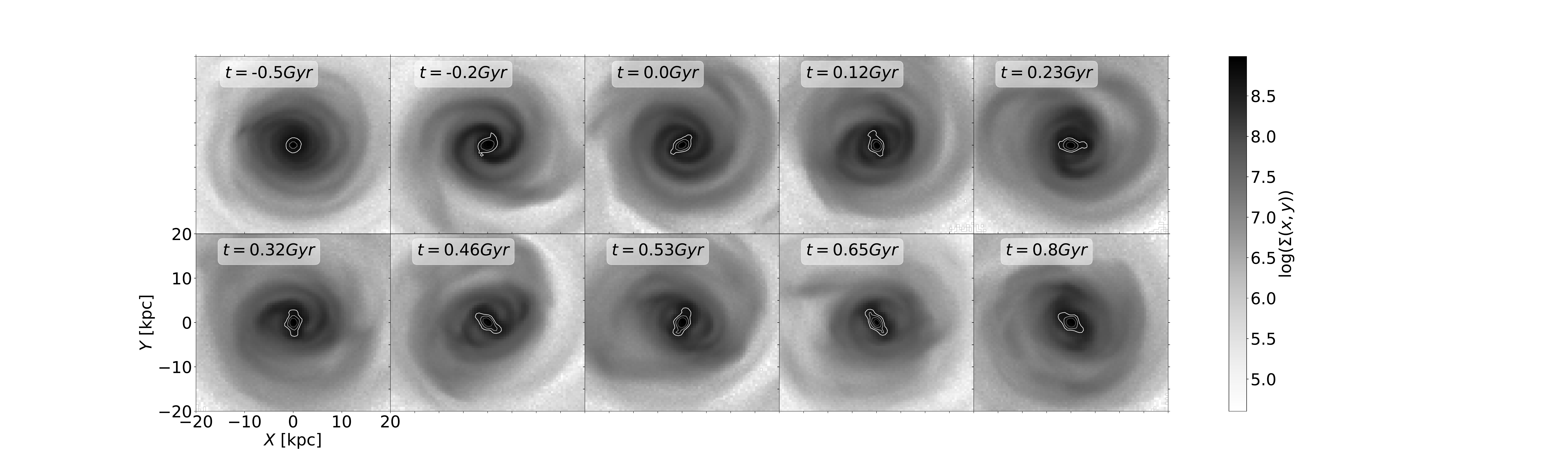}
\caption[]{Surface density evolution of the Galactic disc during the last stage of the Sgr impact on the disc. The last two pericentric passages of Sgr occur at $t\sim-0.5\,\rm{Gyr}$ and $t\sim0.0\,\rm{Gyr}$ which seed the formation of a strong bi-symmetric spiral as well as fast bar with pattern speed $\Omega_{b}=\omega/m\sim65\,\rm{km\,s^{-1}/kpc}$ which we delineate with density contours. Sgr's present-day position in the L2 model is reached at $t\sim0.46\, \rm{Gyr}$ but due to the uncertainty in the orbital period of Sgr \citep[see][]{johnston05,penarrubia10}, we follow the response of the disc for another $\Delta\sim0.34\,\rm{Gyr}$ window.}
\end{figure*}

\begin{figure}
\includegraphics[width=0.5\textwidth,trim=0mm 0mm 0mm 0mm, clip]{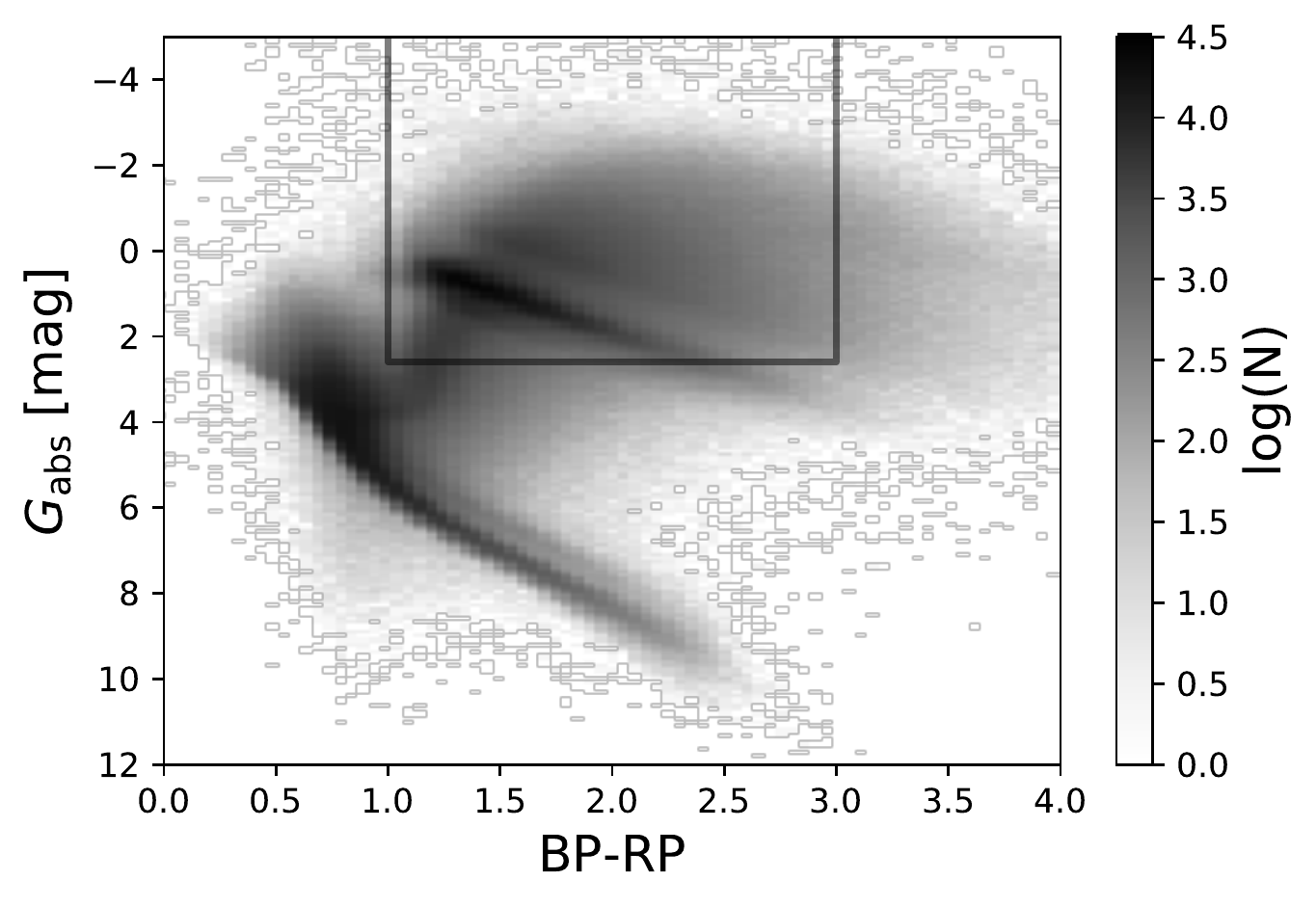}
\caption[]{\bf {\it Gaia} DR2 absolute non-derredened colour-magnitude diagram using parallax distances from the catalog of \cite{bailerjones18}. Our selection of giants for the construction of equivalent Gaia 6D solution velocity field maps is bounded by black box selecting stars with $M_{G}<2.6$ with colours $1<\rm{BP-RP}<3$.}
\end{figure}

\begin{figure*}
\includegraphics[width=1.0\textwidth,trim=70mm 0mm 60mm 0mm, clip]{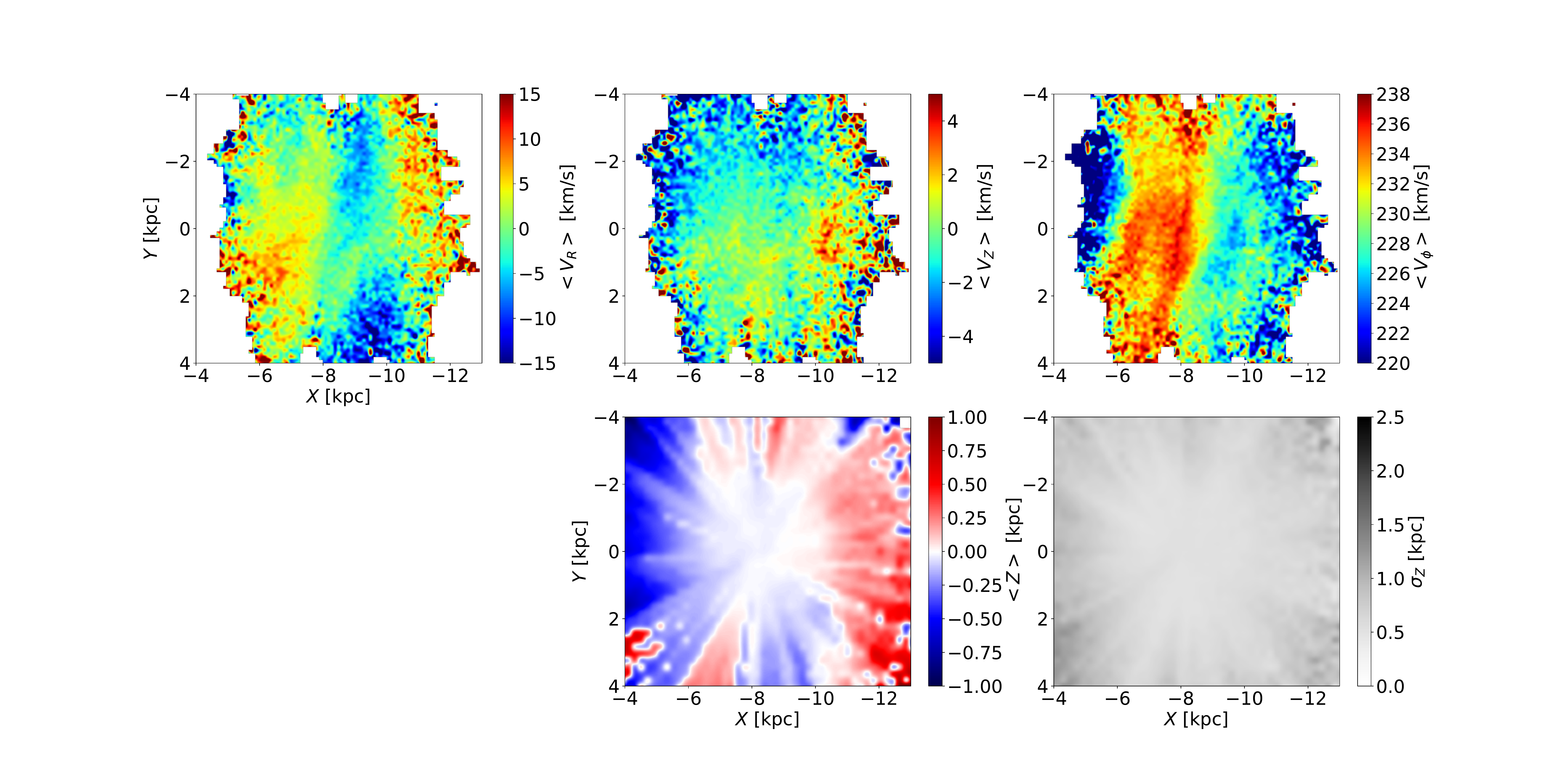}
\caption[]{\bf {\it Gaia} DR2 6D solution median $V_{R}$, $V_{Z}$, $V_{\phi}$ kinematic maps using giants with $M_{G}<2.6$  and $1<$BP-RP$<3$. The maps reproduce {\it all} the same features as those reported in \citep{katz18} but using the revised distances provided by \cite{bailerjones18}.}
\end{figure*}

\begin{figure*}
\includegraphics[width=1.0\textwidth,trim=0mm 0mm 0mm 0mm, clip]{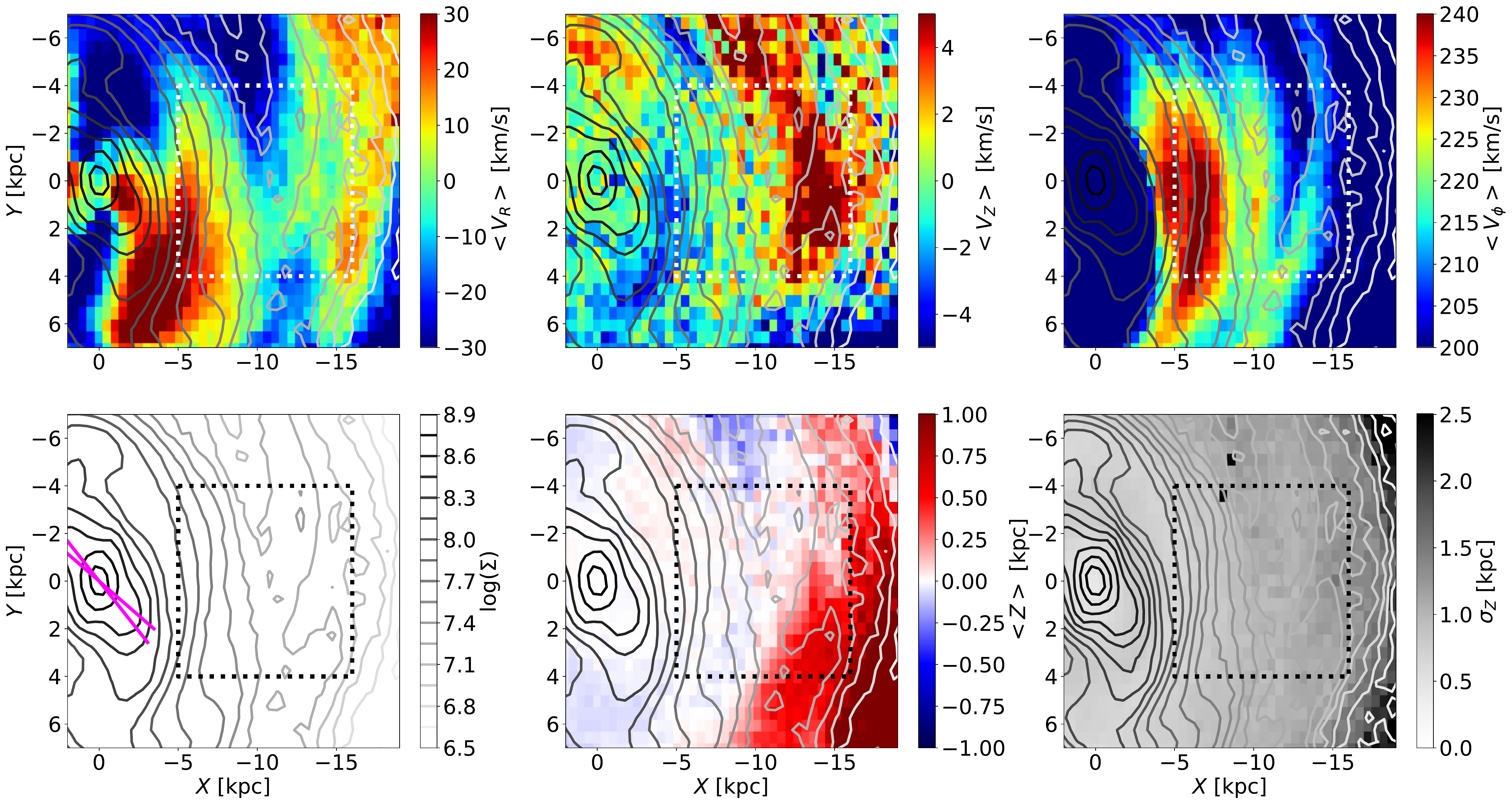}
\caption[]{{\it Top panels}: Simulated L2 model 2D $V_{R}$, $V_{Z}$, $V_{\phi}$ maps in the left, middle and right panels respectively, with density contours overlaid (greyscale lines). The square region corresponds to a spatial section similar to that in Katz et al. (2018) and Poggio et al. (2018). The reported velocity pattern from spiral arms can be clearly seen, albeit with differing amplitudes but with similar wavelengths to the observations. The disc is also warped in a similar orientation to the data. {\it Bottom panels:} Simulated L2 model 2D Surface density contours, median $Z$ and $\sigma_{Z}$ maps in the the left, middle and right panels respectively. The magenta lines mark the lines of 30,40 degrees respectively along which the bar is aligned. The disc shows both signs of flaring and warping (as expected from the complementary $V_{Z}$ map in the top middle panel).}
\end{figure*}

Some of the various quantitative and qualitative successes of the runs in \cite{laporte18b} with observations prior to the {\it Gaia} revolution are listed below:

\begin{itemize}

\item Vertical density and velocity perturbations \citep{gomez13} in the solar neighbourhood in agreement with constraints from \cite{widrow12,williams13}. These being absolute relative changes of 10 \% in the vertical number density and vertical streaming motions below $v_{z}<10 \,\rm{km\,s^{-1}}$.

\item A quantitative match to the morphology of the Monoceros Ring/Galactic Anticenter Stellar Stream \citep[GASS,][]{newberg02,crane03} in terms of stellar disc material kicked to low latitudes $|b|\sim 30^{\circ}$ and above.

\item A qualitative reproduction of distant stellar overdensities in the Anticenter in terms of location on the sky and distances such as A13 \citep{sharma10}, Triangulum Andromeda clouds \citep[TriAnd,][]{martin07} which have been confirmed to have populations and motions that match the stellar disc \citep{price-whelan15,xu15, li17,bergemann18,sheffield18}. 

\item The predicted amplitudes of vertical streaming motions in the outer disc also reproduced those measured in the observations \citep{deboer17,deason18} of $|v_{z}|\sim 50 \,\rm{km\,s^{-1}}$. 

\item  The formation of structures akin to the Anticenter Stream and Eastern Banded Structure \citep[][,ACS, EBS]{grillmair06,grillmair11} with similar lengths ($\Delta l \sim80-180^{\circ}$) and width $\delta\sim3^{\circ}-10^{\circ}$ \citep{laporte18c}. 

\item An overall Galactic flaring consistent with the outer disc structure and Cepheids measurements towards the Galactic Center by \cite{feast14} with $|Z|\sim2\,\rm{kpc}$. We also note in passing that since Gaia DR2, the models have been shown to trace remarkably well the flare of the outer disc as traced by the Blue Stragglers (BS) identified in the CFIS survey \citep{thomas19}.
\end{itemize}

We also identified models which were {\it not} able to reproduce the outer disc structures, yet were consistent with solar neighbourhood constraints\footnote{A significant difference to most studies was that we considered varying the central densities of the Sgr progenitors according to the mass-concentration relation \citep{gao08}.}. This was notably the case of model L1. Based on these earlier findings we argued that the progenitor of Sgr must have had mass $M_{Sgr}>6\times10^{10}\,\rm{M_{\odot}}$ at infall. A fundamental prediction of these models was that they could simultaneously explain the amplitudes of vertical density fluctuations and vertical streaming motions in the solar neighbourhood as well as outer disc structures (e.g. Monoceros, TriAnd amongst others). The hope was that Gaia would bring new constraints for the models and confirm those predictions on intermediate scales beyond the Sun. This has now been fulfilled with its second data release.

With the recent discoveries and updates on many non-axisymmetric features in and around the solar neighbourhood, it becomes evident to use some of these models to interpret the signals in the data and see if any such features can be linked to the orbital mass loss history of Sgr and its imprint on the Galactic disc. This is the aim of this paper. Because these models were not run with the intent to fit observations of the disc's non-axisymmetries, we are mainly interested to study qualitative properties of the Galactic disc subject to its interaction with a Sgr-like object. While full quantitative matches would need to await more sophisticated numerical experiments, an exploration of the qualitative features imprinted on the disc subject to this recent interaction in the last Gyr is evidently timely for interpretative purposes and in order to inform us on the possible scales and role of the perturbations set by Sgr in giving the stellar disc its present structure/kinematics. For this reason, we allow ourselves to analyse the Galaxy in a more general fashion in the spirit of \cite{laporte18c}. If a qualitative match is found, it may still prove interesting to study what this could possibly tell us about the Galaxy in relation to its interaction with Sgr.

\subsection{Adopted analysis of the simulations: L2 model}
In this contribution, we focus on the L2 model which was presented in \cite{laporte18b}. Other more massive progenitor models produced results with amplitude in excess to those of the L2 model and are not presented here for clarity. Similarly, the L1 model was already ruled out in \cite{laporte18b} due to its inability to kick disc material beyond $|b|>15$ at $d\sim10\,\rm{kpc}$. This leaves us with one particularly pertinent realisation.

We also note that our suite of models varied masses and central densities (and as a result the orbits) for Sgr, its last pericentric passage was found to be anywhere between $0.4-1 \,\rm{Gyr}$ in the past. However, because of the simple symmetries imposed by live N-body models the stream and main body location and velocity cannot be exactly simultaneously matched \citep[see][]{dierickx17, laporte18b}. Accurate stellar stream fitting models generally find orbital timescales which are on the longer side for Sgr \citep[e.g.][]{johnston05,penarrubia10,law10} with $t_{\rm{orb}}\sim0.7-1 \,\rm{Gyr}$. In this contribution, we are interested in the recent {\it physical} response and evolution of the Galactic disc during the last pericentric impact of Sgr on the disc which seeds off as we shall see the last major disc perturbations which evolve unhindered. Indeed after the present-day snapshot in the L2 model, the simulated Sgr is stripped down to masses below $M\sim10^{9} \,\rm{M_{\odot}}$ and its later evolution does not affect the disc, which is why we can take the liberty of studying the response for a  longer period of time.

Given this uncertainty window in the orbital timescale of Sgr, we allow ourselves to search for a snapshot where the disc qualitatively matches the Gaia median velocity field with the additional constraint of having the bar facing at 30-40 degrees \citep{dehnen00,minchev07} without changing the orientation of the Sgr orbital plane w.r.t. the Sun. For the L2 model, this allows us to consider the state of the Galaxy within $\Delta t\sim500\,\rm{Myr}$ past the ``present-day'' snapshot $t\sim0.46\,\rm{Gyr}$, where we define $t=0$ as the time of the last pericentric passage. We find such a match at $t\sim0.8\,\rm{Gyr}$ which is in line with estimates of the orbital timescale of Sgr.

Figure 1 shows a series of surface density maps of the MW analogue subject to impacts with Sgr during the last two pericentric passages situated at $t\sim -0.5,\, 0.0 \,\rm{Gyr}$ respectively, and beyond. The orbital plane of Sgr is along the x-axis. We note that during its previous passages, Sgr has excited spiral structure which gradually winds up. During last two pericentric passages a strong bi-symmetric spiral is excited which quickly winds up and the disc goes bar unstable. Sgr reaches its present-day position by $t=0.46 \,\rm{Gyr}$.  At this point, Sgr has $M\sim10^{9} \,\rm{M_{\odot}}$ bound mass and is on its course to full disruption. Subsequent evolution of the disc in the presence of the heavily shredded and low-mass Sgr is thus not affected in any significant fashion. As such we can interpret the later evolution of the Galactic disc as physically pertinent to the response of the disc to a perturbation which was set off $\Delta t\sim800$ Myr ago. Our results comparing to the {\it Gaia} data are all taken from this particular snapshot, unless stated otherwise.

Accurate modeling of the Sgr stream and the disc perturbations set during the last impact are a topic for different studies which requires altogether a different set-up to our live N-body models which followed the influence of Sgr from its infall from the virial radius to the present-day under simplistic symmetry (spherical symmetry) which cannot provide perfect matches to the stream and remnant body final position and velocity \cite[e.g. see][]{dierickx17,laporte18b}. However, much can be already learned from those pre-Gaia DR2 simulations to interpret the recent Gaia DR2 observations.

\section{Response of the disc in the Gaia volume}

\subsection{The {\it Gaia} DR2 data velocity fields}

As part of the official DR2 release papers, \cite{katz18} presented a preliminary analysis of the kinematics of the disc as revealed by the 6-D {\it Gaia} solution in roughly a box bound by $-4<Y/\rm{kpc}<4$ and $-5<X/\rm{kpc}<12$. They noted a series of interesting features in the median $V_{R}(x,y)$, $V_{Z}(x,y)$ and $V_{\phi}(x,y)$ planes (see their Fig. 10).
\begin{itemize}
\item Median radial velocity variations of $-10<V_{R}/\rm{km\,s^{-1}}<10$ interpreted as the existence of successive intertwined spiral arms \citep{reid14} or the 2-spiral arm model from \cite{drimmel00}. Indeed, to a first order, such succession of inward and outward motion patterns can easily be accounted for by spiral arms as shown by the linear perturbation theory study of \cite{monari16} which shows remarkable agreement with the data. 
\item a warp revealed by the median vertical kinematical component of the disc. Increasing from $V_{Z}\sim0\,\rm{km \,s^{-1}}$ to $\sim4\,\rm{km \,s^{-1}}$ in the range $6<R/\rm{kpc}<12$ along the $0^{\circ}$ line in azimuth connecting the Sun to the Galactic Center. This warp was further put into evidence using both old and young stellar tracers \citep{poggio18}. 
\item Median azimuthal velocity variations in the midplane between $\sim 220\,\rm{km \, s^{-1}}$ and $\sim235\, \rm{km\,s^{-1}}$.

\item Median height variations of $|\Delta Z|\sim1\,\rm{kpc}$ in red clump stars across the disc.

\end{itemize}

For clarity and ease of comparison with our simulations, we reproduce similar maps using giants identified the 6D {\it Gaia} DR2 solution. To do this, we queried all stars with measured radial velocities and with parallax measurements $\varpi/\sigma_{\varpi}>5$. We use the distances derived by \citep{bailerjones18} and we transformed from ICRS to Galactocentric coordinates using the {\sc astropy} routines and taking the position of the sun at $R_{\odot}=8.34 \,\rm{kpc}$  and adopting a solar motion of $v=240 \,\rm{km\,s^{-1}}$ and peculiar motion of the Sun with respect to the Local Standard of Rest of $(U_{\odot}, V_{\odot}, W_{\odot})=(11.10, 12.24, 7.25)\,\rm{km\,s^{-1}}$  \citep{schoenrich10}. In Figure 2, we construct the absolute colour-magnitude diagram using and select giants through a cut of $M_{G}<2.6$ and $1<\rm{BP-RP}<3$.

In Figure 3, we present the resulting 2D median {\it Gaia} DR2 velocity maps in $V_{R}, V_{Z}, V_{\phi}$ as well as median height $<Z>$ and standard deviation $\sigma_{Z}$ maps. Although our selection of giants is somewhat different, close inspection and comparison of these non-dereddened maps shows that they are close to identical and capture all the features listed above and presented in \citep{katz18} and \citep{poggio18}.

\begin{figure*}
\includegraphics[width=1.0\textwidth,trim=30mm 0mm 30mm 0mm, clip]{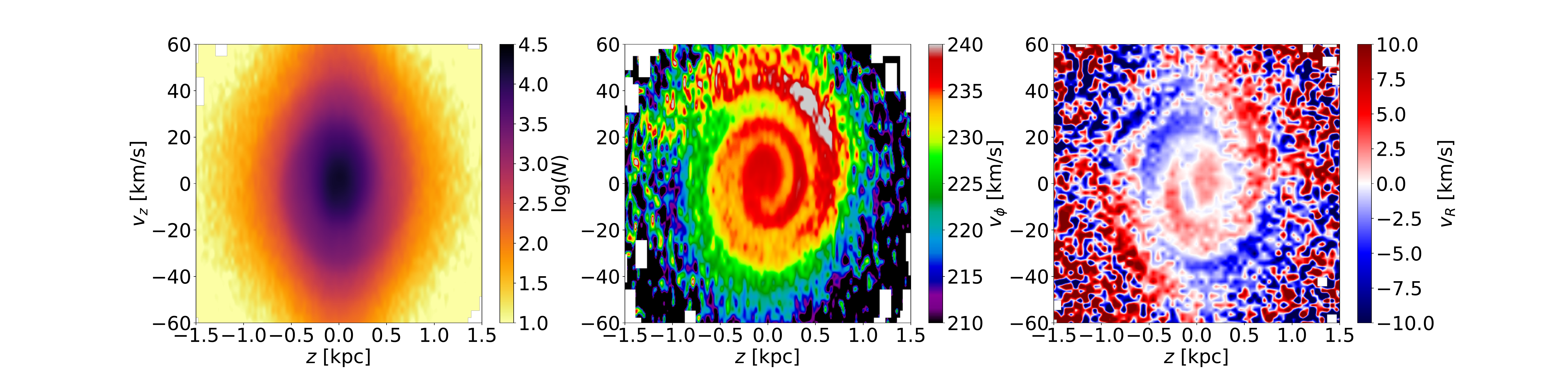}
\caption[]{{\it Gaia} DR2 phase-space spiral and its associated projections in $v_{\phi} (v_{z},z)$, $v_{R} (v_{z},z)$ within a 1 kpc sphere around the Sun. For the $v_{R} (v_{z},z)$ projection we make an additional selection cut on the absolute value of the radial velocities $|v_{R}|<40$.}
\end{figure*}

\subsection{Numerical simulations }

In this section, we compare the median velocity patterns seen in the $x-y$ plane of the {\it Gaia} DR2 data \citep{katz18} to the outcome of the Galactic disc being perturbed by the Sgr, to see if any reasonable match can be found in amplitude, scale and/or orientation. Indeed we noticed in the earlier section that after last pericentric passage, Sgr induces a strong spiral structure which quickly winds up over a period of $\sim 800\,\rm{Myr}$. 

As explained earlier, using our L2 model, we looked for snapshots where a qualitative match is visible in 2D maps of $V_{R}$, $V_{\phi}$ and $V_{Z}$, with the added requirement that the MW bar should be oriented at a thirty degree angle with respect to the observer. We found such a match at $t\sim 0.8 \rm{Gyr}$ (i.e. after $\Delta t\sim500\,\rm{Myr}$), or the equivalent of two rotations of the Sun past $t\sim0.46\,\rm{Gyr}$ which also happens to put the bar at $\sim30-40^{\circ}$ orientation with respect to the $x-$axis connecting the Sun to the Galactic Center which also corresponds to the orbital plane of Sgr\footnote{We have also tried to follow the bar at an angle of 30-40 degrees (i.e. not necessarily with the Sun lying along Sgr's orbital plane intersection with the $x-y$ plane) but were not able to find any match except for the currently presented snapshot.}.

In the top panels of Figure 4, we present such simulated 2D maps of median $V_{R}, V_{Z}, V_{\phi}$ which closely reproduce the features seen in \cite{katz18} (see their Figure 10 and our Figure 3). We also complement these with 2D projections of the surface density contours overlaid on them. In the bottom panels of Figure 4, we show maps of the density contours, median height $<Z>$ and standard deviation in height $\sigma_{Z}$ as a proxy for the flaring of the disc. We see that our median radial velocity map is remarkably similar to that of \cite{katz18} and in the simulation traces tightly wound spirals. In the median $V_{R}(x,y)$ map, we find a similar interchanging succession pattern of radially outward and inward motions between $-20<V_{R}/\rm{km\,s^{-1}}<20$. We also note the existence of a bridge between the positive velocity arcs at $(X,Y)=(-12,2)$, also seen in the DR2 data, which in the numerical models marks a remaining inlet of high-density in the disc. The orientation is also strikingly similar. We note small differences with the data which are below a factor of $\sim2$ in amplitude and scaling (somewhat larger wavelength in the simulations) and orientation of a few degrees ($\sim 10^{\circ}$).

 The corresponding vertical velocity map in the Gaia-like region in the top-middle panel of Figure 4 also traces a warp as noticed in \cite{katz18} and \cite{poggio18}, with amplitude variations that are similar to the data in the range $-5<V_{Z}/\,\rm{km s^{-1}}<5$. However, we caution that this interpretation while reasonable, needs to be tempered by the fact that this does not correspond to a simple warp as generally been advocated in the observations of the HI neutral gas \citep{levine06} with a simple $m=0,1,2$ Fourier series ring decomposition in $Z$. Zooming out, we notice that this is only about half of the bigger picture. Indeed, during its interaction, Sgr excites vertical density waves in the disc which propagate outward \citep{gomez13, laporte18b} with physical wavelengths varying from 5 to 10 kpc (see Figures 8 and 9 in \cite{laporte18b}). It is expected that beyond a Galactocentric radius of $\sim15$ kpc, a dip in $V_{Z}$ should arise which would be a clear signature of a large scale radially propagating bending wave in the disc. We thus caution that a much {\it larger} volume will need to be probed in order to gain a more observationally complete concensus on the structure of the Galaxy rather than adopting simplistic warp decompositions as done in the past \citep[e.g.][]{momany06,reyle09}. This should in principle be testable through the numerous upcoming spectroscopic campaigns in both hemispheres with APOGEE, SDSS V, DESI, WEAVE, PFS and 4MOST. 

Evidently, there are hints in the data that we are not observing just a warp in the disc but most likely corrugations in the disc. If a simple warp was expected most of the signal in median vertical height of the disc would be at 90 degrees offset to the vertical velocities maps. This is not seen what is seen strictly speaking in the {\it Gaia} DR2 data as shown in the red clump median height map in \citep{katz18} and in Figure 2 with the giants. We also note that the level of flaring in the simulations $\sigma_Z$ (bottom right panel in Fig. 4)  is also consistent with that seen for the red giants in the data. The simulated median  $V_{\phi}$ azimuthal velocity map in the top right panel of Figure 4 also reveals a qualitatively similar behaviour to that seen in the Gaia data \citep{katz18} with a decrease from $240 $ to $220 \rm{km\,s^{-1}}$ as one moves in the direction of the Anticenter within the region probed by the data. 

In this section, we have linked the formation of this velocity field in $V_{R}, V_{Z}, V_{\phi}$ to the disc response after the last pericentric passage of Sgr. During this time, the dwarf excites a bi-symmetric spiral (and a fast bar) which tighly winds up after $0.8\rm{Gyr}$ until it gives rise to a velocity field with qualitatively similar features to the Gaia probed volume. Interestingly, this timescale, $\Delta t\sim 0.8\rm{Gyr}$, is also consistent with different models for the last pericentric passage \citep{ostholt12,dierickx17,laporte18b}. Thus, it is tempting to link the present-day structure of the disc as that which resulted from the perturbations which were seeded during to Sgr's last pericentric passage and evolved to the present-day. As we shall see, since the last major disc tidal excitation of Sgr, several generations of bending waves follow but despite the few disc rotations, {\it locally}, the stars still retain much of the memory of this last passage through the formation and winding of a phase-space spiral.

Whether the spiral patterns seen in the data have a solely internal origin due to intrinsic disc instabilities remains an open possibility. However, together with the signs of vertical oscillations in the disc \citep{widrow12,antoja18,monari18}, it is inevitable that the impacts of Sgr on the disc played a role in setting the structural/kinematical asymmetries of the disc in both the radial and vertical direction. Nonetheless, we note that a striking resemblance to the signal in the \cite{katz18} maps can be reproduced in the interaction scenario producing tightly wound spiral structure. This is in agreement with \cite{quillen18a}, who linked low-velocity moving groups in the $U-V$ plane to tightly wound spirals.

\begin{figure}
\includegraphics[width=0.5\textwidth,trim=0mm 0mm 0mm 0mm, clip]{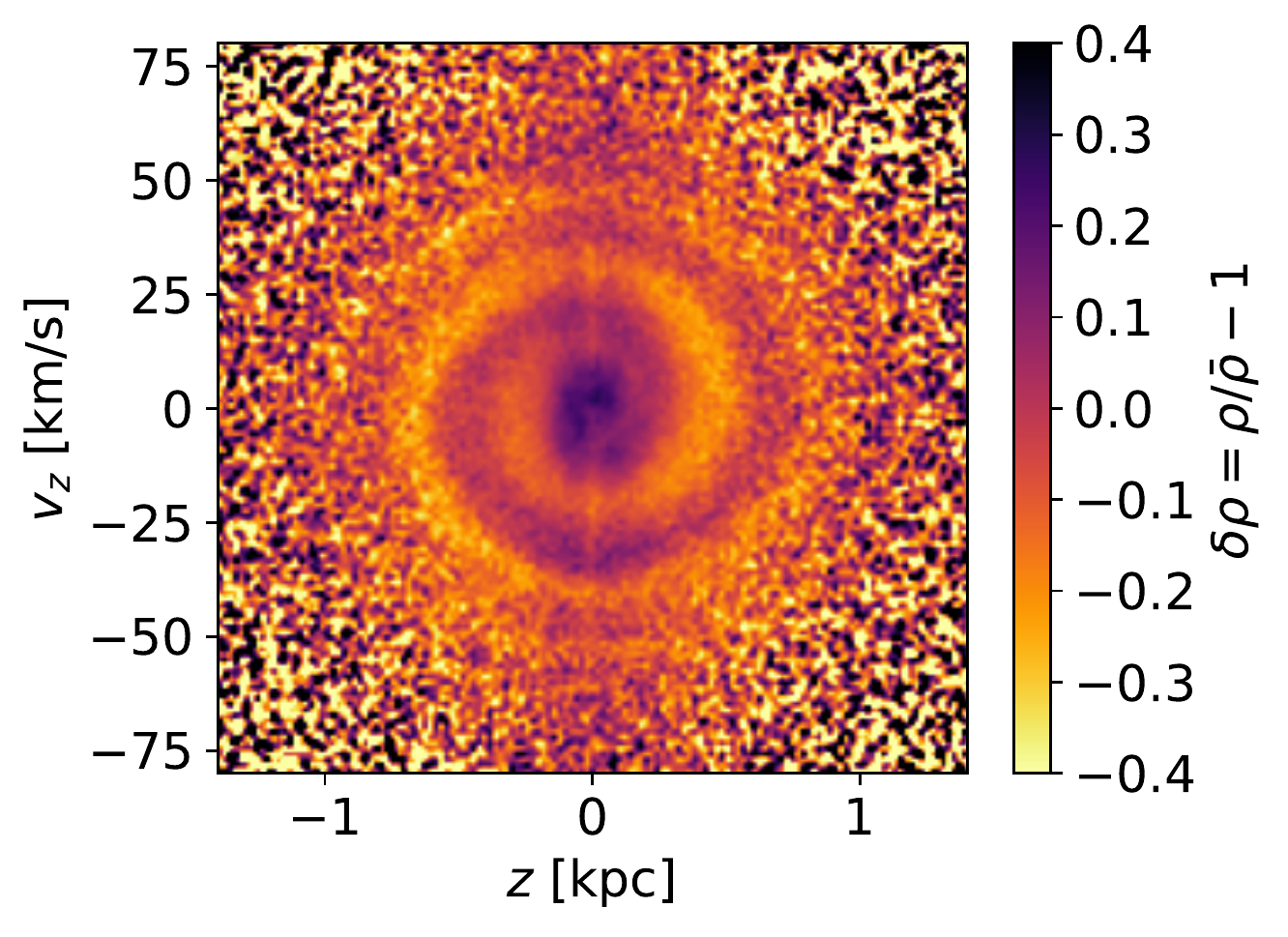}
\includegraphics[width=0.5\textwidth,trim=0mm 0mm 0mm 0mm, clip]{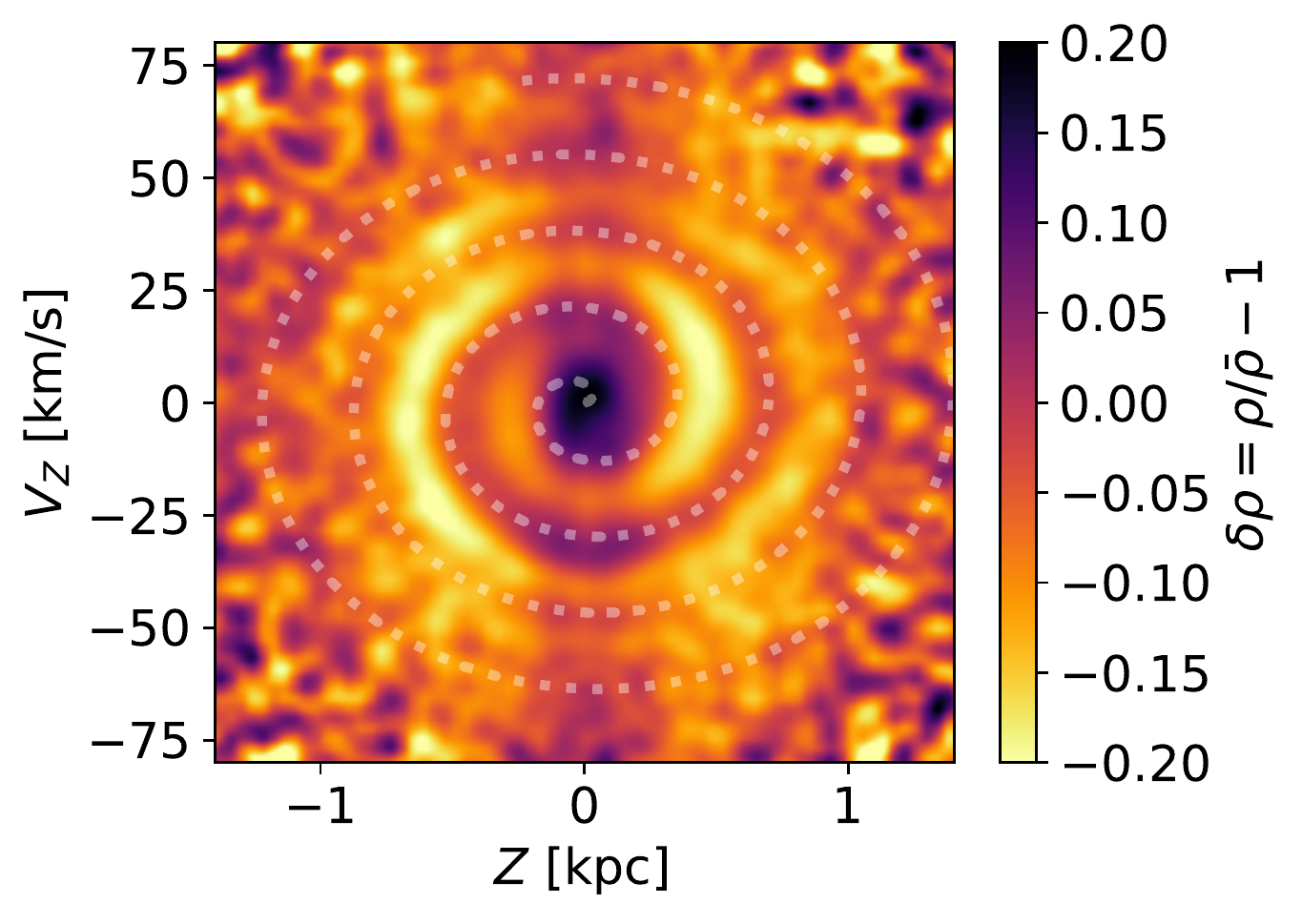}

\caption[]{{\it Gaia} DR2 $\delta\rho(v_{z},z)$ phase-space spiral overdensity using all the stars within 1 kpc from the Sun. This constitutes the clearest visualisation of phase mixing in number counts as opposed to projections of the data in $v_{\phi}(v_{z},z)$ or $v_{R}(v_{z},z)$. It also supports the underlying assumption of modeling the perturbation as a clump diffusing in phase-space by \cite{antoja18} which captured most of the essential aspects of the formation of the snail pattern except for self-gravity or the strength of the perturbation. {\it Bottom:} Same as above except using bins that are twice as large as the top panel and re-smoothed with a gaussian of 1 pixel width, overlaid with a spiral, indeed a clear third wrap is visible which continues up to a total of four wraps. Thus we caution studies solely relying on the $v_{\phi}(v_{z},z)$ morphology to time the onset of the spiral formation. Future Gaia data releases should allow more precise determination of the full morphology of $\delta(v_{z},z)$.}
\end{figure}

\section{Substructure in the solar vicinity}

In this section we inspect the structures imparted on the disc after its interaction with Sgr in the last phases of its evolution inside SNbhd-like volumes using the {\it same} snapshot as identified in the preceding section. Before we begin comparison to our simulations, it is good to present again the structure uncovered by \citep{antoja18} and note a few additional important details missed in prior studies which we uncover for the first time.

\subsection{The {\it Gaia} DR2 6D solution phase-space spiral}

Using the 6D data, \cite{antoja18} presented evidence for ongoing phase-mixing in the stellar disc as unveiled by the Gaia data. Here we reproduce the discovery and its associated projections in $v_{\phi}, v_{R}$ within 1 kpc from the Sun in Figure 6. We also note that a cut with $|v_{R}|<40\,\rm{km\,s^{-1}}$ in producing $v_{R}(v_z,z)$ achieves good results. A tightly wound spiral is clearly evident which is taken as evidence for incomplete phase-mixing in the disc. One can however notice that in the $v_{\phi}(v_z,z)$ map the existence of a branching out of the spiral in the region bound by $-1<z/\rm{kpc}<1$ and $30<v_{z}<60$, suggesting that this projection is not the characteristic one to identify the number of wraps. Given that the perturbed disc distribution function (D.F.) $f(\mathbf{x},\mathbf{v})$ can be thought of as an axisymmetric distribution background term $f_0$ and smaller perturbative terms, we need to get rid of the background signal from the zeroth order distribution function which makes it hard to discern a phase-spiral in number density counts. 

In Figure 6, we show the overdensity phase-plane map defined as:
\begin{equation}
\delta\rho =\rho(v_{z},z)/\bar{\rho}(v_{z},z)-1,
\end{equation}
where $\bar{\rho}(v_{z},z)$ denotes the mean number density map. The pixel size is $\Delta \rm{pix}=0.02\times 1.2 \,(\rm{kpc} \times \rm{km\,s^{-1}})$ and the mean density map is constructed by processing the signal $\rho(v_z,z)$ with a Gaussian filter of width $4\Delta\rm{pix}$. We note that the overdensities of the spiral vary by an absolute value of order $|\delta\rho|\sim\mathcal{O}(15\%)$ as one goes along the $v_{z}=0$ axis, consistent with the wavelength of the density fluctuations from \cite{widrow12}. Indeed our earlier experiments, showed that Sgr could reproduce simultaneously outer disc structures as well as local density fluctuations as a function of height in the solar neighbourhood, which is just the 1-D projection of the $(v_z,z)$ spiral revealed by \cite{antoja18}. We also note that there are additional arching overdensities along the $z=0$ axis which hint to a few more additional wraps of the spiral. Reproducing the same figure using pixels that are twice coarser we are able to uncover up to 4 wraps, which we highlight by overplotting an Archimedian spiral.

Evidently, Figure 6 supports the fact that the earlier experiments of \citep{antoja18} captured all the physics of phase-mixing and their ``clump'' should have been interpreted as the density fluctuation to the disc's distribution function, which would naturally arise for a tidal encounter with a satellite like Sgr as will be shown in Fig. 7 below. One can appreciate at least two self winding wraps in number counts, which was also evident in \cite{antoja18} but perhaps not clear to a few. However, in overdensity, we are able to capture more wraps which are consistent with the fact that we do see a parallel running top ridge in Figure 5 in $v_{\phi}(v_{z},z)$ at $(v_z,z)=(50,0)$. The different projections of the data in $\delta\rho, v_{R},v_{\phi}$ should give us information on the functional form of the underlying potential \citep{darling18}, but also the direction of the pertubation and potentially the mass of the perturber at impact.

\begin{figure*}
\includegraphics[width=1.0\textwidth,trim=0mm 0mm 0mm 0mm, clip]{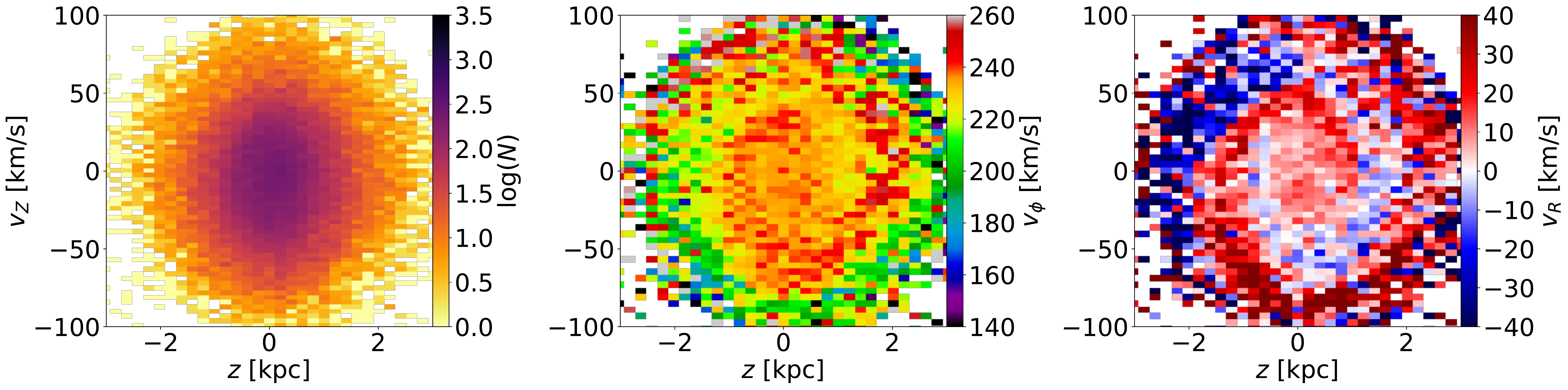}
\caption[]{Phase-space spiral in a solar neighbourhood-like region in the disc following after the impact with Sgr at $t=0.8\,\rm{Gyr}$. {\it Left:} $\rho(v_{z},z)$ map. {\it Middle:} $v_{\phi}(v_z,z)$ snail pattern showing $\sim 2$ wraps . {\it Right:} $v_{R}(v_z,z)$ map showing two consecutive wraps. These maps are in qualitative agreement with those observed in the real data. These are first such maps reproduced by an N-body simulation of the Sgr dwarf impacting the disc. Disagreements on the quantitive side lie in the amplitude of the spiral pattern which is larger by a factor of $\approx1.5$. }
\end{figure*}

\subsection{The phase-space spiral pattern}

In this section, we study phase-space structures imparted by the last phase of interactions of Sgr with the Galactic disc in SNbhd-like volumes. As a starting point, we choose purposefully, the same snapshot for which we found a match in the velocity structures in the midplane in section 3. We place ourselves in a SNbhd-like region at $(X,Y)=(-8,0)$. Given the modest number of particles used in the disc ($N_{\rm{disc}}=5\times10^{6}$), we choose a region bounded by $R_{helio}<3.0$ and $160^{\circ}<\phi<280^{\circ}$. We caution that these cuts are somewhat larger than the volumes probed by \cite{antoja18} or \cite{monari18}, but due to our limited number of N-body particles, such a choice is necessary to better sample the phase-space structure of a SNbhd-like region. We do not think this choice would hamper the overall analysis given that we are mainly interested in looking at qualitative matches (if possible) and whether these can be tied to the accretion history of Sgr. Moreover, we showed in earlier works that the disc responds {\it globally} and thus the phase-space spiral is expected to persist on larger scales than just a $500 \,\rm{pc}$ volume.

In Figure 7, we examine the $(v_{z}-z)$ plane colored by density (left), $v_{R}$ (middle), and $v_{\phi}$ (right) and find a pattern strikingly similar to those observed by \cite{antoja18}. This is {\it the first time this is presented in a full N-body simulation of the impact of Sgr with the Milky Way}. This result is particularly striking because this simulation was run prior to the data release to explain observations on large but also prior observations on small scales scales (e.g. Widrow et al. 2012). While the amplitudes of the vertical perturbations are larger by a factor of 1.5 compared to the observations, the qualitative match to both the $v_{r}(v_{z},z)$ and $v_{\phi}(v_z,)$ maps (middle and right panels) is noteworthy and remarkable. The snail shell in the number counts requires more imagination to see, probably due to the large spatial volume used and low particle numbers. Indeed, thanks to the coupled motions in non-separable potentials, the spirals become more easily identifiable in the space of $v_{R}(v_{z},z)$ or $v_{\phi}(v_{z},z)$ as pointed out in \cite{binney18} and \cite{darling18}.

To understand the origin of this structure, it is useful to consider the evolution of the Galactic disc back in time all the way to the last pericentric passage of Sgr in its last phase of disruption. In Figure 8, we show a time series of the mean vertical velocity of the disc. We note that as Sgr hits the disc, it excites vertical density waves throughout the whole disc, which are maintained on Gyr timescales. One can discern by eye the local impacts of Sgr during each pericentric passages which give local velocity kicks of a few 10s of $\rm{km\,s^{-1}}$ but also set new generations of bending waves for which the disc is {\it globally} perturbed as seen in the snapshot $\sim100\,\rm{Myr}$ after the last pericentric passage. The latter phenomenon is unique to self-gravitating discs and missed in toy models of phase-mixing \citep{minchev09,antoja18,binney18}. Moreover, throughout its interaction Sgr sets off previous generations of bending waves which also gave rise to phase-space spirals.

\begin{figure*}
\includegraphics[width=1.0\textwidth,trim=0mm 0mm 0mm 0mm, clip]{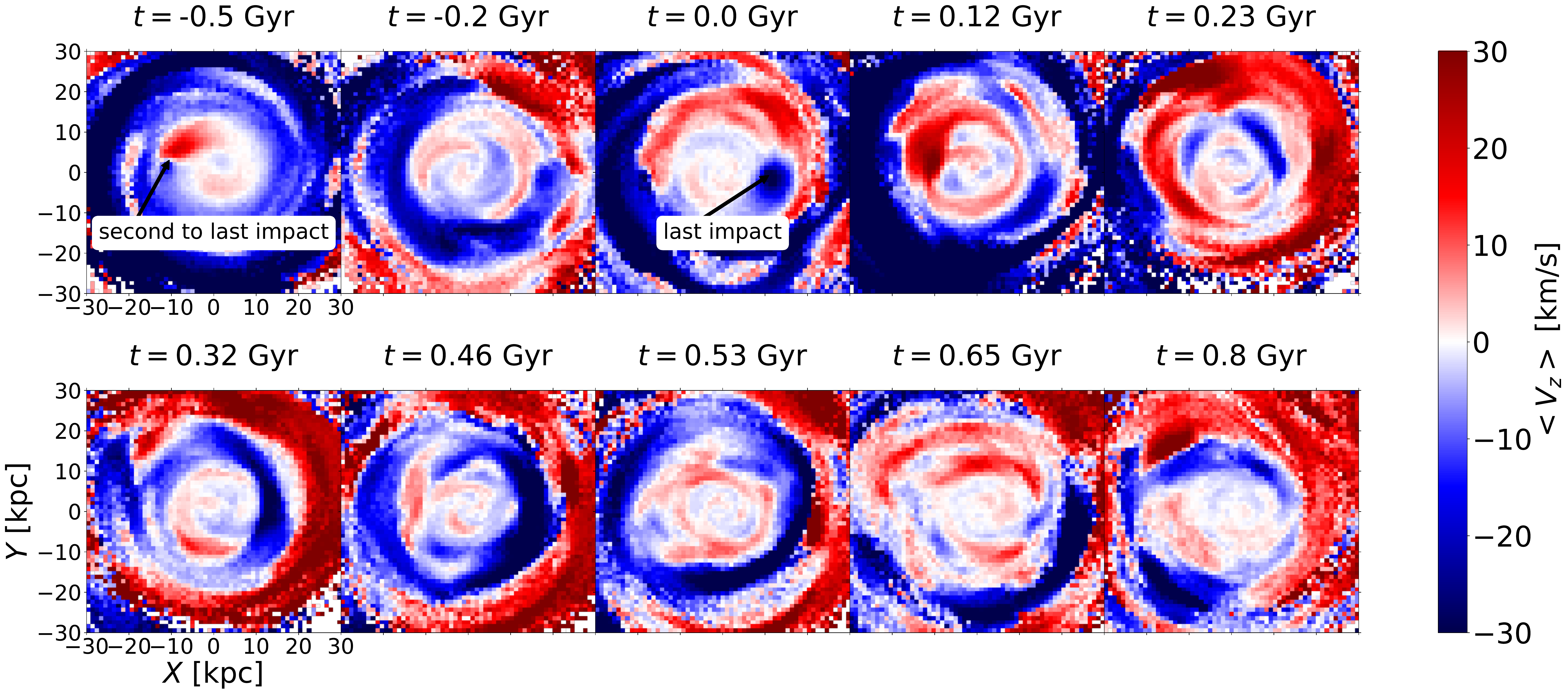}
\caption[]{ Evolution of the mean vertical velocity of the disc with respect to the last pericentric passage of Sgr. Tidal impulses from last two pericentric passages are highlighted at $t=-0.5\,\rm{Gyr}$ and $t=0.0\,\rm{Gyr}$ respectively. These passages maintain the generation of new vertical density perturbations showing that the disc response is {\it global} and not just local.}
\end{figure*}

\begin{figure*}
\includegraphics[width=1.0\textwidth,trim=10mm 0mm 50mm 0mm, clip]{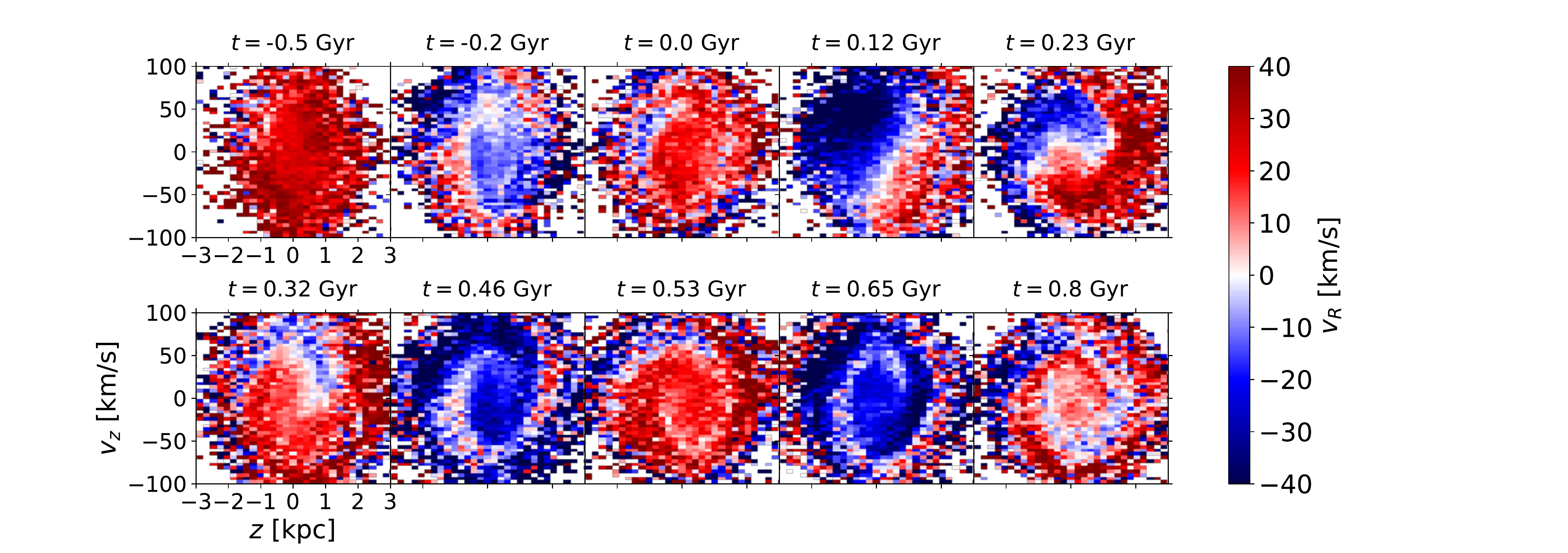}
\caption[]{Time evolution of the snail phase-mixing pattern in the disc. The onset of the spiral is set during Sgr's last last pericentric passage at $t\sim 0.0$ and persists for $\Delta t\sim0.4-0.8 \rm{Gyr}$ during which the pattern winds itself further due to phase-mixing.}
\end{figure*}

\begin{figure*}
\includegraphics[width=1.0\textwidth,trim=10mm 0mm 50mm 0mm, clip]{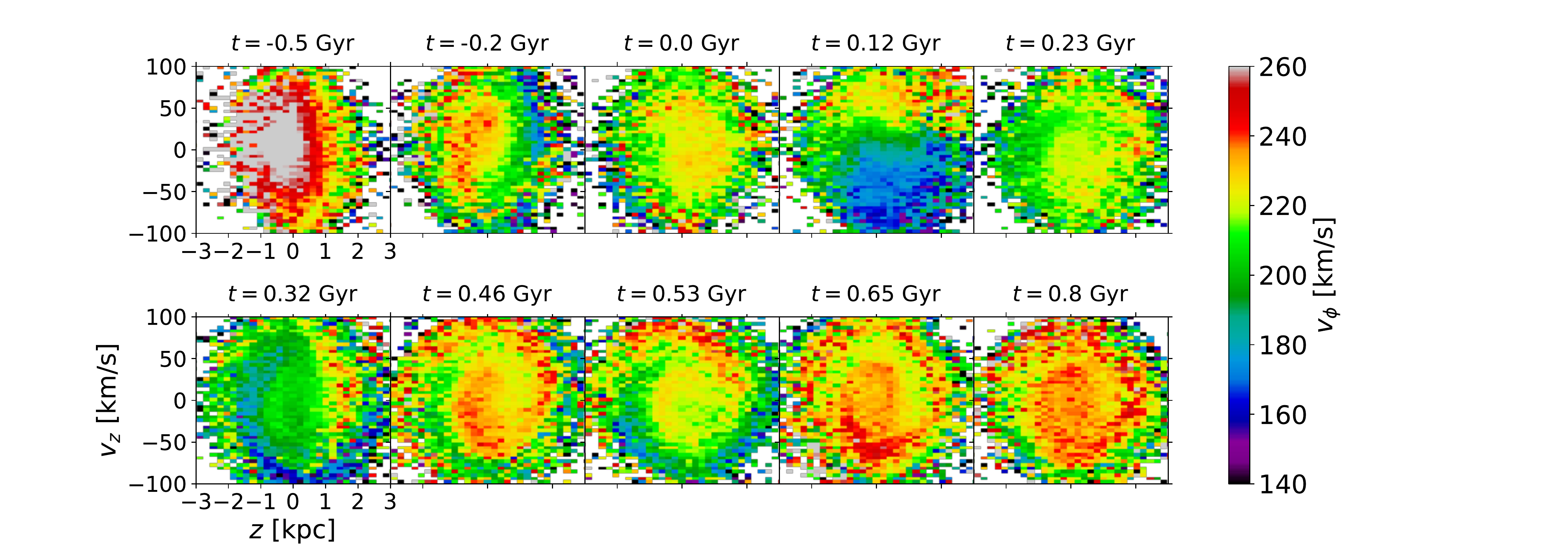}
\caption[]{Similar to Figure 5 but showing maps of $v_{\phi}(v_z,z)$ as a function of time. Previous perturbations set off at $t=-0.5\,\rm{Gyr}$ persist out to $t=0\,\rm{Gyr}$ which then get reset by the newer perturbation.}
\end{figure*}

In Figure 9, we follow the formation of this structure in $v_{R}(v_z,z)$ back in time all the way back to the time prior to the last pericentric passage. Given that the impact with Sgr creates vertical density perturbation for which the whole disc is participating, we just focus on a volume fixed and centered at $(X,Y)=(-8,0)$ tracking the snail pattern back in time up to just before the last pericentric passage. \footnote{We have also tried following the individual particles making up the snail pattern and following a volume with the LSR and found that our results did not vary much.}. The latest perturbation $t=0\,\rm{Gyr}$ clearly dominates over previous ones (e.g. $t=-0.5\,\rm{Gyr}$) which re-sets a snail-like pattern (see second map in top panel of Figure 9) which grows in time wrapping over itself until it goes through two tight consecutive wraps \footnote{Due to the coarse-graining the two wraps could in fact hide others. Higher resolutions will be needed to assess this.}. A similar pattern behaviour is also seen in the $v_{\phi}(v_z,z)$ plane in Figure 10. We note that while the last pericentric passage setting off the onset of the local phase-space spiral happens at $t=0\,\rm{Gyr}$ (see Figure 8), the disc evidently is only able to react $~120\,\rm{Myr}$ for the chosen fixed SNbhd, the equivalent of half a disc rotation. It also becomes evident that the perturbations seeded by the preceding peri at $t\sim-0.5\,\rm{Gyr}$ which developed in to a phase-plane spiral by $t=0\,\rm{Gyr}$, get subsequently re-set and overridden. This last pericentric passage perturbation creates the new phase-space spiral which evolves into a full wrap by the ``present-day'' snapshot $t=0.46\,\rm{Gyr}$ and continues to develop itself unhindered to $t\sim0.8\rm{Gyr}$ because beyond $t=0.46\,\rm{Gyr}$ the shredded Sgr is no longer capable to affect the disc despite the fact that we follow the integration forward. This visibly shows that the present structure of the disc in the observations could very well reflect the later evolution of the disc to Sgr's last pericentric passage, which is consistent with $T_{orb}\sim0.7-1\rm{Gyr}$ \citep{johnston05, penarrubia10,law10} found by detailed Sgr stream fitting studies.

The phase-mixing timescale, ultimately depends on the structure of the MW potential and the strength of the impact, but the number of tightly wound wraps may be used to give a constraint on the time since impact \citep{antoja18}. The {\it Gaia data} seems to show two (possibly four) clear consecutive wraps. This is also the case in our numerical experiments, for which we can trace the origin of the snail back to the time of the last major impact with the Sgr dSph and trace two wraps in $v_{\phi}(v_{z}, z)$ and $v_{R}(v_{z}, z)$ at $\Delta t \sim 0.4-0.8\,\rm{Gyr}$ past the last pericentric passage. This is consistent with the values derived by \cite{antoja18}\footnote{Upon finishing writing this paper \cite{binney18} presented a re-evaluation of the data and derived a lower timescale for the snail pattern of $t_{snail}\sim 0.2-0.4 \,\rm{Gyr}$ which brackets our predicted N-body simulation values from the impact of Sgr with those from \cite{antoja18}. This is not surprising since their toy-model misses the response of the disc which is known to give rise to appreciable differences in the behaviour of self-gravitating systems \citep{weinberg89}, an observation which was re-iterated in \cite{darling18}, but also focused only on the two wraps in $v_{\phi}(v_z,z)$, which closer inspection in overdensity (see Fig. 7) show the existence a few more. This would thus result in longer timescales if re-evaluated properly.}.

However, this seems at face value at odds with the separation of arches in the u-v plane by $\sim20\,\rm{km\,s^{-1}}$, which would correspond to a perturbation $\sim1.9$ Gyr ago according to the models of \cite{minchev09}. On the other hand, given that Sgr impacts the disc multiple times in its lifetime, it could be that because of the longer orbital periods in the radial direction, phase-mixing does not wash away these fine structures as fast as in the vertical direction. Further investigation is warranted but beyond the scope of this paper.

We conclude that Sgr may be the culprit behind the origin of the snail. This is the first time recorded in a N-body simulation of the Sgr dwarf galaxy impact with the Milky Way. It demonstrates that the effect of the Sgr dwarf galaxy in setting up perturbations in the disc from its outer edge (through the excitation of the MW dark matter halo wake) to its inner-most region (through its tides) may be more important than previously appreciated (or ignored). More work will be necessary to be carried out with adapted simulations to properly model the orbit of the dwarf through careful stream fitting and varying initial conditions for the structure of the Galaxy to assess the ranges of outcomes where a full quantitative agreement may be achieved. However, already this model essentially explains all the features in the data over several scale lengths. It is thus difficult to imagine what newer insights a perfect model would bring, but to polish agreements further. Given the remarkable qualitative match between the simulated velocity fields and phase-space spirals with the data, this may after all be within reach especially given the modest ($N_{disc}\sim 5\times10^{6}$) number of particles used in our N-body experiments.

\section{Ridges in the $V_{\phi}-R$ plane}

Using the same simulation time ouput for which we have identified a matching snail-shell pattern and median velocity fields we look at the structure of the $V_{\phi}-R$ ridges. \cite{ramos18} studied moving groups and ridges in the $V_{\phi}-R$ plane through wavelet transforms on the $U-V$ plane. We take a simpler approach but also effective at revealing the ridges in the data by computing the overdensity factor after smoothing the data with a Gaussian kernel. This is shown in Figure 11 where we select {\it Gaia} DR2 stars with in a wedge $160^{\circ}<\phi<220^{\circ}$ and bounded by $|z|<500\,\rm{pc}$. We also note many more features than initially identified by \cite{kawata18}, with ridges along the $R=8\,\rm{kpc}$ axis which are separated by $\sim25-30\,\rm{km\,s^{-1}}$. Earlier signs of these ridges may have existed and manifested themselves as bumps in the MW rotation curve \citep{martinez18}.

In Figure 12, we see some resemblance to the {\it Gaia} DR2 data with similar ridges of similar slopes as in the observations, many of which follow remarkably well lines of constant angular momentum with local slopes at $R\sim8\,\rm{kpc}$ of $25-30\,\rm{km\,s^{-1}/kpc}$. When analysed to larger radii, many more appear with ridges splitting in the outer disc. \cite{hunt18} noted similar structures in their toy model of transient spirals, though the triggering origin of these patterns could not be identified as the perturbations were imposed in an ad-hoc fashion. What is clear is that the Milky Way is a more complicated system than a mere simple disc galaxy being affected by solely internally triggered instabilities (e.g. bar and spiral arms) as often considered. A common struggle of the \cite{hunt18} model and ours is that we have fewer ridges than in the data at a fixed radius. In our case, we suspect this to be likely due to the poorer disc D.F. sampling locally at $R\sim8\rm{kpc}$ than in the data which are different by a factor of more than $100$.

Regardless, given that we only consider one fiducial MW model, it will be worthwhile to study the evolution in isolation of different MW models to identify which ridges can be reproduced by impact models or be reproduced solely by internal instabilities, or the combination of the two. Although this is a topic for future investigations, we do note that \cite{fragkoudi19} presented a recent study on the origin of long-lived prominent ridges and undulations in the $v_{\phi}-R-v_{r}$ plane in relation to the Outer Lindblad Resonance (OLR) of a fast bar \citep{dehnen00} in isolated MW-like disc simulations. Though we note that the pattern speed of the bar is still debated and \cite{monari18b} recently showed that the slow bar model of \cite{portail17} can in fact reproduce Hercules due to orbits trapped at corotation while the OLR would produce a higher azimuthal velocity arch. In fact, in addition to undulations in $v_{R}$ we also see correlations in some of the ridges in $v_{\phi}-R-v_{z}$ plane, which would hint that not all ridges are necessarily produced by a sole mechanism.

\begin{figure}
\includegraphics[width=0.5\textwidth,trim=0mm 0mm 0mm 0mm, clip]{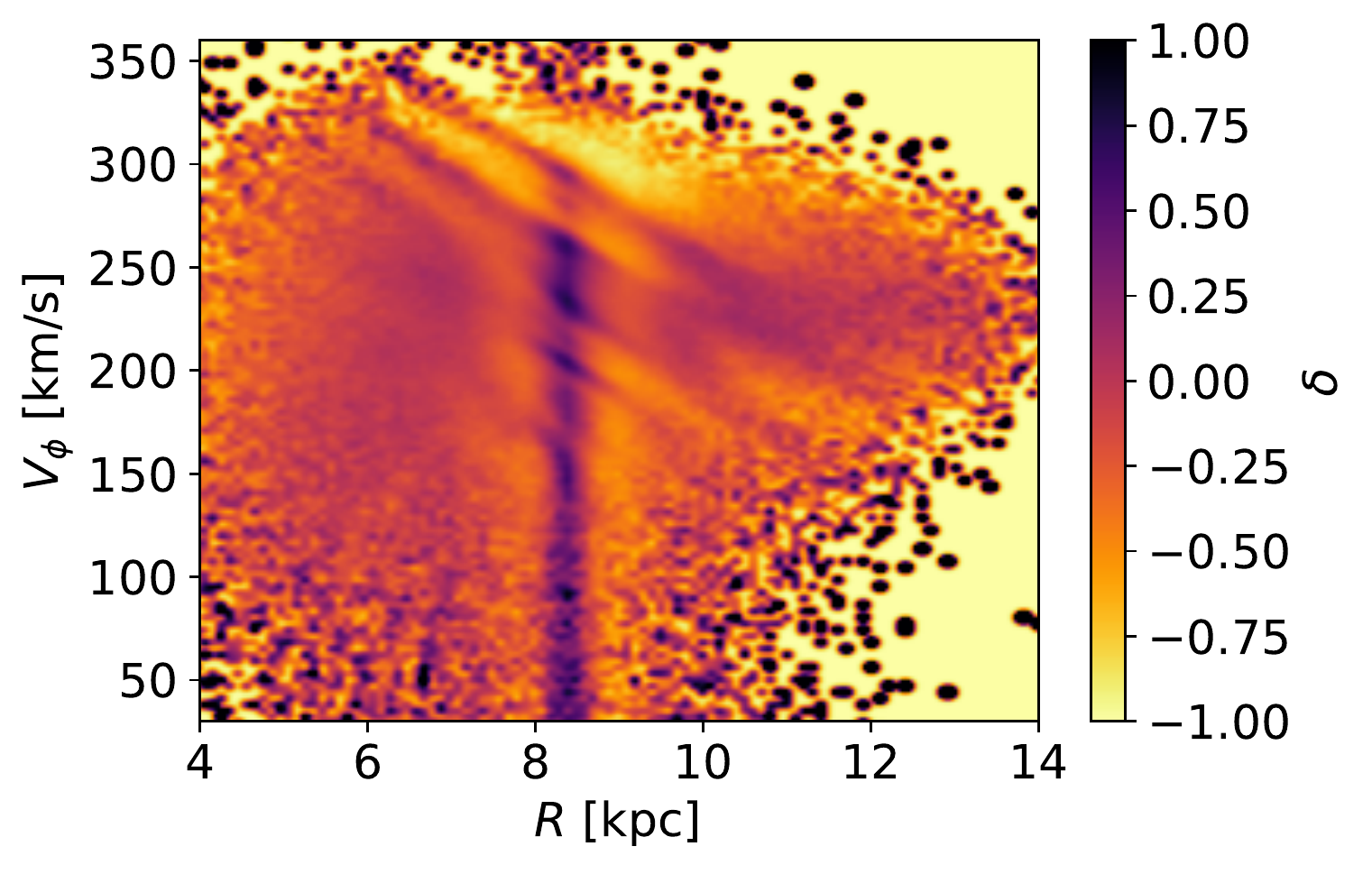}
\caption[]{Ridges in the $v_{phi}-R$ plane in the Gaia DR2 data shown in overdensity. The binsize is $\Delta R\times\Delta v_{\phi}=0.1\,\rm{kpc}\times 3 \,\rm{km\,s^{-1}}$ and we used a Gaussian kernel of width $3\rm{pix}$ for the background $v_{\phi}-R$ map. Most of the density ridges are mostly prominent around the solar radius $R_{\odot}$, because the density of stars is the highest due to selection effects. We see that many of these ridges are separated by $\sim25-30\,\rm{ km\,s^{-1}}$ with a few odd few ridges interspaced between.}
\end{figure}

\begin{figure}
\includegraphics[width=0.5\textwidth,trim=0mm 0mm 0mm 0mm, clip]{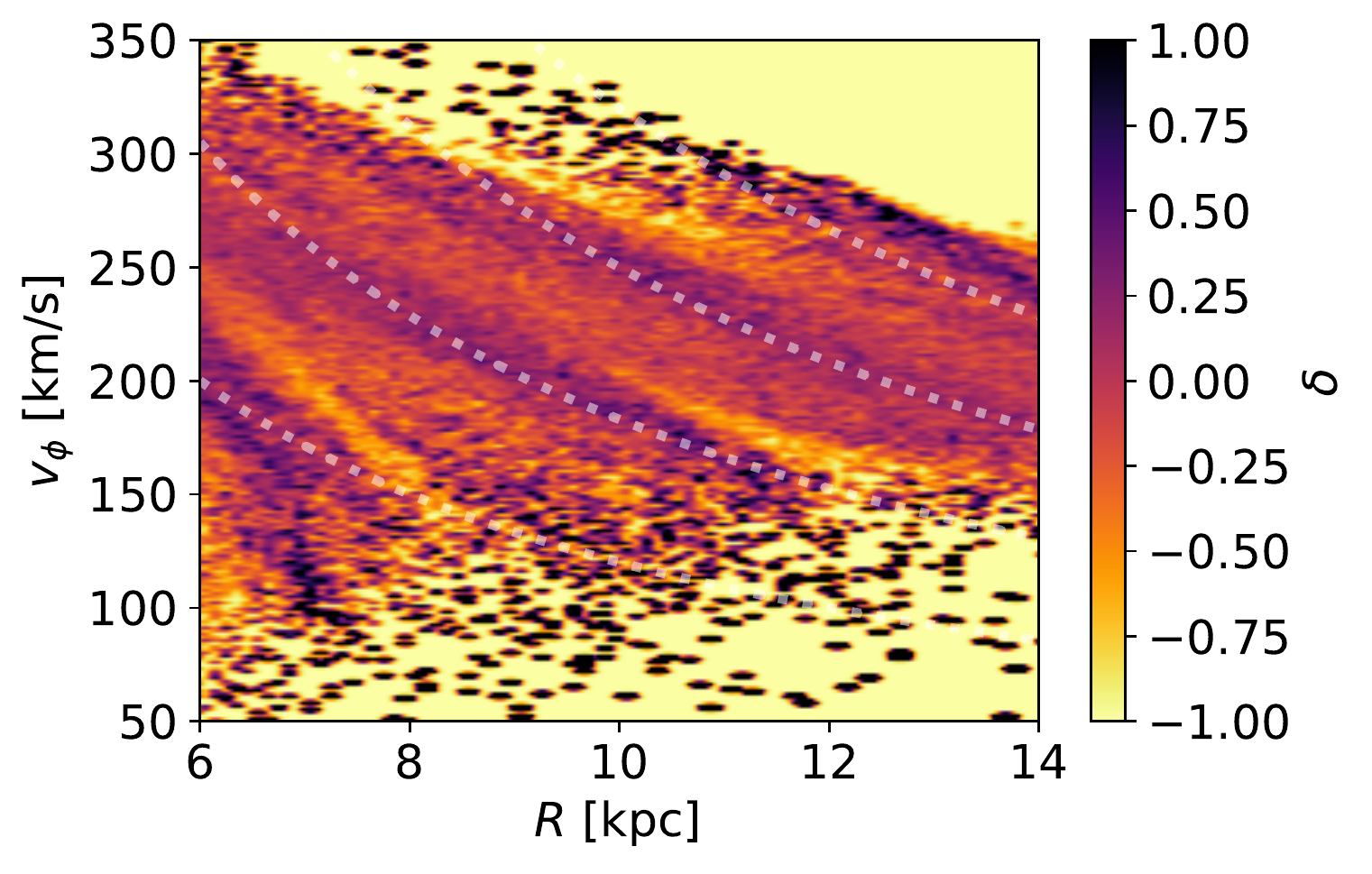}
\includegraphics[width=0.5\textwidth,trim=0mm 0mm 0mm 0mm, clip]{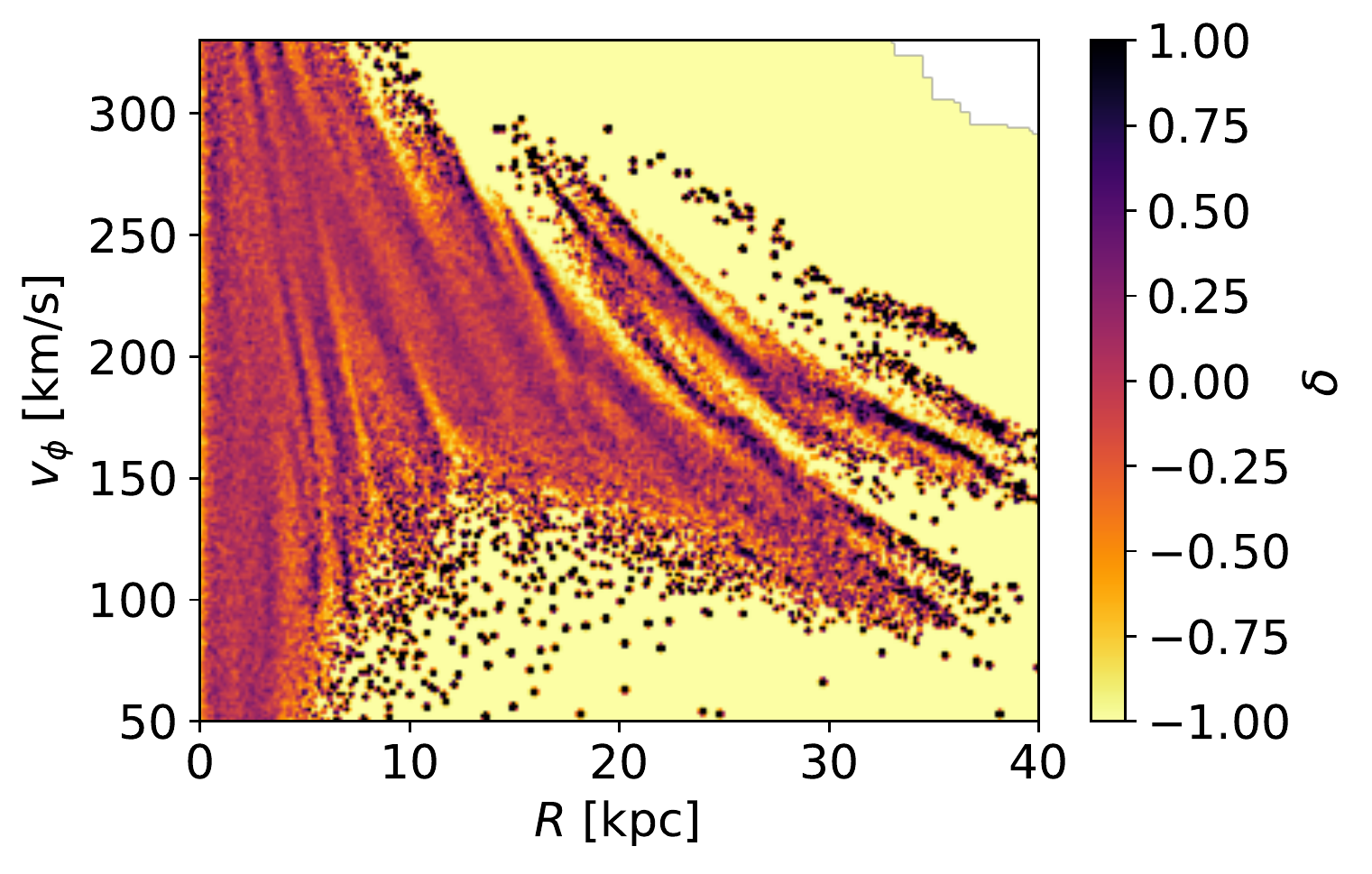}
\caption[]{Ridges in the $v_{phi}-R$ plane, excited by the interaction with Sgr. The winding spiral structure in the Milky Way excited by Sgr is able to reproduce qualitatively similarly ridges to the data seen in \cite{antoja18} with a tilt of $\sim25-30 \,\rm{km\,s^{-1}/kpc}$. Dashed white dotted lines mark curves of constant angular momentum.}
\end{figure}

\begin{figure}
\includegraphics[width=0.45\textwidth,trim=0mm 0mm 0mm 0mm, clip]{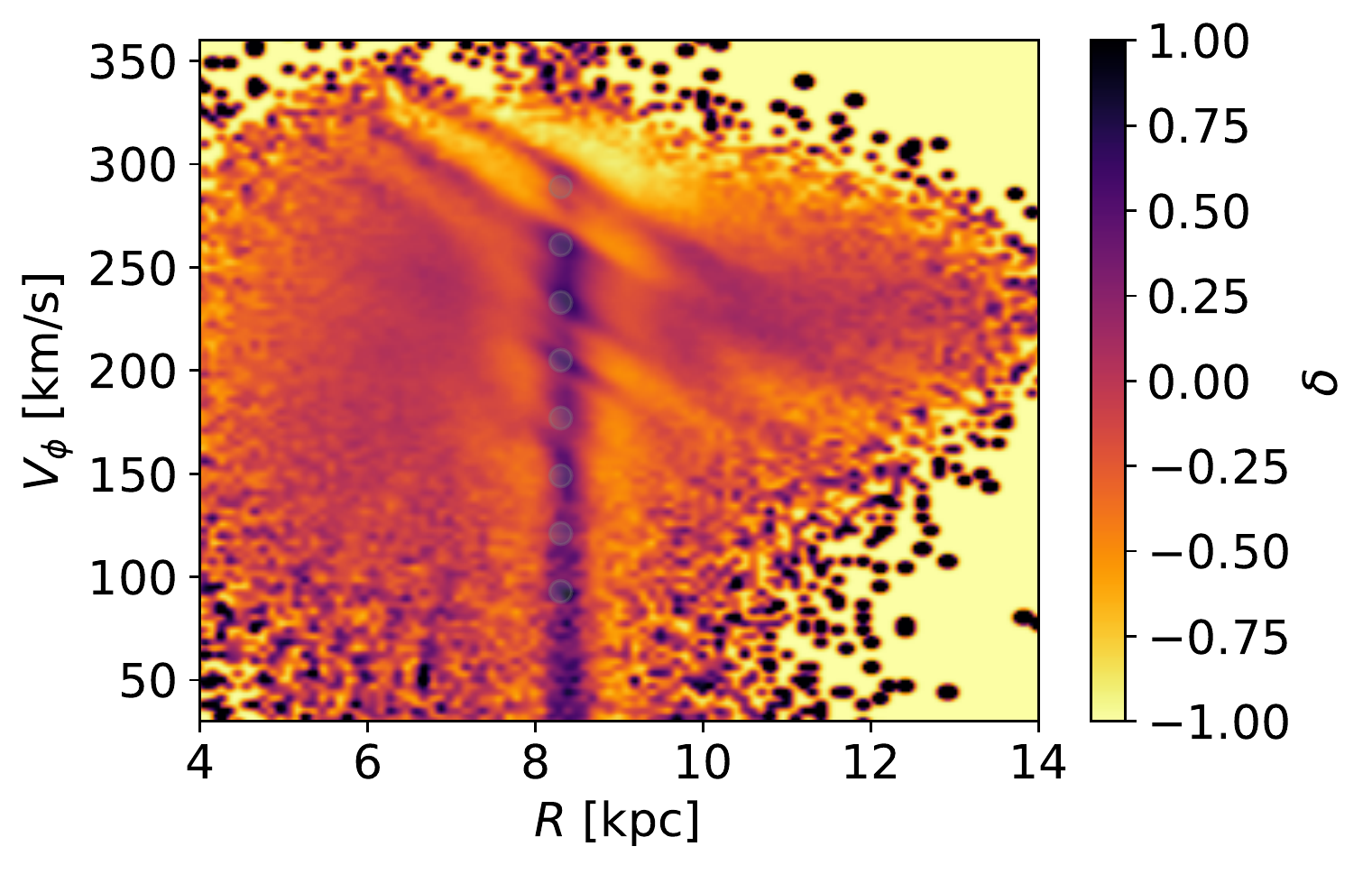}
\includegraphics[width=0.45\textwidth,trim=0mm 0mm 0mm 0mm, clip]{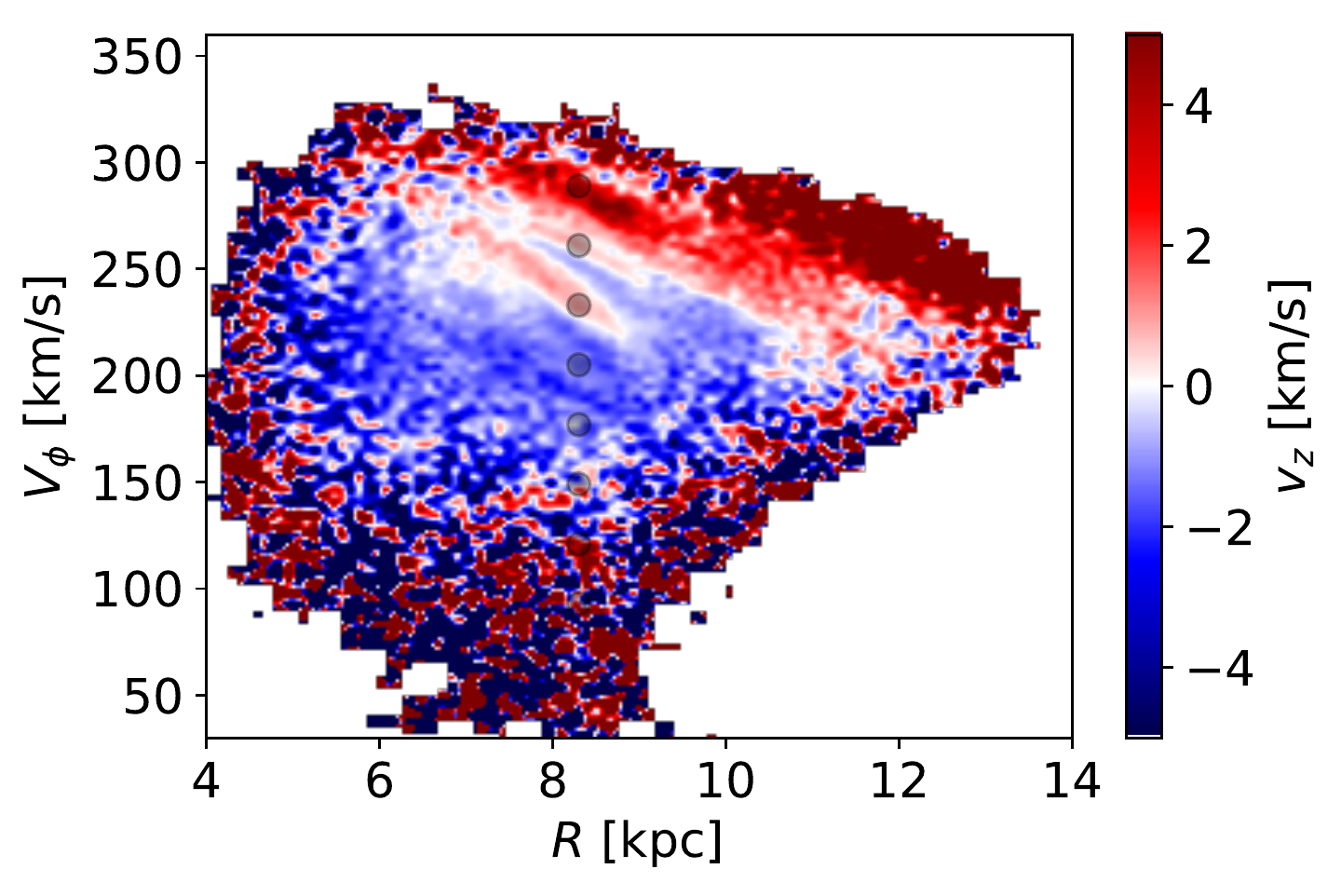}
\includegraphics[width=0.45\textwidth,trim=0mm 0mm 0mm 0mm, clip]{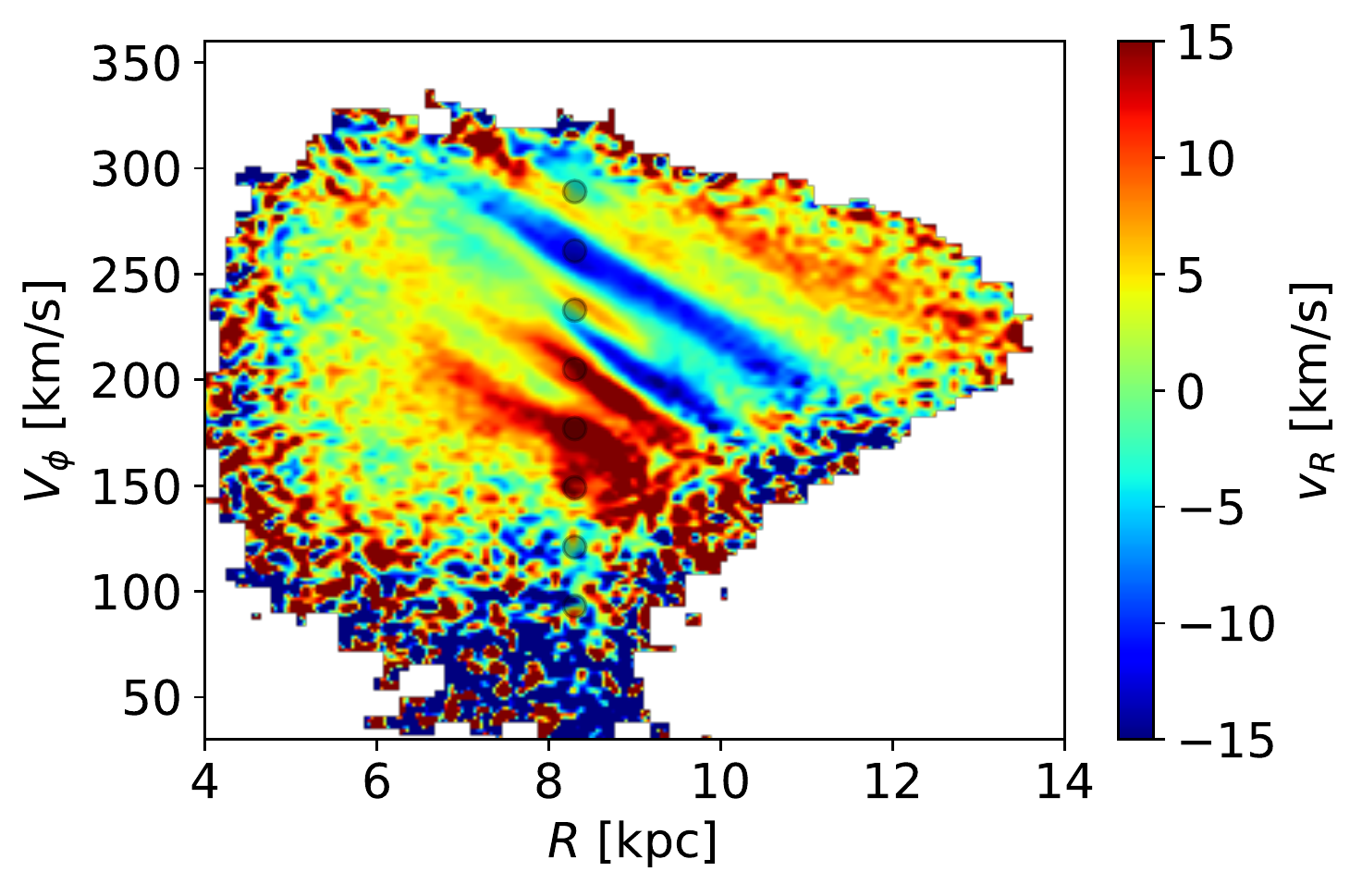}
\caption[]{Ridges in the $v_{\phi}-R$ plane in the Gaia DR2 data shown in overdensity and their projections in $v_{z},v_{\phi}$ in the top, middle and bottom panels respectively. These maps are constructed in the same way as in Fig. 11. We highlight recurrent ridges which are regularly spaced by $\sim28\,\rm{ km\,s^{-1}}$ by the grey and black transparent dots down to $\sim90 \,\rm{km\,s^{-1}}$, noting the presence of an odd few ridges interspaced between. These correspond to the L2 ridge of, Sirius, Hyades, Hercules, L11/L12, Arcuturus recorvered in \cite{ramos18}. We note the presence of a thinner ridge between L2 and Sirius which has not yet been reported. Between Hyades and Hercules we can also see the L7/L8 ridge \citep{ramos18}. The last two highligthed ridges (``Snoop'' and ``Herbie'') are new and coincide with the expected ridges below Arcuturus predicted by \cite{minchev09}.}
\end{figure}

In Figure 13 we show that the ridges $v_{\phi}-R$ from {\it Gaia} DR2 have a lot more fine-grained structure than appreciated in \cite{kawata18}. Our selection is the same as in Figure 11. Indeed we do notice the appearance of distinct ridges which are better visible around the Sun, which are spaced by $\sim28\,\rm{km\,s^{-1}}$ (albeit with a few disparate intervening additional fine ridges), identified by the transparent grey which can be traced down to $v_{\phi}\sim90\,\rm{km\,s^{-1}}$. In this series, we can note the existence of known moving groups L2, Sirius, Hyades, Hercules, L11/L12\footnote{These correspond to projections of HR1614 and the stream identified by \cite{arifyanto06} in the $v_{\phi}-R$ plane.}, Arcuturus \citep[already reported in][see their nomenclature]{ramos18} as well as two more additional ridges below Arcturus. To our knowledge, these have not been reported down to $v_{\phi}=90\,\rm{km\,s^{-1}}$, but predicted to exist from the ringing models of \cite{minchev09}. We note that these ridges also show undulations in $v_{R}$ \citep{fragkoudi19} as well as in $v_{z}$ which we also reveal in this contribution.

\section{The broader picture: Sgr as the dominant perturbing architect in the last 6 Gyrs.}

In this section, we tie what we learned from the pre-/post-{\it Gaia} DR2 eras to connect the predictions from the Sgr infall model with the bigger picture, much in the spirit of Galactic archeology studies for our closest neighbour, the Andromeda Galaxy \citep[e.g.][]{bernard12, dorman15}. 

\subsection{Global view of the phase space spiral from DR2}

In earlier works and this contribution, we showed that the response of the disc to the perturbing Sgr is global and that the local phase-space spiral of \cite{antoja18} is a projection of it. Thus, we should be able to trace the phase-space spiral across the whole disc. In Figure 14, we use the {\it Gaia} DR2 data alone to show that this phase-space spiral is present at different radii \footnote{After our submission, a paper by \cite{jbh18} claim that with the combination of GALAH and {\it Gaia} DR2 that they are able to dissect the Gaia phase-spiral to larger distances, which corroborated our work. But  the range covered was restricted to $R_{\odot}-0.5<R/\rm{kpc}<R_{\odot}+0.5$, i.e. within less than a disc scale length.}, corroborating our interpretation using simulations. We do so by tracing the shape of the phase-space spiral in overdensity $\delta\rho(vz, z)$ as well as its projection in $v_{\phi}(vz, z)$ at different locations $R\sim 6.6, 8, 10 \,\rm{kpc}$. As expected, from the general form of the Galactic potential \citep{mcmillan11}, we see a change in shape and amplitude of the phase-space spiral which is due to differences in the local underlying potential. At large radii, where the self-gravity of the disc is least strong and dynamical timescales are longer the phase-space spiral should be more elongated along the $z$-axis and less wound, whereas at smaller radii the phase-space spiral should be more squashed along the $z$-axis \citep{BT08}, which the DR2 data confirms beautifully (see left and right sides of Figure 14 w.r.t. the middle panel). This is also naturally seen in the simulations.

Complementing the observed Gaia DR2 maps of the phase-space spiral as a function of radius, in Figure 15, we show a series of volumes selected along the $y=0$ axis in the $(x,y)$ plane bounded by Galactocentric radii  $R_{i}-1<R/\,{\rm{kpc}}<R_{i}+1$ and $160^{\circ}<\phi<220^{\circ}$, where $R_{i}=6,8,10,14\,\rm{kpc}$. We see that as we go out towards the outer disc, the phase-space spiral shows a decrease of woundedness due to the longer orbital timescales in those regions. Moreover, as we go further into the inner disc, the spiral becomes more squashed in the $z-$direction due to the strong restoring force of the disc.

The combination of both projections of the data is particularly useful to discern features which may not be instantaneously obvious in solely one projection. This is particularly true at $R\sim 6 \,\rm{kpc}$, where the $v_{\phi}$-projection shows a clear spiral whereas in overdensity it is not obvious, likely due to the poorer density sampling.

\begin{figure*}
\includegraphics[width=1.0\textwidth,trim=0mm 0mm 0mm 0mm, clip]{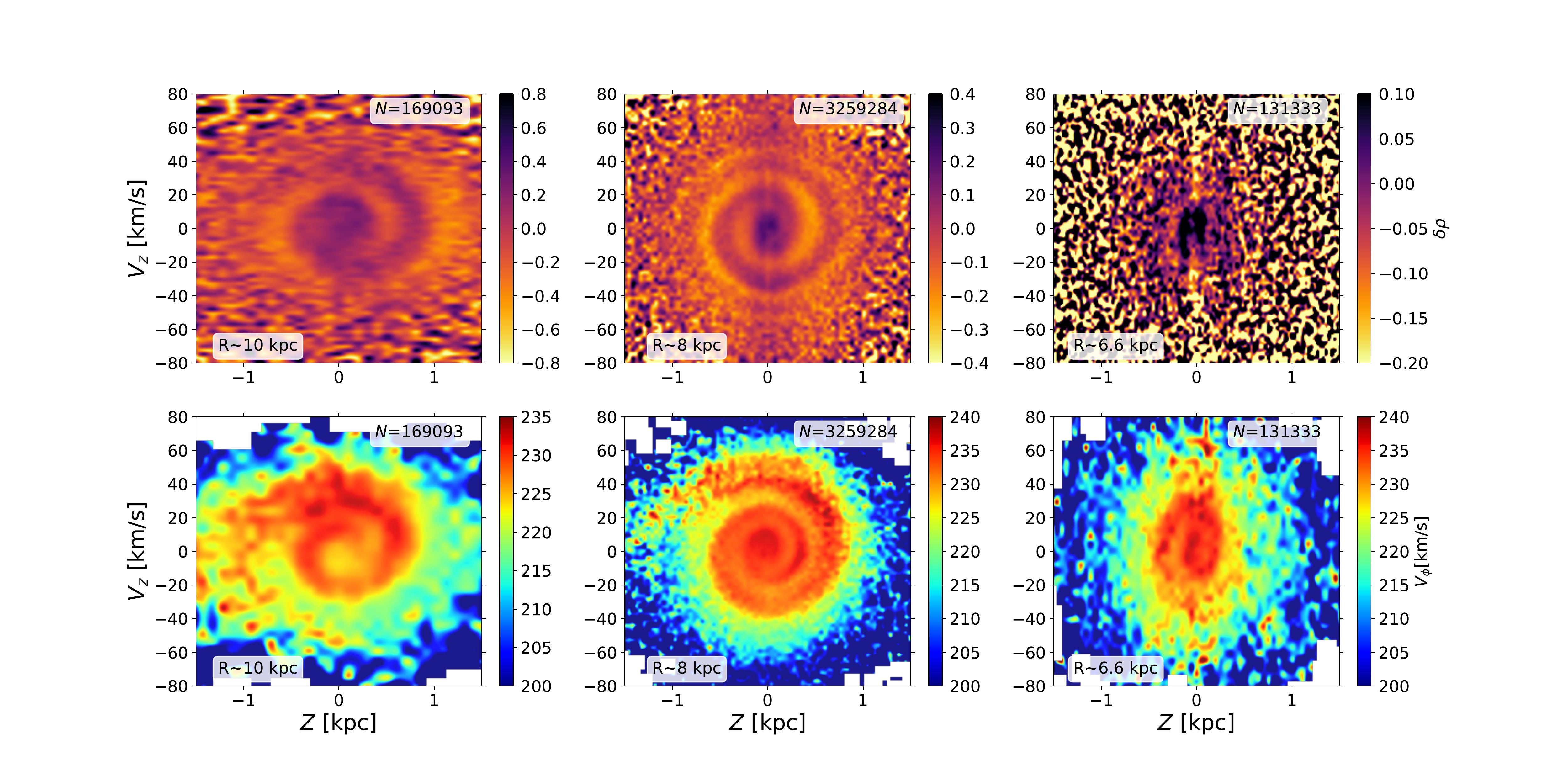}
\caption[]{ {\it Gaia} DR2 $(vz,z)$ phase-space spiral as a function of Galactocentric radius in star counts parametrised through the overdensity $\delta\rho(v_{z},z)$ and its projection in azimuthal velocity $v_{\phi}(v_{z},z)$ (top and bottom panels respectively). The stars selected in each panels are selected in cylinders of radii $R_{cyl}=0.5,1.0, 1.0 \,\rm{kpc}$ centered at $R=6.6, 8, 10\,\rm{kpc}$ respectively along the axis connecting the Galactic Center to the Sun (i.e. $y=0$). As expected in a MW-like potential consisting of an exponential disc and halo + bulge components as in \cite{mcmillan11} in which a global corrugating perturbation is present, the phase-space spiral should grow in $z$-amplitude because of the decreasing restoring force in the midplane \citep[see][]{gomez13,laporte18b}. This is shown explicitly in the data for the first time.}
\end{figure*}

\begin{figure*}
\includegraphics[width=1.0\textwidth,trim=0mm 0mm 0mm 0mm, clip]{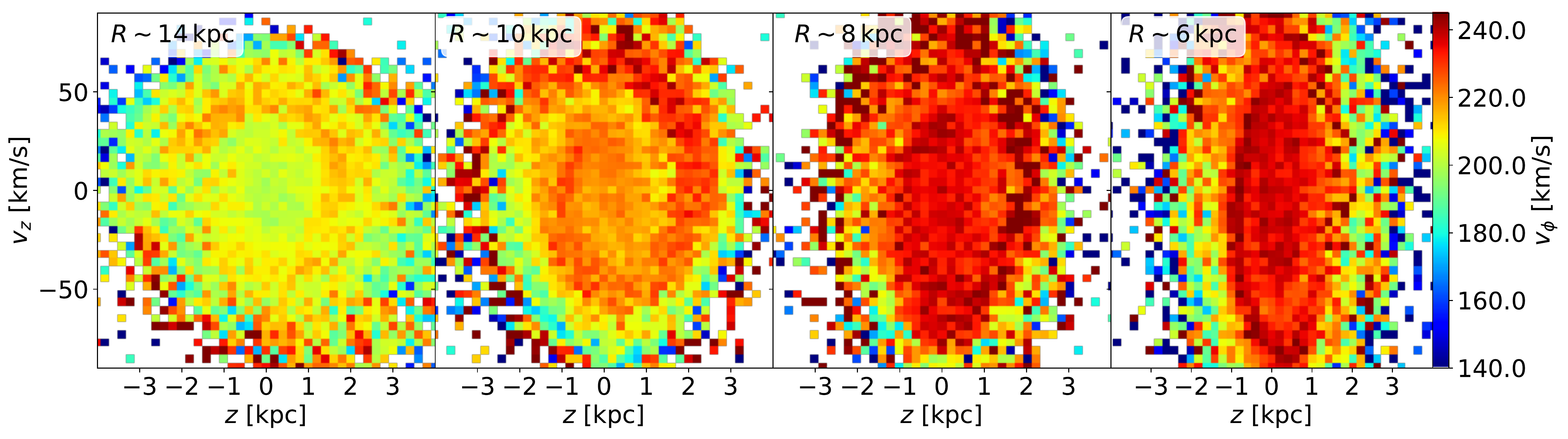}
\caption[]{ Simulated L2 model phase-space spirals at $R\sim6,8,10,14 \,\rm{kpc}$ at $t=0.8\, \rm{Gyr}$}
\end{figure*}

\subsection{Further expectations from the Sgr impact model: outer disc to star formation histories}

The strength of our model is that it makes a number of clear predictions on the structure/kinematics of the disc spread over about 10 scale lengths, which can all be tied to a principal culprit, setting simultaneously coupled vertical and radial motions in the disc, while also leaving room for better agreements with the data through fine-tuning. Given the multiple passages of Sgr around the Galaxy, we discuss other predictions which will soon be within reach, if not already with the current DR2 data.

\begin{itemize}
\item {\bf Outer disc studies of the phase-space spiral:}

For the moment, the DR2 6D data alone lets us securely probe a volume out to $R\sim12 \,\rm{kpc}$. But what would the phase-space spiral look like in the outer disc under the Sgr model? Indeed if the disc grew inside-out, which observations and numerical simulations of galaxy formation suggest \citep[e.g.][]{bird13,minchev15,hayden15,ma17,navarro18} and that Sgr was as massive as now favoured by different lines of evidence from chemical abundance measurements in the Sgr stream \citep{deboer14} to stream velocity dispersion \citep{gibbons16} and disc dynamics \citep{laporte18b}, then we expect there to be a sweetspot Galactocentric radius where perturbations from different passages may overlap due to the longer orbital timescales $\mathcal{O}(\rm{Gyr})$. This should be visible in the outer disc as already pointed out in \cite{laporte18c}. 

Because in the outer disc, $T_{\kappa} \sim T_{\nu} \sim \mathcal{O}(\rm{Gyr})$ the phase-space diagram of the disc should exhibit a large number of fine grained structures. In \cite{laporte18b} we presented that in addition to exciting large scale oscillations of the disc (e.g. Monoceros Ring), Sgr also excited tidal tails in the outer disc for which we see its present-day remnants as thin stream-like structures in the Anticenter (e.g. ACS/EBS). If one were to make a $(v_{z},z)$ phase-space map at $R\sim16-18 \,\rm{kpc}$, one would not recover a simple spiral anymore but a superposition of overdensities from prior generations of bending waves. Furthermore, structures like the ACS/EBS would also appear as detached chunks of phase-space spirals. In Figure 16, we show and example of what such maps could look like. Not surprisingly, a more disturbed disc $(v_{z},z)$ phase-plane is evident and one can discern in the $v_{\phi}$ plane the remains of a prior perturbation and a much large scale one that is being setup. Future surveys such as WEAVE, 4MOST, SDSS V, PFS will allow us to peer further in and further out in the stellar disc of the Milky Way to hopefully uncover such diagrams which will add valuable information on the {\it perturbation history} of the disc. Such maps, will ultimately provide a window into the orbital mass-loss history of the Sgr and potentially a track record of its repeated impacts on the disc.

\item {\bf Star formation histories:} While we have presented numerous dynamical predictions for our model, it is worthwhile to also think of other tests to be explored beyond the current dynamical signatures on the disc. Indeed, Gaia samples the phase-space distribution of stars in the Galaxy but also characterises the colour magnitude diagram of the Galaxy, allowing us to derive star formation histories at different locations, giving us another window into the Galaxy's past evolutionary path. A massive Sgr progenitor galaxy, may even be able to modulate the star formation history of the Galactic disc which could be associated with prior pericentric passages, or separated by the orbital timescales between 2 and 0.5 Gyr \citep[e.g. see][as dynamical friction effectively gradually decreases the orbital timescales of Sgr]{laporte18b}. While speculative, some preliminary intriguing results show modulations in the more recent SFH of the Milky way (Gallart private communication).
\end{itemize}

\begin{figure*}
\includegraphics[width=1.0\textwidth,trim=0mm 0mm 0mm 0mm, clip]{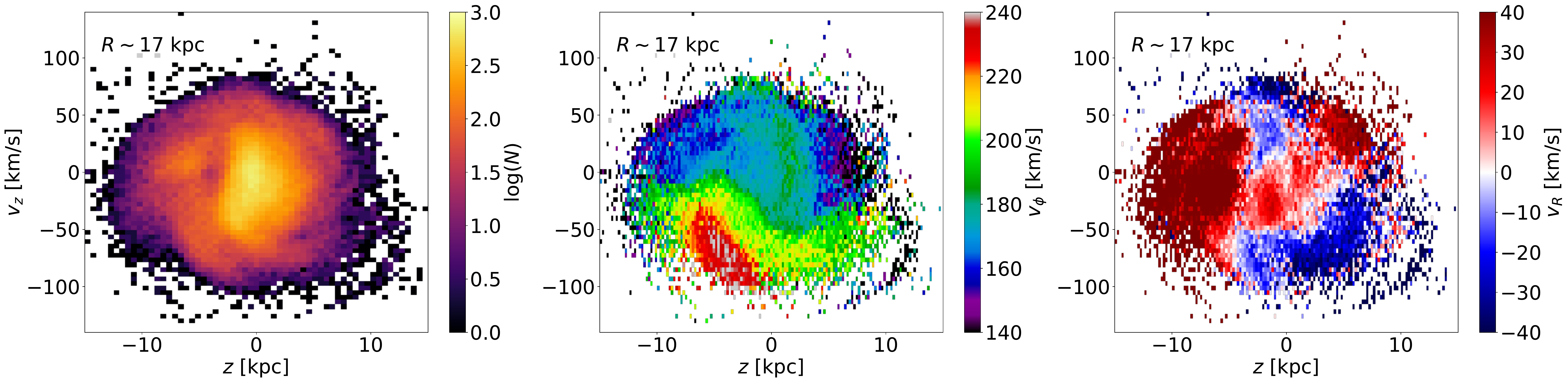}
\caption[]{An example of outer disc phase-plane from the Sgr impact model in $\rho(v_z,z), v_{\phi}(v_z,z), v_{R}(v_z,z),$ at a Galactocentric distance of $R\sim17\,\rm{kpc}$. The phase-plane is now much more sheared and substructures in number counts are more clearly visible than in the inner disc. We also notice the superposition of different generations of perturbations, particularly in the $v_{\phi}(v_z,z)$ plane. Such similar structures would be identifiable with future spectroscopic surveys (e.g. SDSS V, PSF, WEAVE, 4MOST).}
\end{figure*}

\subsection{Signatures in Stellar populations: a recent disc-wide phenomenon and cautionary tale}

\cite{tian18} presented a study dissecting the phase-space spiral using ages from LAMOST spectra and deduced that the age of the perturbation could be dated back to at least 500 Myr. This seems rather convincing although the small number of stars made the visibility of the spiral at small ages somehow difficult in the $v_\phi-$projection. The advantage of having different surveys and spectra means that their combination could be particularly powerful with the {\it Gaia} DR2 data. In this section we make use of the data compilation catalog provided by \cite{sanders18} who derived isochrone ages by combining all the major spectroscopic surveys (APOGEE, LAMOST, RAVE, GALAH).  This undoubtedly increases the samples for which we present in Figure 17 an age decomposition in overdensity in $\delta\rho(v_{z},z)$ and $v_{\phi}(v_{z},z)$. These maps are constructed by considering all stars within a distance of $d=2$ kpc from the Sun and using the ``0'' quality flag of \cite{sanders18} and binning in ages.

 Although the signal gets weaker with decreasing number of stars used in each panels, we can see that the phase-space spiral is present at {\it all ages}. This corroborates \cite{tian18} findings, with the difference that here we are actually able to clearly detect the phase-space spiral at ages larger than $\tau>6\,\rm{Gyr}$. While the ages determinations are different between \cite{tian18} and  \cite{sanders18}, these are recognised to be of the order of 30\% percent which is too low to explain why we detect phase-space spiral at older ages. The likely cause for this difference is due to the fact that \cite{tian18} used bins which were too small to measure a median $v_{\phi}$, although their large sample exceeding $N>100,000$ would have allowed them to do so. Our detection of the phase-space spiral at all ages however is a clear indication that this is not physical and cautions the use of isochrone determined ages to date the time of the perturbation. Indeed, we note that we are able to depict the phase-space spiral down to ages of 100 Myr. This is not physical as there is simply not enough time for the Milky Way to respond on such short timescales. This serves as a reminder that we need to treat isochrone ages as relative ages and not absolute ones. \cite{queiroz18} shows that for younger stars the spread in error is in fact larger. Nonetheless, the idea of \cite{tian18} is worth pursuing further particularly in the outer disc as we argued in section 6.2. With upcoming future large spectroscopic surveys such as WEAVE, 4MOST, PFS, SDSS V, age decomposition of the phase-space $(z,v_{z})$ could potentially probe perturbations from different passages as a function of Galactocentric radius. The compilation from \cite{sanders18} demonstrates the power of combining {\it all} surveys in the northern and southern to interpret localised disturbances, which allowed us to probe the phase-space spiral at all ages.

\begin{figure*}
\includegraphics[width=1.0\textwidth,trim=0mm 0mm 0mm 0mm, clip]{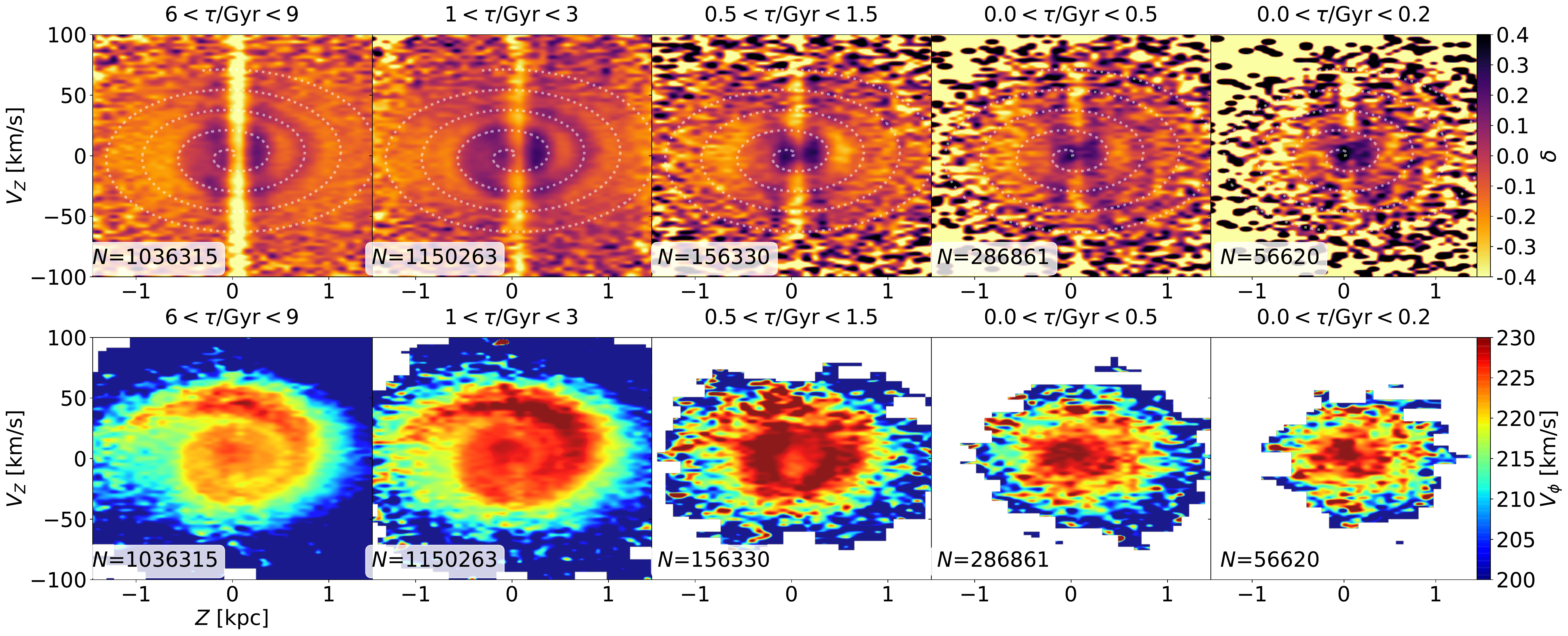}
\caption[]{ Age dissection of the $(v_{z},z)$ spiral in number counts (through the overdensity parameter) and its $v_{\phi}(v_{z},z)$ projection using the \cite{sanders18} compilation catalog. The binsize used here is $\Delta z \times\Delta v_{z}=0.08\,\rm{kpc}\times2 \,\rm{km\,s^{-1}}$. An archimedian spiral is overplotted to help the reader discern the second wrap arching in the positive $z-$quadrant towards $z=0.7\,\rm{kpc}$. We note that the spiral can be traced at {\it all} ages, showing that all disc populations are reacting to the recent perturbation. The detection at very young ages with $\tau<0.5\,\rm{Gyr}$ is not physical signaling the limitation of isochrone ages to date perturbations.}
\end{figure*}

\subsection{Other perturbers?}

Finally, in this section we discuss the possibility of other potential perturbers put forward to set the phase-space spiral and current velocity field seen in the DR2 data. Given the wealth of data at hand we can already rule out all these alternatives.

\begin{itemize}
\item {\bf The Magellanic Clouds:}
The LMC has long been suggested as a perturber to the Milky Way disc through the wake response of halo it leaves \citep{weinberg98} which would explain the origin of the HI warp \citep{weinberg06}. This has been recently revisited in \cite{laporte18a} who presented live N-body simulations of the LMC in a first infall scenario \citep{besla07} as suggested by recent proper motion measurements \citep{kallivayalil13}. Because of its larger than previously thought mass, the LMC causes a shift in the center of mass of the MW \citep{gomez15}, causing over/underdensities of order $|\delta\rho|\sim50\%$ in MW's dark matter halo which torque the disc \citep{laporte18a,garavito-camargo19}. While this mechanism is able to penetrate to the center of the halo and warp the disc to produce a simple warp consisting of 3 Fourier terms in its vertical decomposition with lines of nodes coinciding those from the HI maps \citep{levine06,kalberla09}, the mechanism does not have enough time to operate to give rise to vertical corrugations. In Figure 18, we show the same volume used in Figure 7 for the present-day snapshot of the MW-LMC N-body models of \cite{laporte18a}. Indeed, the disc phase-plane in $v_{\phi}(v_z,z)$ seems stretched as a result of the recent response of the disc but given that the LMC is just past its pericentric passage, no spiral has been able to form. Thus we can rule out a Magellanic origin to the perturbations. We already noted in \cite{laporte18b} where we presented models considering {\it both} satellites (Sgr, LMC) that the LMC's effect, while not negligible in the outer disc, only acts as an extra modulation term to the response of the disc to Sgr.
\begin{figure}
\includegraphics[width=.50\textwidth,trim=0mm 0mm 0mm 0mm, clip]{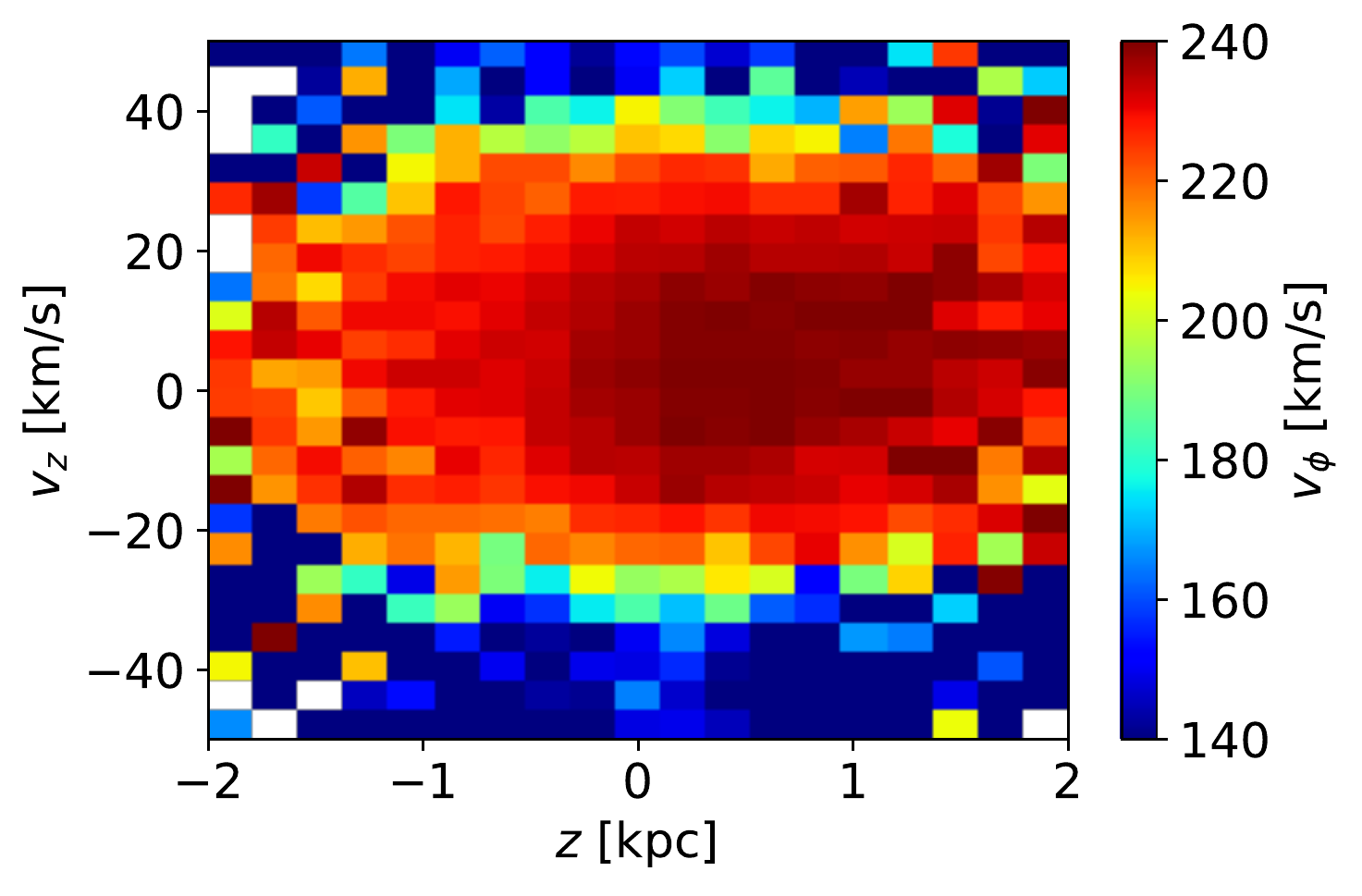}
\caption[]{Perturbation due to the LMC on the Galactic disc in the $v_\phi(v_z,z)$ plane. Indeed, as shown already in \citep{laporte18a} the LMC on a first infall passage does not affect the solar neighbourhood appreciably because there has not been enough time elapsed for the disc to react, only a simple warp forms in the outer disc but no corrugations in the solar vicinity. }
\end{figure}

\item {\bf Purely dark subhalos:}
Smaller mass perturbers from dark matter halos are also not capable of forming the phase-space spiral seen in the Milky Way as the perturbations are only kinematic in nature and would not affect the density profile of the disc globally. This is ruled out by the {\it Gaia} DR2 data as we showed that the phase-space spiral to be localised at different radii from 6 kpc to 10 kpc. 

While idealised numerical experiments, have been performed to study the effect of purely dark subhalos \citep{feldmann15} on the stability of the disc, these small mass perturbers do not produce strong density fluctuations as seen in the data \citep{widrow12}, but only weak kinematic perturbations which shear and get damped on $\mathcal{O}(100)\,\rm{Myr}$ timescales. However, in $\Lambda$CDM, Galactic halos are highly substructured with a clear mass function predicted by dark matter only simulations \citep[e.g.][]{Springel2008}.  Their collective effect of the subhalo mass function has been proposed to excite bending waves in the disc in numerical experiments by \cite{chequers18}, however we note that these used initial conditions which did not model this effect correctly. Indeed such experiments have used velocity distribution functions similar to that of the underlying parent halo, which is not representative of cosmologically formed Galactic dark matter halos \citep{sawala17}. Moreover, under full cosmological contexts, the subhalo mass function in the vicinity of the disc gets severely depleted \citep{d'onghia10, errani17, SGK17}. This has in fact been studied in very high resolution N-body simulations not prone to artificial disruption of subhalos in \citep{errani17}. So we do not expect any major perturbations from purely dark subhalos to set the currently observed global signs of phase mixing in the disc. While these may have a collective effect if studied in isolation and with realistic setups informed by cosmological models, their signal will nonetheless be completely dwarfed by the effect of Sgr on the disc, which recent independent lines of evidence place its progenitor mass at least $6\times 10^{10}\,\rm{M_{\odot}}$ \citep{deboer14,gibbons16,laporte18b}.

\item {\bf An ancient merger:}
Recently, with the Gaia DR1 and DR2 releases, renewed interest in an ancient merger previously known to be associated with the progenitor galaxy of $\omega$Cen \citep[e.g.][]{meza05,navarro11} that built most of the inner stellar halo has come into focus \citep[e.g.][]{belokurov18,helmi18}, situating the merger around $\sim10\,\rm{Gyr}$ ago. While this event must have been violent since many chemical thick disc stars are now populating the stellar halo, this could not possibly have formed the phase-space spiral that we see at present. First, because of the long timescales, the phase-space spiral would need to be more tightly wound and almost non discernible by today. Second, we observe the phase-space spiral at {\it all} stellar ages. This means that all the stellar populations in the disc responded to a single {\it recent} major perturbation, ruling out the possibility of an ancient merger origin.

\item {\bf The remains of a bar buckling instability:} Another type of perturbations which also occurs in our simulations but dwarfed by the continuous tides of Sgr is that of the bar buckling. \cite{khoperskov18} present a high-resolution run of an isolated MW-like galaxy undergoing bar buckling instability. This induces the generation of bending waves and not surprisingly exhibits also similar shaped spiral patterns in $(v_{z},z)$. While interesting in its own merit, we note that the amplitudes of the perturbations are low compared to the DR2 data. Moreover, a drawback to this model is that such a mechanism would not explain perturbations beyond the Gaia 6D volume, and would require invoking separate explanations for the large-scale velocity fields, even if it were to be fine tuned to reproduce the \cite{antoja18} spiral. Indeed, our model also went through a bar buckling instability, but the bending waves associated with it are dwarfed by those excited by the interaction with Sgr which resets the phase-space spiral during the last Gyrs of its evolution after each passage. A Sgr origin is indeed capable of reconciling both the amplitude of density fluctuation in the solar neighbourhood and the outer disc {\it simultaneously} \citep{laporte18b} and as we have shown here in this contribution, also the Gaia 6D bulk velocities as well as the origin of the phase-space spiral. While the proposed scenario could be applicable to other galaxies, its relevance to our own Milky Way seems vague and no results have been presented on the local velocity field. Given the current evidence in the data that all stellar populations have recently responded to the perturbation in the same manner, as demonstrated by our overdensity $\delta(v_z,z)$ age decompositions (see the top panels in Figure 17), the proposed scenario does not seem favoured by the data and possibly ruled out by our current analysis. We note that isochrone ages below 0.5 Gyr cannot put a limit on the onset of the perturbation, but \citep{Cheng19} show using a sample of OB stars that the phase-space spiral is not visible as expected in the Sgr pertubation scenario.
 \end{itemize}

\section{Discussion}

Only a few models considering the self-gravitating response of the disc to the Sgr dSph exist \citep{purcell11,gomez13,laporte18a} and quantitative successes have only recently been achieved in the latest experiments of \cite{laporte18b} which considered more realistic intial conditions in order to explain the origin of outer disc overdensities, thin stellar streams towards the Anticenter \citep{laporte18c} while satisfying pre-{\it Gaia} constraints on disc asymmetries in the solar neighbourhood \citep{widrow12}. 

All these past experiments considered a fiducial disc model to study its response to radial and vertical density perturbations. In this regard, the range of possibilities of outcomes on the amplitude of the fluctuations in the radial and vertical directions are somewhat restricted, leaving room for additional complex phenomenology\footnote{In \cite{laporte18b}, we considered a disc with a Toomre $Q>2$ everywhere because we were interested in studying the response of a disc which did not develop a bar or spiral arms through internal instabilities when evolved in isolation, for ease of interpretation. We caution that this was a purely numerical choice.}. For now, we note that many features revealed in the Gaia data can be qualitatively reproduced by a model of the Milky Way disc interacting with the Sgr dSph and that the structure of the velocity field and the final morphology of the disc points to a close link between the interaction of the Milky Way with its most recently almost engulfed massive satellite galaxy. We note that our quantitative mismatches in the z-amplitude of the snail pattern, or radial velocity maps (e.g. Fig 8, top left panel Fig 4 ) are below a factor of $\sim1.5$ in the amplitude which can easily be accounted for with fine tuned models by varying the progenitor mass of the perturber and disc internal structure. 

Our study clearly highlights to the importance of the role of Sgr in exciting many disequilibrium features in the Milky Way. In section 5.3, we considered other potential challengers to the Sgr impact model. We showed in Figure 19 that the LMC could not produce a phase-space spiral in the solar neighbourhood because it is just past their first pericentric passage \citep{besla07,kallivayalil13} leaving not enough time for it to form \citep{laporte18a}. Ancient merger events are also ruled out by our Figure 19 which shows an identical overdensity phase-space spiral at all stellar ages. While dark subhalos have been proposed to excite vertical density waves \citep{widrow14,feldmann15, chequers18}, their amplitudes are systematically too low to be consistent with the data and any signal would be buried by Sgr's own effect on the disc. In fact, we checked that our L1 model of Sgr with $6\times 10^{10}\,\rm{M_{\odot}}$ but lower central density could not set off perturbations seen in the data, leaving a massive Sgr as the only presently available culprit. Other more distant past merger events are also ruled out because of the identical shape of the phase-space spiral overdensity $\delta(v_{z},z,\tau)$ as a function of age $\tau$. While it has been suggested that the bar alone could produce structures reminiscent of the \cite{antoja18} snail pattern, we note that the amplitudes are not well reproduced and would require chance alignment for the phase-space spiral to match the data \citep{khoperskov18}. Moreover another drawback to this model is that it alone cannot predict the structures in the outer disc (e.g. Monoceros, TriAnd, ACS). If anything, we expect the bar buckling signal to be buried within the signal from Sgr as our own models also produce X-shaped bars. On the other hand, the future is bright as our model makes clear predictions in the outer disc which are still to be probed with upcoming surveys SDSS V, WEAVE, 4MOST, PFS. In \cite{laporte18b}, we argued that because of the different timescales present in the Galactic disc, the outer disc structures of the Milky Way together with those in the inner-Galaxy may probe different ``time regimes'' of the orbital mass-loss history of Sagittarius. The existence of the phase-space spiral \cite{antoja18} and excitation of tidal tails of the Milky Way \citep{laporte18c} in the outer disc will probe different events from a few 100 Myr to Gyrs. By combining the constraints at different Galactocentric radii, we may hope to be able to both uncover the origin of various non-axisymmetries and their relation to the interaction history of the Milky Way with Sgr in the last $\sim6\,\rm{Gyrs}$.

\section{Summary \& Conclusion}
Many of the signs of perturbations in the Milky Way such as the vertical disturbances in the MW within the solar neighbourhood, the spiral patterns in the solar vicinity, ridges in $v_{\phi}-R$, the Monoceros Ring and other overdensities such as EBS, ACS and TriAnd can {\it all be understood as part of the interaction of the disc with the Sgr dwarf galaxy}, with the condition that the progenitor was more massive than previously assumed with $M\sim6\times10^{10} \rm{M_{\odot}}$, a picture that has recently gained support on different but complementary grounds \citep[see][]{deboer14,gibbons16,laporte18b}. In summary, we show that:

\begin{enumerate}

\item The velocity fields seen in the DR2 data as in \cite{katz18} can be reproduced qualitatively by an interaction of Sgr with the Milky Way disc which excites radial and vertical motions in the disc. These signals get translated in velocity space as a succession of inward and radial ridges in $V_{R}$ as well as kinematic warping of the disc in $V_{Z}$. The radial motion pattern is related to a bi-symmetric spiral which Sgr excites during its last pericentric passage which gets tightly wound at present. This also gives rise to more complex structures such as the ``bridge'' seen in the median radial velocity field. However, the amplitudes in the models considered are larger notably in $V_{R}$ by a factor of $\sim2$. This may be related to the orbital mass loss history of Sgr (which would affect the resulting tides acting on the Galactic disc) and/or the underlying disc model assumed for the MW.

\item The interaction with Sgr leaves signs of ongoing phase-mixing around the Sun, with patterns remarkably similar to those recently observed in the $Gaia$ DR2 data \citep{antoja18}. The disagreements are only within a factor of $\sim1.5$ in the $z$ amplitude of the signal, leaving room for more detailed exploration, using different numerical setups specifically tailored to stream fitting Sgr and studying the disc response during the last pericentric passage while varying the Galactic disc structure. 

\item The interaction model also produces asymmetric moving groups reminiscent of structures seen in the solar neighbourhood such as Coma Berenices (see Appendix).

\item Ridges similar to \cite{antoja18} are produced continuously by spiral arms excited by the interaction with Sgr throughout the evolution of the Milky Way. The current patterns seen today in $v_{\phi}-R$  have similar slopes to those excited in the simulations by Sgr, suggesting some possible link between the excitations of the ridges with the interaction with the Sgr. We do not get the same number of observed ridges, but other models of transient spirals do not either \citep{hunt18}. Some prominent ridges can be explained by resonant processes with the bar \citep{fragkoudi19}, thus future explorations will be needed to disentangle which ridges could be associated with a Sgr excitation origin. Indeed the spacing of these ridges every $\sim30 km\,s^{-1}$ in the {\it Gaia} DR2 data bares resemblance with the earlier studies of phase-mixing in the disc in the $U-V$ plane by \cite{minchev09} due to recent mergers. 

\item We find two previously unknown ridges at $v_\phi\sim120, 90\,\rm{km\,s^{-1}}$ below Arcuturus, confirming predictions of semi-analytic models of Galactic ringing \citep{minchev09}.

\item Another prediction from the Sgr impact model is that the phase-space spiral in the Anticenter should harbour signs of past perturbations due to the longer timescales sustained in the disc at large Galactocentric distances, with large amplitudes in the $z$-direction of order  $\Delta Z\gg 1 \rm{kpc}$.

\item We presented the first map of the phase-space spiral in overdensity $\delta\rho(v_z,z)$, showing clearly more than two wraps (Fig. 7). This 
would naturally increase the timescales of the onset of the perturbation derived through the use of toy-models by \cite{binney18} which missed this aspect of the data, putting them in better agreement with the accepted orbital timescales for Sgr from stream fitting \citep[e.g.][]{johnston05,law10} of $t\sim0.8\rm{Gyr}$.

\item We show that the phase-space spiral has the same shape in all stellar age populations in overdensity $\delta(v_z,z,\tau)$. This means that: 1) all the stars responded the same and 2) that the onset of the perturbation must have been recent, the latter point corroborating \cite{tian18}'s findings. This at face value rules out a bar buckling origin for it in our Galaxy.

\item The response of the disc is global and we trace the \cite{antoja18} phase-space spiral for the first time using the DR2 data alone out to appreciably small and larger Galactocentric distances where their shape changes as expected from the underlying potential, a behaviour that is not surprisingly captured in the numerical model. These maps hold crucial information which will be invaluable for any attempts trying to model the zeroth order distribution function of the Milky Way or infer the mass of Sgr during its last pericentric passage.

\end{enumerate}

We note that \cite{quillen09,purcell11} argued that the Sgr dwarf could be an architect of spiral structure in the Milky Way. At the time, not many constraints around the Sun were available. From our new models, we conclude that the Sgr dwarf may be more closely linked to the disc than previously thought, generating the outer disc structures \citep{laporte18b,laporte18c} all the way to shaping the central part of the Galaxy and the phase-space structure around the the Sun as revealed by the {\it Gaia} satellite. We showed that the impact of Sgr can explain the origins of many different non-equilibrium features, presenting some qualitative/quantitative agreements. Requiring matches between the models and the data (e.g. Sgr stream and disc asymmetries) could tell us a lot about the disc structure and orbital mass-loss history of Sgr. Finally, a number of predictions of the Sgr impact model are awaiting to be confirmed with upcoming spectroscopic surveys which will complement {\it Gaia} which should further open up the potential of dating even more ancient encounters.

\section{Acknowledgments}
Simulations are fully available to anyone who request them to the first author and will be made public shortly. This work has made use of data from the European Space Agency (ESA) mission {\it Gaia} (\url{https://www.cosmos.esa.int/gaia}), processed by the {\it Gaia} Data Processing and Analysis Consortium (DPAC, \url{https://www.cosmos.esa.int/web/gaia/dpac/consortium}). Funding for the DPAC has been provided by national institutions, in particular the institutions participating in the {\it Gaia} Multilateral Agreement.  This work made use of {\tt numpy, scipy} and {\tt matplotlib} \citep{numpy, scipy, hunter07} as well as the {\tt astropy} package \citep{astropy1, astropy2}. We thank Volker Springel for giving us access to the {\sc gadget-3} code. CL is supported by a CITA National Fellow award. This work used the Extreme Science and Engineering Discovery Environment (XSEDE), which is supported by National Science Foundation grant number OCI-1053575. We also acknowledge use of computing facilities at the Rechenzentrum Garching (RZG) and the Max Planck Institute for Astrophysics (MPA). IM acknowledges support by the Deutsche Forschungsgemeinschaft under the grant MI 2009/1-1. KVJ's contributions were supported by NSF grants AST-1312196 and AST-1614743. FAG acknowledges support from Fondecyt Regular 1181264, and funding from the Max Planck Society through a €œPartner Group€ grant. CL thanks David W. Hogg for inviting him at the {\it Gaia} DR2 release meeting hosted at the Center for Computation Astrophysics (CCA), where this project was initiated. The new ridges are named after Snoop Dogg and Herbie Hancock. CL acknowledges interesting discussions with Carme Gallart, Ronald Drimmel, Hamish Silverwood, Teresa Antoja and thanks Vasily Belokurov and Martin Smith for teaching him data mining 9 years ago in Cambridge. CL thanks James Francies and Maurice Brown for the jam session at the 55 bar during a work trip and dedicates this first Gaia paper to Federico Gonz\'alez Pe\~na for his teachings, patience and opportunity to learn. 

\bibliographystyle{mn2e}
\bibliography{master2.bib}{}

\begin{thebibliography}{}

\bibitem[\protect\citeauthoryear{{The merger that led to the formation of the
  Milky Way's inner stellar halo and thick disk}}{kho}{}]{khoperskov18}


\bibitem[\protect\citeauthoryear{{Antoja}, {Helmi}, {Romero-G{\'o}mez}, {Katz},
  {Babusiaux}, {Drimmel}, {Evans}, {Figueras}, {Poggio}, {Reyl{\'e}}, {Robin},
  {Seabroke} \& {Soubiran}}{{Antoja} et~al.}{2018}]{antoja18}
{Antoja} T.,  {Helmi} A.,  {Romero-G{\'o}mez} M.,  {Katz} D.,  {Babusiaux} C.,
  {Drimmel} R.,  {Evans} D.~W.,  {Figueras} F.,  {Poggio} E.,  {Reyl{\'e}} C.,
  {Robin} A.~C.,  {Seabroke} G.,    {Soubiran} C.,  2018, \nat, 561, 360

\bibitem[\protect\citeauthoryear{{Arifyanto} \& {Fuchs}}{{Arifyanto} \&
  {Fuchs}}{2006}]{arifyanto06}
{Arifyanto} M.~I.,  {Fuchs} B.,  2006, \aap, 449, 533

\bibitem[\protect\citeauthoryear{{Astropy Collaboration}, {Price-Whelan},
  {Sip{\H o}cz}, {G{\"u}nther}, {Lim}, {Crawford}, {Conseil}, {Shupe}, {Craig},
  {Dencheva} \& {Astropy Contributors}}{{Astropy Collaboration}
  et~al.}{2018}]{astropy2}
{Astropy Collaboration} {Price-Whelan} A.~M.,  {Sip{\H o}cz} B.~M.,
  {G{\"u}nther} H.~M.,  {Lim} P.~L.,  {Crawford} S.~M.,  {Conseil} S.,  {Shupe}
  D.~L.,  {Craig} M.~W.,  {Dencheva} N.,    {Astropy Contributors} 2018, \aj,
  156, 123

\bibitem[\protect\citeauthoryear{{Astropy Collaboration}, {Robitaille},
  {Tollerud}, {Greenfield}, {Droettboom}, {Bray}, {Aldcroft}, {Davis},
  {Ginsburg}, {Price-Whelan}, {Kerzendorf} \& {Conley}}{{Astropy Collaboration}
  et~al.}{2013}]{astropy1}
{Astropy Collaboration} {Robitaille} T.~P.,  {Tollerud} E.~J.,  {Greenfield}
  P.,  {Droettboom} M.,  {Bray} E.,  {Aldcroft} T.,  {Davis} M.,  {Ginsburg}
  A.,  {Price-Whelan} A.~M.,  {Kerzendorf} W.~E.,    {Conley} 2013, \aap, 558,
  A33

\bibitem[\protect\citeauthoryear{{Bailer-Jones}, {Rybizki}, {Fouesneau},
  {Mantelet} \& {Andrae}}{{Bailer-Jones} et~al.}{2018}]{bailerjones18}
{Bailer-Jones} C.~A.~L.,  {Rybizki} J.,  {Fouesneau} M.,  {Mantelet} G.,
  {Andrae} R.,  2018, \aj, 156, 58

\bibitem[\protect\citeauthoryear{{Bailin}}{{Bailin}}{2004}]{bailin04}
{Bailin} J.,  2004, PhD thesis, The University of Arizona, Arizona, USA

\bibitem[\protect\citeauthoryear{{Belokurov}, {Erkal}, {Evans}, {Koposov} \&
  {Deason}}{{Belokurov} et~al.}{2018}]{belokurov18}
{Belokurov} V.,  {Erkal} D.,  {Evans} N.~W.,  {Koposov} S.~E.,    {Deason}
  A.~J.,  2018, \mnras, 478, 611

\bibitem[\protect\citeauthoryear{{Bergemann}, {Sesar}, {Cohen}, {Serenelli},
  {Sheffield}, {Li}, {Casagrande}, {Johnston}, {Laporte}, {Price-Whelan},
  {Schoenrich} \& {Gould}}{{Bergemann} et~al.}{2018}]{bergemann18}
{Bergemann} M.,  {Sesar} B.,  {Cohen} J.~G.,  {Serenelli} A.~M.,  {Sheffield}
  A.,  {Li} T.~S.,  {Casagrande} L.,  {Johnston} K.~V.,  {Laporte} C.~F.~P.,
  {Price-Whelan} A.~M.,  {Schoenrich} R.,    {Gould} A.,  2018, ArXiv e-prints

\bibitem[\protect\citeauthoryear{{Bernard}, {Ferguson}, {Barker}, {Hidalgo},
  {Ibata}, {Irwin}, {Lewis}, {McConnachie}, {Monelli} \& {Chapman}}{{Bernard}
  et~al.}{2012}]{bernard12}
{Bernard} E.~J.,  {Ferguson} A.~M.~N.,  {Barker} M.~K.,  {Hidalgo} S.~L.,
  {Ibata} R.~A.,  {Irwin} M.~J.,  {Lewis} G.~F.,  {McConnachie} A.~W.,
  {Monelli} M.,    {Chapman} S.~C.,  2012, \mnras, 420, 2625

\bibitem[\protect\citeauthoryear{{Besla}, {Kallivayalil}, {Hernquist},
  {Robertson}, {Cox}, {van der Marel} \& {Alcock}}{{Besla}
  et~al.}{2007}]{besla07}
{Besla} G.,  {Kallivayalil} N.,  {Hernquist} L.,  {Robertson} B.,  {Cox} T.~J.,
   {van der Marel} R.~P.,    {Alcock} C.,  2007, \apj, 668, 949

\bibitem[\protect\citeauthoryear{{Binney} \& {Sch{\"o}nrich}}{{Binney} \&
  {Sch{\"o}nrich}}{2018}]{binney18}
{Binney} J.,  {Sch{\"o}nrich} R.,  2018, \mnras, 481, 1501

\bibitem[\protect\citeauthoryear{{Binney} \& {Tremaine}}{{Binney} \&
  {Tremaine}}{2008}]{BT08}
{Binney} J.,  {Tremaine} S.,  2008, {Galactic Dynamics: Second Edition}.
Princeton University Press

\bibitem[\protect\citeauthoryear{{Bird}, {Kazantzidis}, {Weinberg}, {Guedes},
  {Callegari}, {Mayer} \& {Madau}}{{Bird} et~al.}{2013}]{bird13}
{Bird} J.~C.,  {Kazantzidis} S.,  {Weinberg} D.~H.,  {Guedes} J.,  {Callegari}
  S.,  {Mayer} L.,    {Madau} P.,  2013, \apj, 773, 43

\bibitem[\protect\citeauthoryear{{Bland-Hawthorn}, {Sharma}, {Tepper-Garcia},
  {Binney}, {Freeman}, {Hayden}, {Kos}, {De Silva}, {Ellis}, {Lewis}, {Asplund}
  \& {Buder}}{{Bland-Hawthorn} et~al.}{2019}]{jbh18}
{Bland-Hawthorn} J.,  {Sharma} S.,  {Tepper-Garcia} T.,  {Binney} J.,
  {Freeman} K.~C.,  {Hayden} M.~R.,  {Kos} J.,  {De Silva} G.~M.,  {Ellis} S.,
  {Lewis} G.~F.,  {Asplund} M.,    {Buder} S.,  2019, \mnras

\bibitem[\protect\citeauthoryear{{Blumenthal}, {Faber}, {Flores} \&
  {Primack}}{{Blumenthal} et~al.}{1986}]{blumenthal86}
{Blumenthal} G.~R.,  {Faber} S.~M.,  {Flores} R.,    {Primack} J.~R.,  1986,
  \apj, 301, 27

\bibitem[\protect\citeauthoryear{{Carlin}, {DeLaunay}, {Newberg}, {Deng},
  {Gole}, {Grabowski}, {Jin}, {Liu}, {Liu}, {Luo}, {Yuan}, {Zhang}, {Zhao} \&
  {Zhao}}{{Carlin} et~al.}{2013}]{carlin13}
{Carlin} J.~L.,  {DeLaunay} J.,  {Newberg} H.~J.,  {Deng} L.,  {Gole} D.,
  {Grabowski} K.,  {Jin} G.,  {Liu} C.,  {Liu} X.,  {Luo} A.-L.,  {Yuan} H.,
  {Zhang} H.,  {Zhao} G.,    {Zhao} Y.,  2013, \apjl, 777, L5

\bibitem[\protect\citeauthoryear{{Carrillo}, {Minchev}, {Kordopatis},
  {Steinmetz}, {Binney}, {Anders}, {Bienaym{\'e}}, {Bland-Hawthorn}, {Famaey},
  {Freeman}, {Gilmore} \& {Gibson}}{{Carrillo} et~al.}{2018}]{carrillo18}
{Carrillo} I.,  {Minchev} I.,  {Kordopatis} G.,  {Steinmetz} M.,  {Binney} J.,
  {Anders} F.,  {Bienaym{\'e}} O.,  {Bland-Hawthorn} J.,  {Famaey} B.,
  {Freeman} K.~C.,  {Gilmore} G.,    {Gibson} B.~K.,  2018, \mnras, 475, 2679

\bibitem[\protect\citeauthoryear{{Cheng}, {Liu}, {Mao} \& {Cui}}{{Cheng}
  et~al.}{2019}]{Cheng19}
{Cheng} X.,  {Liu} C.,  {Mao} S.,    {Cui} W.,  2019, \apjl, 872, L1

\bibitem[\protect\citeauthoryear{{Chequers}, {Widrow} \& {Darling}}{{Chequers}
  et~al.}{2018}]{chequers18}
{Chequers} M.~H.,  {Widrow} L.~M.,    {Darling} K.,  2018, \mnras, 480, 4244

\bibitem[\protect\citeauthoryear{{Crane}, {Majewski}, {Rocha-Pinto},
  {Frinchaboy}, {Skrutskie} \& {Law}}{{Crane} et~al.}{2003}]{crane03}
{Crane} J.~D.,  {Majewski} S.~R.,  {Rocha-Pinto} H.~J.,  {Frinchaboy} P.~M.,
  {Skrutskie} M.~F.,    {Law} D.~R.,  2003, \apjl, 594, L119

\bibitem[\protect\citeauthoryear{{Darling} \& {Widrow}}{{Darling} \&
  {Widrow}}{2019}]{darling18}
{Darling} K.,  {Widrow} L.~M.,  2019, \mnras, 484, 1050

\bibitem[\protect\citeauthoryear{{de Boer}, {Belokurov}, {Beers} \& {Lee}}{{de
  Boer} et~al.}{2014}]{deboer14}
{de Boer} T.~J.~L.,  {Belokurov} V.,  {Beers} T.~C.,    {Lee} Y.~S.,  2014,
  \mnras, 443, 658

\bibitem[\protect\citeauthoryear{{de Boer}, {Belokurov} \& {Koposov}}{{de Boer}
  et~al.}{2018}]{deboer17}
{de Boer} T.~J.~L.,  {Belokurov} V.,    {Koposov} S.~E.,  2018, \mnras, 473,
  647

\bibitem[\protect\citeauthoryear{{de la Vega}, {Quillen}, {Carlin},
  {Chakrabarti} \& {D'Onghia}}{{de la Vega} et~al.}{2015}]{delavega15}
{de la Vega} A.,  {Quillen} A.~C.,  {Carlin} J.~L.,  {Chakrabarti} S.,
  {D'Onghia} E.,  2015, \mnras, 454, 933

\bibitem[\protect\citeauthoryear{{Deason}, {Belokurov} \& {Koposov}}{{Deason}
  et~al.}{2018}]{deason18}
{Deason} A.~J.,  {Belokurov} V.,    {Koposov} S.~E.,  2018, \mnras, 473, 2428

\bibitem[\protect\citeauthoryear{{Dehnen}}{{Dehnen}}{1998}]{dehnen98}
{Dehnen} W.,  1998, \aj, 115, 2384

\bibitem[\protect\citeauthoryear{{Dehnen}}{{Dehnen}}{2000}]{dehnen00}
{Dehnen} W.,  2000, \aj, 119, 800

\bibitem[\protect\citeauthoryear{{Dierickx} \& {Loeb}}{{Dierickx} \&
  {Loeb}}{2017}]{dierickx17}
{Dierickx} M.~I.~P.,  {Loeb} A.,  2017, \apj, 836, 92

\bibitem[\protect\citeauthoryear{{D'Onghia}, {Springel}, {Hernquist} \&
  {Keres}}{{D'Onghia} et~al.}{2010}]{d'onghia10}
{D'Onghia} E.,  {Springel} V.,  {Hernquist} L.,    {Keres} D.,  2010, \apj,
  709, 1138

\bibitem[\protect\citeauthoryear{{Dorman}, {Guhathakurta}, {Seth}, {Weisz},
  {Bell}, {Dalcanton}, {Gilbert}, {Hamren}, {Lewis}, {Skillman}, {Toloba} \&
  {Williams}}{{Dorman} et~al.}{2015}]{dorman15}
{Dorman} C.~E.,  {Guhathakurta} P.,  {Seth} A.~C.,  {Weisz} D.~R.,  {Bell}
  E.~F.,  {Dalcanton} J.~J.,  {Gilbert} K.~M.,  {Hamren} K.~M.,  {Lewis} A.~R.,
   {Skillman} E.~D.,  {Toloba} E.,    {Williams} B.~F.,  2015, \apj, 803, 24

\bibitem[\protect\citeauthoryear{{Drimmel}}{{Drimmel}}{2000}]{drimmel00}
{Drimmel} R.,  2000, \aap, 358, L13

\bibitem[\protect\citeauthoryear{{Eggen}}{{Eggen}}{1969}]{eggen69}
{Eggen} O.~J.,  1969, \pasp, 81, 553

\bibitem[\protect\citeauthoryear{{Errani}, {Pe{\~n}arrubia}, {Laporte} \&
  {G{\'o}mez}}{{Errani} et~al.}{2017}]{errani17}
{Errani} R.,  {Pe{\~n}arrubia} J.,  {Laporte} C.~F.~P.,    {G{\'o}mez} F.~A.,
  2017, \mnras, 465, L59

\bibitem[\protect\citeauthoryear{{Feast}, {Menzies}, {Matsunaga} \&
  {Whitelock}}{{Feast} et~al.}{2014}]{feast14}
{Feast} M.~W.,  {Menzies} J.~W.,  {Matsunaga} N.,    {Whitelock} P.~A.,  2014,
  \nat, 509, 342

\bibitem[\protect\citeauthoryear{{Feldmann} \& {Spolyar}}{{Feldmann} \&
  {Spolyar}}{2015}]{feldmann15}
{Feldmann} R.,  {Spolyar} D.,  2015, \mnras, 446, 1000

\bibitem[\protect\citeauthoryear{{Fragkoudi}, {Katz}, {White}, {Di Matteo},
  {Trick}, {Sormani}, {Khoperskov}, {Haywood}, {Hall{\'e}} \&
  {G{\'o}mez}}{{Fragkoudi} et~al.}{2019}]{fragkoudi19}
{Fragkoudi} F.,  {Katz} D.,  {White} S.~D.~M.,  {Di Matteo} P.,  {Trick} W.,
  {Sormani} M.~C.,  {Khoperskov} S.,  {Haywood} M.,  {Hall{\'e}} A.,
  {G{\'o}mez} A.,  2019, arXiv e-prints

\bibitem[\protect\citeauthoryear{{Frinchaboy}, {Majewski}, {Mu{\~n}oz}, {Law},
  {{\L}okas}, {Kunkel}, {Patterson} \& {Johnston}}{{Frinchaboy}
  et~al.}{2012}]{frinchaboy12}
{Frinchaboy} P.~M.,  {Majewski} S.~R.,  {Mu{\~n}oz} R.~R.,  {Law} D.~R.,
  {{\L}okas} E.~L.,  {Kunkel} W.~E.,  {Patterson} R.~J.,    {Johnston} K.~V.,
  2012, \apj, 756, 74

\bibitem[\protect\citeauthoryear{{Gaia Collaboration}, {Katz}, {Antoja},
  {Romero-G{\'o}mez}, {Drimmel}, {Reyl{\'e}}, {Seabroke}, {Soubiran},
  {Babusiaux}, {Di Matteo} \& et al.}{{Gaia Collaboration}
  et~al.}{2018}]{katz18}
{Gaia Collaboration} {Katz} D.,  {Antoja} T.,  {Romero-G{\'o}mez} M.,
  {Drimmel} R.,  {Reyl{\'e}} C.,  {Seabroke} G.~M.,  {Soubiran} C.,
  {Babusiaux} C.,  {Di Matteo} P.,    et al. 2018, \aap, 616, A11

\bibitem[\protect\citeauthoryear{{Gao}, {et} \& {al.}}{{Gao}
  et~al.}{2008}]{gao08}
{Gao} L.,  {et}   {al.} 2008, \mnras, 387, 536

\bibitem[\protect\citeauthoryear{{Garavito-Camargo}, {Besla}, {Laporte},
  {Johnston}, {G{\'o}mez} \& {Watkins}}{{Garavito-Camargo}
  et~al.}{2019}]{garavito-camargo19}
{Garavito-Camargo} N.,  {Besla} G.,  {Laporte} C.~F.~P.,  {Johnston} K.~V.,
  {G{\'o}mez} F.~A.,    {Watkins} L.~L.,  2019, arXiv e-prints

\bibitem[\protect\citeauthoryear{{Garrison-Kimmel}, {Wetzel}, {Bullock},
  {Hopkins}, {Boylan-Kolchin}, {Faucher-Gigu{\`e}re}, {Kere{\v s}}, {Quataert},
  {Sanderson}, {Graus} \& {Kelley}}{{Garrison-Kimmel} et~al.}{2017}]{SGK17}
{Garrison-Kimmel} S.,  {Wetzel} A.,  {Bullock} J.~S.,  {Hopkins} P.~F.,
  {Boylan-Kolchin} M.,  {Faucher-Gigu{\`e}re} C.-A.,  {Kere{\v s}} D.,
  {Quataert} E.,  {Sanderson} R.~E.,  {Graus} A.~S.,    {Kelley} T.,  2017,
  \mnras, 471, 1709

\bibitem[\protect\citeauthoryear{{Gibbons}, {Belokurov} \& {Evans}}{{Gibbons}
  et~al.}{2017}]{gibbons16}
{Gibbons} S.~L.~J.,  {Belokurov} V.,    {Evans} N.~W.,  2017, \mnras, 464, 794

\bibitem[\protect\citeauthoryear{{G{\'o}mez}, {Besla}, {Carpintero},
  {Villalobos}, {O'Shea} \& {Bell}}{{G{\'o}mez} et~al.}{2015}]{gomez15}
{G{\'o}mez} F.~A.,  {Besla} G.,  {Carpintero} D.~D.,  {Villalobos} {\'A}.,
  {O'Shea} B.~W.,    {Bell} E.~F.,  2015, \apj, 802, 128

\bibitem[\protect\citeauthoryear{{G{\'o}mez}, {Minchev}, {O'Shea}, {Beers},
  {Bullock} \& {Purcell}}{{G{\'o}mez} et~al.}{2013}]{gomez13}
{G{\'o}mez} F.~A.,  {Minchev} I.,  {O'Shea} B.~W.,  {Beers} T.~C.,  {Bullock}
  J.~S.,    {Purcell} C.~W.,  2013, \mnras, 429, 159

\bibitem[\protect\citeauthoryear{{G{\'o}mez}, {Minchev}, {O'Shea}, {Lee},
  {Beers}, {An}, {Bullock}, {Purcell} \& {Villalobos}}{{G{\'o}mez}
  et~al.}{2012}]{gomez12}
{G{\'o}mez} F.~A.,  {Minchev} I.,  {O'Shea} B.~W.,  {Lee} Y.~S.,  {Beers}
  T.~C.,  {An} D.,  {Bullock} J.~S.,  {Purcell} C.~W.,    {Villalobos} {\'A}.,
  2012, \mnras, 423, 3727

\bibitem[\protect\citeauthoryear{{G{\'o}mez}, {Minchev}, {Villalobos}, {O'Shea}
  \& {Williams}}{{G{\'o}mez} et~al.}{2012}]{gomez12a}
{G{\'o}mez} F.~A.,  {Minchev} I.,  {Villalobos} {\'A}.,  {O'Shea} B.~W.,
  {Williams} M.~E.~K.,  2012, \mnras, 419, 2163

\bibitem[\protect\citeauthoryear{{Grillmair}}{{Grillmair}}{2006}]{grillmair06}
{Grillmair} C.~J.,  2006, \apjl, 651, L29

\bibitem[\protect\citeauthoryear{{Grillmair}}{{Grillmair}}{2011}]{grillmair11}
{Grillmair} C.~J.,  2011, \apj, 738, 98

\bibitem[\protect\citeauthoryear{{Hayden}, {Bovy}, {Holtzman}, {Nidever},
  {Bird}, {Weinberg}, {Andrews}, {Majewski}, {Allende Prieto}, {Anders},
  {Beers}, {Bizyaev} \& {Chiappini}}{{Hayden} et~al.}{2015}]{hayden15}
{Hayden} M.~R.,  {Bovy} J.,  {Holtzman} J.~A.,  {Nidever} D.~L.,  {Bird} J.~C.,
   {Weinberg} D.~H.,  {Andrews} B.~H.,  {Majewski} S.~R.,  {Allende Prieto} C.,
   {Anders} F.,  {Beers} T.~C.,  {Bizyaev} D.,    {Chiappini} C.,  2015, \apj,
  808, 132

\bibitem[\protect\citeauthoryear{{Helmi}, {Babusiaux}, {Koppelman}, {Massari},
  {Veljanoski} \& {Brown}}{{Helmi} et~al.}{2018}]{helmi18}
{Helmi} A.,  {Babusiaux} C.,  {Koppelman} H.~H.,  {Massari} D.,  {Veljanoski}
  J.,    {Brown} A.~G.~A.,  2018, \nat, 563, 85

\bibitem[\protect\citeauthoryear{{Hunt}, {Hong}, {Bovy}, {Kawata} \&
  {Grand}}{{Hunt} et~al.}{2018}]{hunt18}
{Hunt} J.~A.~S.,  {Hong} J.,  {Bovy} J.,  {Kawata} D.,    {Grand} R.~J.~J.,
  2018, \mnras, 481, 3794

\bibitem[\protect\citeauthoryear{Hunter}{Hunter}{2007}]{hunter07}
Hunter J.~D.,  2007, Computing In Science \& Engineering, 9, 90

\bibitem[\protect\citeauthoryear{{Ibata}, {Gilmore} \& {Irwin}}{{Ibata}
  et~al.}{1994}]{ibata94}
{Ibata} R.~A.,  {Gilmore} G.,    {Irwin} M.~J.,  1994, \nat, 370, 194

\bibitem[\protect\citeauthoryear{{Ibata} \& {Razoumov}}{{Ibata} \&
  {Razoumov}}{1998}]{ibata98}
{Ibata} R.~A.,  {Razoumov} A.~O.,  1998, \aap, 336, 130

\bibitem[\protect\citeauthoryear{{Johnston}, {Law} \& {Majewski}}{{Johnston}
  et~al.}{2005}]{johnston05}
{Johnston} K.~V.,  {Law} D.~R.,    {Majewski} S.~R.,  2005, \apj, 619, 800

\bibitem[\protect\citeauthoryear{Jones, Oliphant, Peterson et~al.,}{Jones
  et~al.}{01  }]{scipy}
Jones E.,  Oliphant T.,  Peterson P.,    et~al.,, 2001--, {SciPy}: Open source
  scientific tools for {Python}

\bibitem[\protect\citeauthoryear{{Kalberla} \& {Kerp}}{{Kalberla} \&
  {Kerp}}{2009}]{kalberla09}
{Kalberla} P.~M.~W.,  {Kerp} J.,  2009, \araa, 47, 27

\bibitem[\protect\citeauthoryear{{Kallivayalil}, {van der Marel}, {Besla},
  {Anderson} \& {Alcock}}{{Kallivayalil} et~al.}{2013}]{kallivayalil13}
{Kallivayalil} N.,  {van der Marel} R.~P.,  {Besla} G.,  {Anderson} J.,
  {Alcock} C.,  2013, \apj, 764, 161

\bibitem[\protect\citeauthoryear{{Kawata}, {Baba}, {Ciuc{\v a}}, {Cropper},
  {Grand}, {Hunt} \& {Seabroke}}{{Kawata} et~al.}{2018}]{kawata18}
{Kawata} D.,  {Baba} J.,  {Ciuc{\v a}} I.,  {Cropper} M.,  {Grand} R.~J.~J.,
  {Hunt} J.~A.~S.,    {Seabroke} G.,  2018, \mnras, 479, L108

\bibitem[\protect\citeauthoryear{{Laporte}, {G{\'o}mez}, {Besla}, {Johnston} \&
  {Garavito-Camargo}}{{Laporte} et~al.}{2018a}]{laporte18a}
{Laporte} C.~F.~P.,  {G{\'o}mez} F.~A.,  {Besla} G.,  {Johnston} K.~V.,
  {Garavito-Camargo} N.,  2018, \mnras, 473, 1218

\bibitem[\protect\citeauthoryear{{Laporte}, {Johnston}, {G{\'o}mez},
  {Garavito-Camargo} \& {Besla}}{{Laporte} et~al.}{2018b}]{laporte18b}
{Laporte} C.~F.~P.,  {Johnston} K.~V.,  {G{\'o}mez} F.~A.,  {Garavito-Camargo}
  N.,    {Besla} G.,  2018, \mnras

\bibitem[\protect\citeauthoryear{{Laporte}, {Johnston} \&
  {Tzanidakis}}{{Laporte} et~al.}{2019}]{laporte18c}
{Laporte} C.~F.~P.,  {Johnston} K.~V.,    {Tzanidakis} A.,  2019, \mnras, 483,
  1427

\bibitem[\protect\citeauthoryear{{Law} \& {Majewski}}{{Law} \&
  {Majewski}}{2010}]{law10}
{Law} D.~R.,  {Majewski} S.~R.,  2010, \apj, 714, 229

\bibitem[\protect\citeauthoryear{{Levine}, {Blitz} \& {Heiles}}{{Levine}
  et~al.}{2006}]{levine06}
{Levine} E.~S.,  {Blitz} L.,    {Heiles} C.,  2006, \apj, 643, 881

\bibitem[\protect\citeauthoryear{{Li}, {Sheffield}, {Johnston}, {Marshall},
  {Majewski}, {Price-Whelan}, {Damke}, {Beaton}, {Bernard}, {Richardson},
  {Sharma} \& {Sesar}}{{Li} et~al.}{2017}]{li17}
{Li} T.~S.,  {Sheffield} A.~A.,  {Johnston} K.~V.,  {Marshall} J.~L.,
  {Majewski} S.~R.,  {Price-Whelan} A.~M.,  {Damke} G.~J.,  {Beaton} R.~L.,
  {Bernard} E.~J.,  {Richardson} W.,  {Sharma} S.,    {Sesar} B.,  2017, \apj,
  844, 74

\bibitem[\protect\citeauthoryear{{Ludlow}, {Navarro}, {Angulo},
  {Boylan-Kolchin}, {Springel}, {Frenk} \& {White}}{{Ludlow}
  et~al.}{2014}]{ludlow14}
{Ludlow} A.~D.,  {Navarro} J.~F.,  {Angulo} R.~E.,  {Boylan-Kolchin} M.,
  {Springel} V.,  {Frenk} C.,    {White} S.~D.~M.,  2014, \mnras, 441, 378

\bibitem[\protect\citeauthoryear{{Ma}, {Hopkins}, {Wetzel}, {Kirby},
  {Angl{\'e}s-Alc{\'a}zar} \& {Faucher-Gigu{\`e}re}}{{Ma} et~al.}{2017}]{ma17}
{Ma} X.,  {Hopkins} P.~F.,  {Wetzel} A.~R.,  {Kirby} E.~N.,
  {Angl{\'e}s-Alc{\'a}zar} D.,    {Faucher-Gigu{\`e}re} C.-A.,  2017, \mnras,
  467, 2430

\bibitem[\protect\citeauthoryear{{Majewski}, {Skrutskie}, {Weinberg} \&
  {Ostheimer}}{{Majewski} et~al.}{2003}]{majewski03}
{Majewski} S.~R.,  {Skrutskie} M.~F.,  {Weinberg} M.~D.,    {Ostheimer} J.~C.,
  2003, \apj, 599, 1082

\bibitem[\protect\citeauthoryear{{Martin}, {Ibata} \& {Irwin}}{{Martin}
  et~al.}{2007}]{martin07}
{Martin} N.~F.,  {Ibata} R.~A.,    {Irwin} M.,  2007, \apjl, 668, L123

\bibitem[\protect\citeauthoryear{{Martinez-Medina}, {Pichardo}, {Peimbert} \&
  {Valenzuela}}{{Martinez-Medina} et~al.}{2018}]{martinez18}
{Martinez-Medina} L.~A.,  {Pichardo} B.,  {Peimbert} A.,    {Valenzuela} O.,
  2018, arXiv e-prints

\bibitem[\protect\citeauthoryear{{McMillan}}{{McMillan}}{2011}]{mcmillan11}
{McMillan} P.~J.,  2011, \mnras, 414, 2446

\bibitem[\protect\citeauthoryear{{Meza}, {Navarro}, {Abadi} \&
  {Steinmetz}}{{Meza} et~al.}{2005}]{meza05}
{Meza} A.,  {Navarro} J.~F.,  {Abadi} M.~G.,    {Steinmetz} M.,  2005, \mnras,
  359, 93

\bibitem[\protect\citeauthoryear{{Minchev}, {Martig}, {Streich}, {Scannapieco},
  {de Jong} \& {Steinmetz}}{{Minchev} et~al.}{2015}]{minchev15}
{Minchev} I.,  {Martig} M.,  {Streich} D.,  {Scannapieco} C.,  {de Jong} R.~S.,
     {Steinmetz} M.,  2015, \apjl, 804, L9

\bibitem[\protect\citeauthoryear{{Minchev}, {Nordhaus} \& {Quillen}}{{Minchev}
  et~al.}{2007}]{minchev07}
{Minchev} I.,  {Nordhaus} J.,    {Quillen} A.~C.,  2007, \apjl, 664, L31

\bibitem[\protect\citeauthoryear{{Minchev}, {Quillen}, {Williams}, {Freeman},
  {Nordhaus}, {Siebert} \& {Bienaym{\'e}}}{{Minchev} et~al.}{2009}]{minchev09}
{Minchev} I.,  {Quillen} A.~C.,  {Williams} M.,  {Freeman} K.~C.,  {Nordhaus}
  J.,  {Siebert} A.,    {Bienaym{\'e}} O.,  2009, \mnras, 396, L56

\bibitem[\protect\citeauthoryear{{Momany}, {Zaggia}, {Gilmore}, {Piotto},
  {Carraro}, {Bedin} \& {de Angeli}}{{Momany} et~al.}{2006}]{momany06}
{Momany} Y.,  {Zaggia} S.,  {Gilmore} G.,  {Piotto} G.,  {Carraro} G.,  {Bedin}
  L.~R.,    {de Angeli} F.,  2006, \aap, 451, 515

\bibitem[\protect\citeauthoryear{{Monari}, {Famaey}, {Minchev}, {Antoja} \&
  {Bienaym{\'e}}}{{Monari} et~al.}{2018a}]{monari18}
{Monari} G.,  {Famaey} B.,  {Minchev} I.,  {Antoja} T.,    {Bienaym{\'e}} 2018,
  Research Notes of the American Astronomical Society, 2, 32

\bibitem[\protect\citeauthoryear{{Monari}, {Famaey} \& {Siebert}}{{Monari}
  et~al.}{2016}]{monari16}
{Monari} G.,  {Famaey} B.,    {Siebert} A.,  2016, \mnras, 457, 2569

\bibitem[\protect\citeauthoryear{{Monari}, {Famaey}, {Siebert}, {Wegg} \&
  {Gerhard}}{{Monari} et~al.}{2018b}]{monari18b}
{Monari} G.,  {Famaey} B.,  {Siebert} A.,  {Wegg} C.,    {Gerhard} O.,  2018,
  arXiv e-prints

\bibitem[\protect\citeauthoryear{{Navarro}, {Abadi}, {Venn}, {Freeman} \&
  {Anguiano}}{{Navarro} et~al.}{2011}]{navarro11}
{Navarro} J.~F.,  {Abadi} M.~G.,  {Venn} K.~A.,  {Freeman} K.~C.,    {Anguiano}
  B.,  2011, \mnras, 412, 1203

\bibitem[\protect\citeauthoryear{{Navarro}, {Yozin}, {Loewen},
  {Ben{\'{\i}}tez-Llambay}, {Fattahi}, {Frenk}, {Oman}, {Schaye} \&
  {Theuns}}{{Navarro} et~al.}{2018}]{navarro18}
{Navarro} J.~F.,  {Yozin} C.,  {Loewen} N.,  {Ben{\'{\i}}tez-Llambay} A.,
  {Fattahi} A.,  {Frenk} C.~S.,  {Oman} K.~A.,  {Schaye} J.,    {Theuns} T.,
  2018, \mnras, 476, 3648

\bibitem[\protect\citeauthoryear{{Newberg}, {Yanny}, {Rockosi}, {Grebel},
  {Rix}, {Brinkmann}, {Csabai}, {Hennessy}, {Hindsley}, {Ibata}, {Ivezi{\'c}},
  {Lamb}, {Nash}, {Odenkirchen}, {Rave}, {Schneider}, {Smith}, {Stolte} \&
  {York}}{{Newberg} et~al.}{2002}]{newberg02}
{Newberg} H.~J.,  {Yanny} B.,  {Rockosi} C.,  {Grebel} E.~K.,  {Rix} H.-W.,
  {Brinkmann} J.,  {Csabai} I.,  {Hennessy} G.,  {Hindsley} R.~B.,  {Ibata} R.,
   {Ivezi{\'c}} Z.,  {Lamb} D.,  {Nash} E.~T.,  {Odenkirchen} M.,  {Rave}
  H.~A.,  {Schneider} D.~P.,  {Smith} J.~A.,  {Stolte} A.,    {York} D.~G.,
  2002, \apj, 569, 245

\bibitem[\protect\citeauthoryear{{Niederste-Ostholt}, {Belokurov} \&
  {Evans}}{{Niederste-Ostholt} et~al.}{2012}]{ostholt12}
{Niederste-Ostholt} M.,  {Belokurov} V.,    {Evans} N.~W.,  2012, \mnras, 422,
  207

\bibitem[\protect\citeauthoryear{{Pe{\~n}arrubia}, {Belokurov}, {Evans},
  {Mart{\'{\i}}nez-Delgado}, {Gilmore}, {Irwin}, {Niederste-Ostholt} \&
  {Zucker}}{{Pe{\~n}arrubia} et~al.}{2010}]{penarrubia10}
{Pe{\~n}arrubia} J.,  {Belokurov} V.,  {Evans} N.~W.,
  {Mart{\'{\i}}nez-Delgado} D.,  {Gilmore} G.,  {Irwin} M.,
  {Niederste-Ostholt} M.,    {Zucker} D.~B.,  2010, \mnras, 408, L26

\bibitem[\protect\citeauthoryear{{Poggio}, {Drimmel}, {Lattanzi}, {Smart},
  {Spagna}, {Andrae}, {Bailer-Jones}, {Fouesneau} \& {Antoja}}{{Poggio}
  et~al.}{2018}]{poggio18}
{Poggio} E.,  {Drimmel} R.,  {Lattanzi} M.~G.,  {Smart} R.~L.,  {Spagna} A.,
  {Andrae} R.,  {Bailer-Jones} C.~A.~L.,  {Fouesneau} M.,    {Antoja} T.,
  2018, \mnras, 481, L21

\bibitem[\protect\citeauthoryear{{Portail}, {Gerhard}, {Wegg} \&
  {Ness}}{{Portail} et~al.}{2017}]{portail17}
{Portail} M.,  {Gerhard} O.,  {Wegg} C.,    {Ness} M.,  2017, \mnras, 465, 1621

\bibitem[\protect\citeauthoryear{{Price-Whelan}, {Johnston}, {Sheffield},
  {Laporte} \& {Sesar}}{{Price-Whelan} et~al.}{2015}]{price-whelan15}
{Price-Whelan} A.~M.,  {Johnston} K.~V.,  {Sheffield} A.~A.,  {Laporte}
  C.~F.~P.,    {Sesar} B.,  2015, \mnras, 452, 676

\bibitem[\protect\citeauthoryear{{Purcell}, {Bullock}, {Tollerud}, {Rocha} \&
  {Chakrabarti}}{{Purcell} et~al.}{2011}]{purcell11}
{Purcell} C.~W.,  {Bullock} J.~S.,  {Tollerud} E.~J.,  {Rocha} M.,
  {Chakrabarti} S.,  2011, \nat, 477, 301

\bibitem[\protect\citeauthoryear{{Queiroz}, {Anders}, {Santiago}, {Chiappini},
  {Steinmetz}, {Dal Ponte}, {Stassun}, {da Costa}, {Maia}, {Crestani} \&
  {Beers}}{{Queiroz} et~al.}{2018}]{queiroz18}
{Queiroz} A.~B.~A.,  {Anders} F.,  {Santiago} B.~X.,  {Chiappini} C.,
  {Steinmetz} M.,  {Dal Ponte} M.,  {Stassun} K.~G.,  {da Costa} L.~N.,  {Maia}
  M.~A.~G.,  {Crestani} J.,    {Beers} T.~C.,  2018, \mnras, 476, 2556

\bibitem[\protect\citeauthoryear{{Quillen}, {Carrillo}, {Anders}, {McMillan},
  {Hilmi}, {Monari}, {Minchev}, {Chiappini}, {Khalatyan} \&
  {Steinmetz}}{{Quillen} et~al.}{2018}]{quillen18a}
{Quillen} A.~C.,  {Carrillo} I.,  {Anders} F.,  {McMillan} P.,  {Hilmi} T.,
  {Monari} G.,  {Minchev} I.,  {Chiappini} C.,  {Khalatyan} A.,    {Steinmetz}
  M.,  2018, \mnras, 480, 3132

\bibitem[\protect\citeauthoryear{{Quillen}, {De Silva}, {Sharma}, {Hayden},
  {Freeman}, {Bland-Hawthorn}, {{\v Z}erjal} \& {Asplund}}{{Quillen}
  et~al.}{2018}]{quillen18}
{Quillen} A.~C.,  {De Silva} G.,  {Sharma} S.,  {Hayden} M.,  {Freeman} K.,
  {Bland-Hawthorn} J.,  {{\v Z}erjal} M.,    {Asplund} M.,  2018, \mnras, 478,
  228

\bibitem[\protect\citeauthoryear{{Quillen} \& {Minchev}}{{Quillen} \&
  {Minchev}}{2005}]{quillen05}
{Quillen} A.~C.,  {Minchev} I.,  2005, \aj, 130, 576

\bibitem[\protect\citeauthoryear{{Quillen}, {Minchev}, {Bland-Hawthorn} \&
  {Haywood}}{{Quillen} et~al.}{2009}]{quillen09}
{Quillen} A.~C.,  {Minchev} I.,  {Bland-Hawthorn} J.,    {Haywood} M.,  2009,
  \mnras, 397, 1599

\bibitem[\protect\citeauthoryear{{Ramos}, {Antoja} \& {Figueras}}{{Ramos}
  et~al.}{2018}]{ramos18}
{Ramos} P.,  {Antoja} T.,    {Figueras} F.,  2018, \aap, 619, A72

\bibitem[\protect\citeauthoryear{{Reid}, {Menten}, {Brunthaler}, {Zheng},
  {Dame}, {Xu}, {Wu}, {Zhang}, {Sanna}, {Sato}, {Hachisuka}, {Choi}, {Immer},
  {Moscadelli}, {Rygl} \& {Bartkiewicz}}{{Reid} et~al.}{2014}]{reid14}
{Reid} M.~J.,  {Menten} K.~M.,  {Brunthaler} A.,  {Zheng} X.~W.,  {Dame} T.~M.,
   {Xu} Y.,  {Wu} Y.,  {Zhang} B.,  {Sanna} A.,  {Sato} M.,  {Hachisuka} K.,
  {Choi} Y.~K.,  {Immer} K.,  {Moscadelli} L.,  {Rygl} K.~L.~J.,
  {Bartkiewicz} A.,  2014, \apj, 783, 130

\bibitem[\protect\citeauthoryear{{Reyl{\'e}}, {Marshall}, {Robin} \&
  {Schultheis}}{{Reyl{\'e}} et~al.}{2009}]{reyle09}
{Reyl{\'e}} C.,  {Marshall} D.~J.,  {Robin} A.~C.,    {Schultheis} M.,  2009,
  \aap, 495, 819

\bibitem[\protect\citeauthoryear{{Sanders} \& {Das}}{{Sanders} \&
  {Das}}{2018}]{sanders18}
{Sanders} J.~L.,  {Das} P.,  2018, \mnras, 481, 4093

\bibitem[\protect\citeauthoryear{{Sawala}, {Pihajoki}, {Johansson}, {Frenk},
  {Navarro}, {Oman} \& {White}}{{Sawala} et~al.}{2017}]{sawala17}
{Sawala} T.,  {Pihajoki} P.,  {Johansson} P.~H.,  {Frenk} C.~S.,  {Navarro}
  J.~F.,  {Oman} K.~A.,    {White} S.~D.~M.,  2017, \mnras, 467, 4383

\bibitem[\protect\citeauthoryear{{Schuster}, {Moitinho}, {M{\'a}rquez},
  {Parrao} \& {Covarrubias}}{{Schuster} et~al.}{2006}]{schuster06}
{Schuster} W.~J.,  {Moitinho} A.,  {M{\'a}rquez} A.,  {Parrao} L.,
  {Covarrubias} E.,  2006, \aap, 445, 939

\bibitem[\protect\citeauthoryear{Schönrich, Binney \& Dehnen}{Schönrich
  et~al.}{2010}]{schoenrich10}
Schönrich R.,  Binney J.,    Dehnen W.,  2010, Monthly Notices of the Royal
  Astronomical Society, 403, 1829

\bibitem[\protect\citeauthoryear{{Sharma}, {Johnston}, {Majewski}, {Mu{\~n}oz},
  {Carlberg} \& {Bullock}}{{Sharma} et~al.}{2010}]{sharma10}
{Sharma} S.,  {Johnston} K.~V.,  {Majewski} S.~R.,  {Mu{\~n}oz} R.~R.,
  {Carlberg} J.~K.,    {Bullock} J.,  2010, \apj, 722, 750

\bibitem[\protect\citeauthoryear{{Sheffield}, {Price-Whelan}, {Tzanidakis},
  {Johnston}, {Laporte} \& {Sesar}}{{Sheffield} et~al.}{2018}]{sheffield18}
{Sheffield} A.~A.,  {Price-Whelan} A.~M.,  {Tzanidakis} A.,  {Johnston} K.~V.,
  {Laporte} C.~F.~P.,    {Sesar} B.,  2018, \apj, 854, 47

\bibitem[\protect\citeauthoryear{{Springel}, {Wang}, {Vogelsberger}, {Ludlow},
  {Jenkins}, {Helmi}, {Navarro}, {Frenk} \& {White}}{{Springel}
  et~al.}{2008}]{Springel2008}
{Springel} V.,  {Wang} J.,  {Vogelsberger} M.,  {Ludlow} A.,  {Jenkins} A.,
  {Helmi} A.,  {Navarro} J.~F.,  {Frenk} C.~S.,    {White} S.~D.~M.,  2008,
  \mnras, 391, 1685

\bibitem[\protect\citeauthoryear{{Tepper-Garc{\'{\i}}a} \&
  {Bland-Hawthorn}}{{Tepper-Garc{\'{\i}}a} \&
  {Bland-Hawthorn}}{2018}]{tepper18}
{Tepper-Garc{\'{\i}}a} T.,  {Bland-Hawthorn} J.,  2018, \mnras, 478, 5263

\bibitem[\protect\citeauthoryear{{Thomas}, {Laporte}, {McConnachie}, {Famaey},
  {Ibata}, {Martin}, {Starkenburg}, {Carlberg}, {Malhan} \& {Venn}}{{Thomas}
  et~al.}{2019}]{thomas19}
{Thomas} G.~F.,  {Laporte} C.~F.~P.,  {McConnachie} A.~W.,  {Famaey} B.,
  {Ibata} R.,  {Martin} N.~F.,  {Starkenburg} E.,  {Carlberg} R.,  {Malhan} K.,
     {Venn} K.,  2019, \mnras, 483, 3119

\bibitem[\protect\citeauthoryear{{Tian}, {Liu}, {Wu}, {Xiang} \&
  {Zhang}}{{Tian} et~al.}{2018}]{tian18}
{Tian} H.-J.,  {Liu} C.,  {Wu} Y.,  {Xiang} M.-S.,    {Zhang} Y.,  2018, \apjl,
  865, L19

\bibitem[\protect\citeauthoryear{Van Der~Walt, Colbert \& Varoquaux}{Van
  Der~Walt et~al.}{2011}]{numpy}
Van Der~Walt S.,  Colbert S.~C.,    Varoquaux G.,  2011, Computing in Science
  \& Engineering, 13, 22

\bibitem[\protect\citeauthoryear{{Weinberg}}{{Weinberg}}{1989}]{weinberg89}
{Weinberg} M.~D.,  1989, \mnras, 239, 549

\bibitem[\protect\citeauthoryear{{Weinberg}}{{Weinberg}}{1991}]{weinberg1991}
{Weinberg} M.~D.,  1991, \apj, 373, 391

\bibitem[\protect\citeauthoryear{{Weinberg}}{{Weinberg}}{1998}]{weinberg98}
{Weinberg} M.~D.,  1998, \mnras, 299, 499

\bibitem[\protect\citeauthoryear{{Weinberg} \& {Blitz}}{{Weinberg} \&
  {Blitz}}{2006}]{weinberg06}
{Weinberg} M.~D.,  {Blitz} L.,  2006, \apjl, 641, L33

\bibitem[\protect\citeauthoryear{{Widrow}, {Barber}, {Chequers} \&
  {Cheng}}{{Widrow} et~al.}{2014}]{widrow14}
{Widrow} L.~M.,  {Barber} J.,  {Chequers} M.~H.,    {Cheng} E.,  2014, \mnras,
  440, 1971

\bibitem[\protect\citeauthoryear{{Widrow}, {Gardner}, {Yanny}, {Dodelson} \&
  {Chen}}{{Widrow} et~al.}{2012}]{widrow12}
{Widrow} L.~M.,  {Gardner} S.,  {Yanny} B.,  {Dodelson} S.,    {Chen} H.-Y.,
  2012, \apjl, 750, L41

\bibitem[\protect\citeauthoryear{{Williams}, {Steinmetz}, {Binney}, {Siebert},
  {Enke}, {Famaey}, {Minchev}, {de Jong}, {Boeche}, {Freeman} \&
  {Bienaym{\'e}}}{{Williams} et~al.}{2013}]{williams13}
{Williams} M.~E.~K.,  {Steinmetz} M.,  {Binney} J.,  {Siebert} A.,  {Enke} H.,
  {Famaey} B.,  {Minchev} I.,  {de Jong} R.~S.,  {Boeche} C.,  {Freeman} K.~C.,
     {Bienaym{\'e}} 2013, \mnras, 436, 101

\bibitem[\protect\citeauthoryear{{Xu}, {Newberg}, {Carlin}, {Liu}, {Deng},
  {Li}, {Sch{\"o}nrich} \& {Yanny}}{{Xu} et~al.}{2015}]{xu15}
{Xu} Y.,  {Newberg} H.~J.,  {Carlin} J.~L.,  {Liu} C.,  {Deng} L.,  {Li} J.,
  {Sch{\"o}nrich} R.,    {Yanny} B.,  2015, \apj, 801, 105

\end{thebibliography}

\appendix


\section{Spatially asymmetric substructures in the $V_{\phi}-V_{R}$ \& $V_{\phi}-V_{Z}$ planes}

It is also interesting to ask whether some of the moving groups identified in the $U-V$ plane are asymmetric about the midplane. Recently, it has been suggested that the moving group Coma Berenices may have been excited by the vertical perturbations from Sagittarius \citep[see][with GALAH and RAVE respectively]{quillen18,monari18}.

In this section, we explicitly choose a different snapshots to the previous sections, so the features presented here {\it do not} coincide with those presented earlier. This is because the morphology of the $U-V$/$U-W$ planes are sensitive to the location of the SNbhd-like volume and the probability of coincidentally reproducing the same moving groups features as seen in the {\it Gaia} data is highly improbable, since the simulations are were not tuned to this effect.

In Figure A2, we look for such Coma Berenices analogues of asymmetric moving groups and identify such an instance at $t=0.4\,\rm{Gyr}$ in a volume centered around $x=-8\,\rm{kpc}$. We notice the existence of a moving group with $V_{\phi}\sim300\,\rm{km\,s^{-1}}$ which is clearly visible The structure and bulk motion of the moving group is of course {\it not} identical to Coma Berenices in shape or velocity, but it should be appreciated that this structure is reminiscent of the Galactic disc and is only present on one side of the midplane. Indeed, as Sgr orbits around the Galaxy, it excites multiple generations of vertical density waves which propagate to the outer disc \citep{gomez13,laporte18b}. More generally, neglecting the self-gravity of the disc, as a result of the perturbations (through DM halo wake and tides from the Sgr dSph body), stars are expected to be simultaneously excited moving on new orbits which then undergo phase-mixing \citep{delavega15}. This holds remarkably well in the outer disc where the self-gravity is negligible and where disc perturbations seeded by Sgr are able to produce ACS/EBS-like stream structures (``feathers") \citep{laporte18c}. It could be that Coma Berenices would be an analogue of the ACS but in the solar-neighbourhood. Chemical tagging of the moving group would be an interesting pursuit to uncover its origin in the disc.

\begin{figure}
\includegraphics[width=0.5\textwidth,trim=0mm 0mm 0mm 0mm, clip]{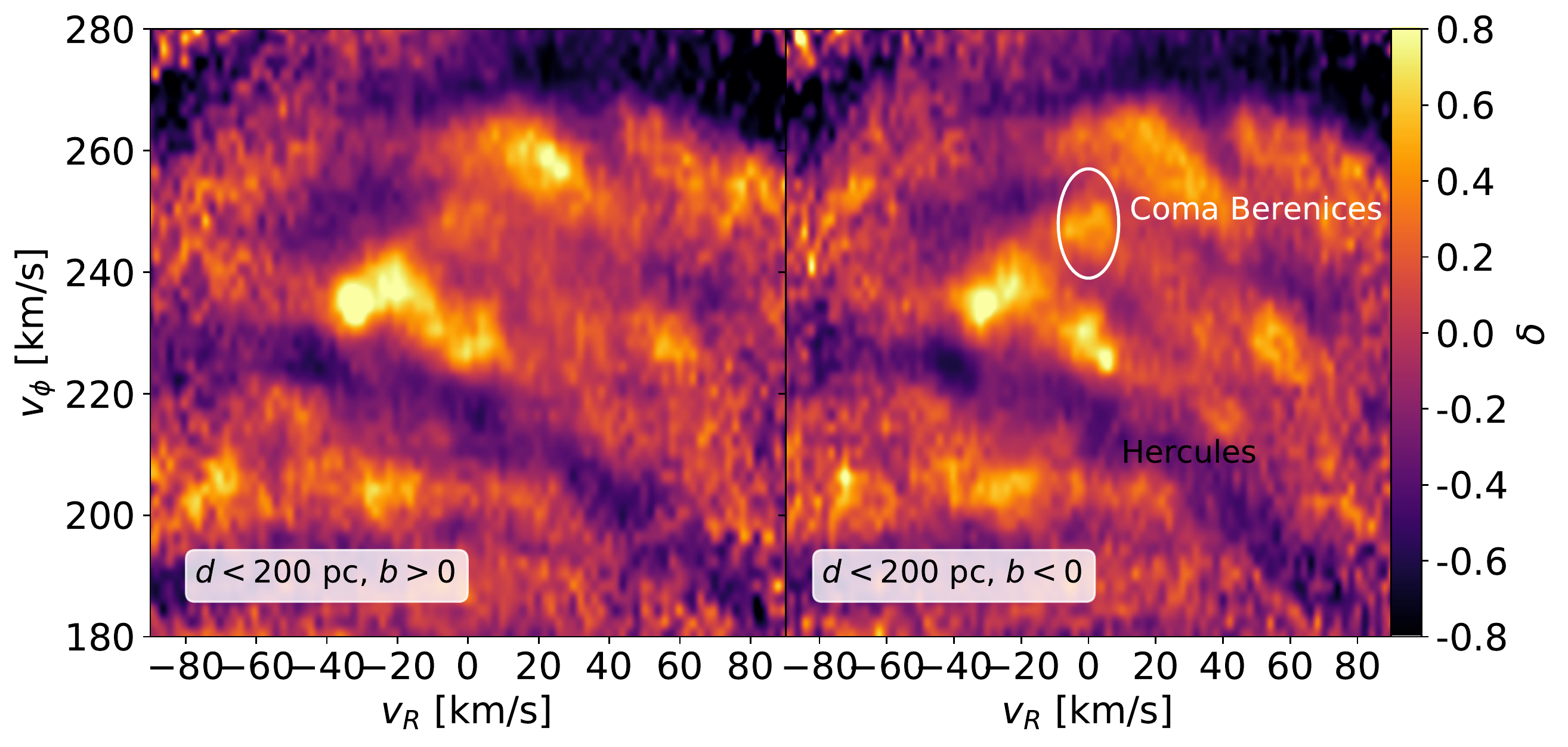}
\includegraphics[width=0.5\textwidth,trim=0mm 0mm 0mm 0mm, clip]{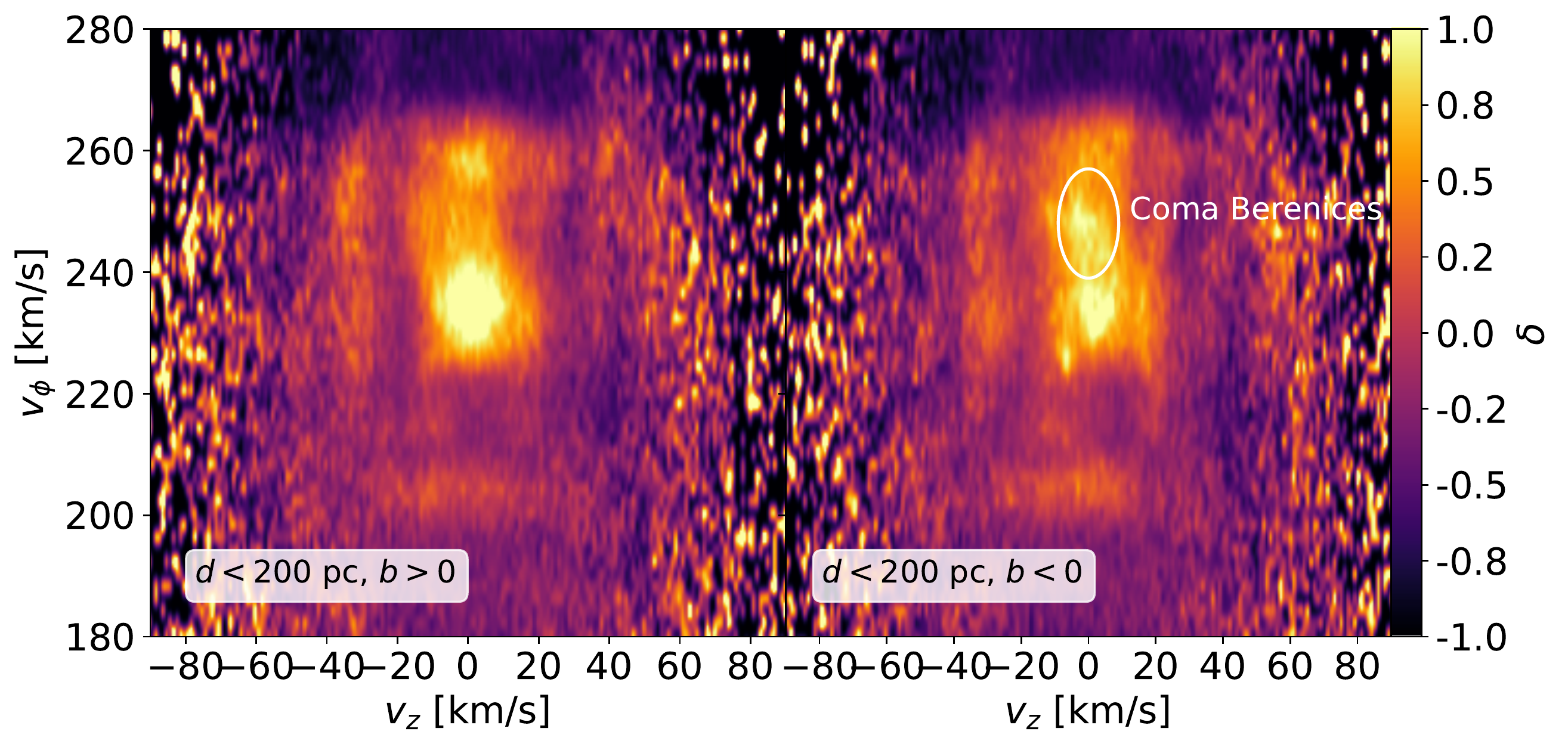}

\caption[]{Gaia DR2 $v_{\phi}-v_{R}$, $v_{\phi}-v_{Z}$ planes for stars selected within $d<200\,\rm{pc}$ around the sun, above and below the midplane. As shown already by \cite{monari18}, Coma Berenices appear more prominent in the south.}
\end{figure}

\begin{figure}
\includegraphics[width=0.5\textwidth,trim=0mm 0mm 0mm 0mm, clip]{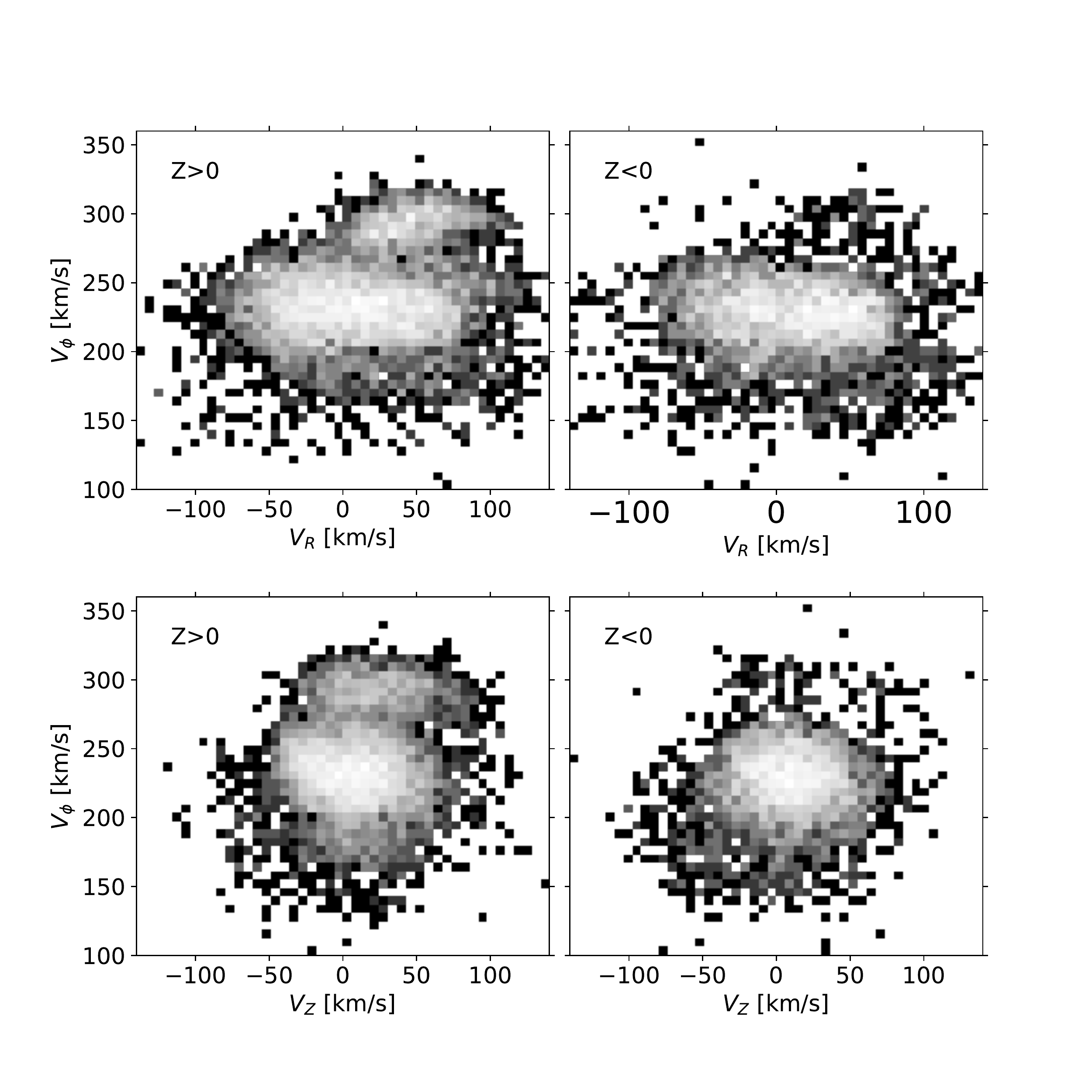}
\caption[]{Simulated $v_{\phi}-v_{R}$, $v_{\phi}-v_{Z}$ planes for stars selected at $t=0.4\,\rm{Gyr}$ in a solar neighbourhood-like region in a cylindrical wedge centered around the Sun with $R_{cyl}<1$ and $150^{\circ}<\phi<210^{\circ}$, above, below the midplane and using all the stars. We also make the distinction between these planes above the midplane and below it (left and middple panels respectively).}
\end{figure}

\label{lastpage}
\end{document}